\documentclass[12pt,preprint]{aastex}
\shorttitle{2-D stellar evolution code}
\newcommand{\p}[2]{\frac{\partial #1}{\partial #2}}
\newcommand{\pl}[2]{{\partial #1}/{\partial #2}}
\newcommand{\pp}[2]{\frac{\partial^2 #1}{\partial #2^2}}
\newcommand{\dd}[2]{\frac{d^2 #1}{d #2^2}}
\newcommand{\od}[2]{\frac{d #1}{d #2}}
\newcommand{\s}[1]{\mbox{\scriptsize{#1}}}
\newcommand{\vb}[1]{{\bf #1}}

\newcommand{\h}[1]{{\cal#1}}

\newcommand{\be}{\begin{equation}}
\newcommand{\ee}{\end{equation}}
\newcommand{\ba}{\begin{eqnarray}}
\newcommand{\ea}{\end{eqnarray}}
\newcommand{\Ba}{\begin{eqnarray*}}
\newcommand{\Ea}{\end{eqnarray*}}

\newcommand{\pa}[2]{\left(\frac{\partial #1}{\partial #2}\right)_m}

\newcommand{\pb}[2]{\left(\frac{\partial #1}{\partial #2}\right)_r}

\newcommand{\nob}{\nonumber\\}

\begin{document}

\title{2-D Stellar Evolution Code Including Arbitrary Magnetic Fields. I. 
Mathematical Techniques and Test Cases}

\author{L. H. Li \altaffilmark{1}, P. Ventura\altaffilmark{2}, 
S. Basu\altaffilmark{1}, S. Sofia\altaffilmark{1}, and P. Demarque\altaffilmark{1}}
\altaffiltext{1}{Department of Astronomy, Yale University, P.O. Box 208101, New Haven, CT 06520-8101}
\email{li@astro.yale.edu} 
\altaffiltext{2}{INAF, Osservatorio Astronomico di Roma, 00040 Monteporzio Catone (RM), Italy}

\begin{abstract}
A high-precision two-dimensional stellar evolution code has been 
developed for studying solar variability due to structural changes produced 
by varying internal magnetic fields of arbitrary configurations. Specifically, 
we are interested in modeling the effects of a dynamo-type field on the detailed
internal structure and on the global parameters of the Sun.    
The high precision is required both to model very small solar changes (of order of $10^{-4}$)
and short time scales (or order of one year). It is accomplished by
using the mass coordinate to replace the radial coordinate, by using fixed and adjustable 
time steps, a realistic stellar atmosphere, elements diffusion, and by adjusting the grid 
points. We have also built into the code the potential to subsequently include 
rotation and turbulence. The current code has been tested for several cases, 
including its ability to reproduce the 1-D results.
\end{abstract}

\section{Introduction}\label{sect:s1}

Modern standard solar models are known to yield the  solar structure 
to an amazing degree of precision (see e.g., Guenther \&\ Demarque 
1997; Basu, Pinsonneault \& Bahcall 2000; Winnick et al. 2002). 
These models, however, cannot explain the solar cycle, and other 
solar-cycle related variability. The reason for this shortcoming 
is that these models do not include the dynamo magnetic fields 
and relevant temporal variability. 

Following the suggestion by Sofia et al. (1979) that any change 
in the solar luminosity, $L$, must be accompanied by a change in 
the radius, $R$, a number of theoretical investigations have 
attempted to establish the relationship between these changes 
(denoted as $W=\Delta\ln R/\Delta\ln L$), by including internal 
processes designed to mimic the effects of dynamo fields.
We classify them into three broad categories:
\begin{description}
\item{(i)} Perturbation calculation. See Endal et al (1985) for 
a review of the early work, and Balmforth et al. (1996) 
for subsequent work.
\item{(ii)} Approximation analysis. See Spruit (1991, 2000) 
for reference.
\item{(iii)} Stellar evolution with magnetic fields. This method was 
initiated by Lydon \&\ Sofia (1995), updated by Li \&\ Sofia (2001), 
generalized to include turbulence by Li et al. (2002), and further 
generalized to include the interaction between turbulence and 
magnetic fields by Li et al. (2003).
\end{description}

The first two are illustrative, but not conclusive. The third can 
model the effects of arbitrary magnetic field configurations. 
Li et al. (2003) attempted to produce the observed cycle variations 
of all global solar parameters and the p-mode oscillation frequencies. 
The result is promising (e.g., Sofia et al. 2005), but it is not 
final both because the one-dimensional approximation is 
utilized, and because not all global parameter data exist 
for the same time span. The 1-D approximation only allows 
us to use a shell-like magnetic field configuration. This 
approximation is relatively limiting. For example, 
in one-dimensional codes the energy flux can only advance 
to the surface by penetrating the magnetic 
field shell. If the magnetic field were toroidal, like most 
dynamo models require, energy flow could circumvent the field.
The aim of this paper is to describe a mathematical technique 
that can model arbitrary magnetic field configurations by generalizing 
our one-dimensional technique into the two-dimensional case.

In order to match the observed variations of solar global 
parameters and helioseismic frequencies, two-dimensional solar 
models should fulfill at least the following precision requirements:
\begin{description}
 \item{(a)} A luminosity resolution equal or better than $10^{-2}\%$ 
per year because the observed cyclic variation of total solar 
irradiance is about $0.1\%$ per cycle,
 \item{(b)} A radius resolution equal or better than $10^{-5}\%$ 
per year  because the observed cyclic variation 
of solar radius may be as small as $10^{-4}\%$ per cycle,
 \item{(c)} A realistic atmosphere model because the helioseismic 
frequencies are sensitive to it, 
 \item{(d)} Suitable boundary conditions because the model is sensitive to them,
 \item{(e)} Elements diffusion because the helioseismic frequencies are 
sensitive to composition,
 \item{(f)} A magnetic field because there is no cyclic variation 
without magnetic field,
 \item{(g)} Turbulence because helioseismic observations require it,
 \item{(h)} The interaction between turbulence and magnetic fields 
because helioseismic observations require it.
\end{description}
Our one-dimensional code, which is based on the Yale Stellar 
Evolution Code YREC (Guenther et al. 1992), meets all these 
requirements, a nontrivial accomplishment. It is difficult 
to modify the other existing two- or three-dimensional codes 
(e.g., Deupree 1990; Turcotte et al. 2001) since each of them 
was developed with specific objectives not requiring this degree 
of precision.

We attempted to include magnetic fields in Deupree's two-dimensional 
stellar evolution code (1990), but we were unable to compare the model results 
with solar observations and our one-dimensional results, probably because
\begin{description}
 \item{(1)} the two-dimensional model has different 
center and surface boundary conditions than the one-dimensional model,
 \item{(2)} the two-dimensional model does not include an atmosphere model,
 \item{(3)} the numerical accuracy is not high enough 
to match the solar observations.
\end{description}
This experience convinced us that it would be easier to develop 
a high precision two-dimensional stellar structure and evolution 
code by straightforward generalizing our one-dimensional code 
rather than modifying an existing two-dimensional code. Our 
experience shows that this conviction was well founded.

The highest precision requirement is that the cyclic variation of solar 
radius should be better than $10^{-5}\%$ per year because the observed cyclic 
variation of solar radius may be as small as $10^{-4}\%$ per cycle.
There are various uncertainties in the input physics (e.g., Boothroyd \&\ Sackmann 2003; 
Sackmann \&\ Boothroyd 2003). Though these uncertainties affect the interior structure 
of the Sun, they have little influence on the cyclic variations of solar global 
parameters such as solar radius, solar luminosity, and solar effective temperature 
because of calibration and subtraction of the same parameter at two different times, 
which remove various possible uncertainties in 
the cyclic variations of global solar parameters. Such a high precision for the cyclic 
variations of global solar parameters is thus achievable.

We outline here the basic schematic of the method in order to prevent 
the readers from getting lost in the detailed derivations.

As it is a common practice, the starting points are the 
conservation laws of mass, momentum, energy, and composition, 
as well as the Newtonian universal gravitational law. Both 
momentum conservation equations and the Poisson equation 
are second-order differential equations. We use the 
radiation transport equation to relate the temperature 
gradient to the energy flux in the radiative zone, and 
use the mixing length theory to calculate the temperature 
gradient in the convective zone. We include magnetic fields in 
this paper, and include in the code the potential to subsequently 
include turbulence and rotation.

The main relation is the coordinate transformation 
from the radial coordinate $r$ to the mass coordinate $m$. 
Regarding mass, we should specify the spatial range that 
the mass occupies. We use the equipotential surface $S_\Phi$ 
on which
\be
  \Phi(r,\theta;t) = \Phi_c  \label{eq:cphi}
\ee
to indicate the spatial range, where we have assumed that the 
system is azimuthally-symmetric or axi-symmetric and that 
$\Phi_c$ may vary with time. The time 
coordinate $t$ is taken as a parameter. Solving Eq.~(\ref{eq:cphi}) 
for $r$, we obtain the equipotential surface:
\be
  r=R(\Phi_c,\theta;t). \label{eq:rr}
\ee
This equipotential surface encloses volume $V_{\Phi}$, 
which is defined by
\be
  V_{\Phi}\equiv \left\{ \begin{array}{l}
      \phi \in [0,2\pi] \\
      \theta\in[0,\pi] \\
      r\in[0,R(\Phi_c,\theta;t)]. \\
 \end{array} \right.
\ee
The mass contained in $V_{\Phi}$ is defined by
\ba
m &=& m(\Phi_c;t)\equiv \int^{V_{\Phi}} \rho dV =\int_0^{2\pi}d\phi\int_0^\pi d\theta\sin\theta
\int_0^{R(\Phi_c,\theta;t)}ds\rho(s,\theta;t)s^2 \nonumber \\
&=& 2\pi \int_0^\pi d\theta\sin\theta\int_0^{R(\Phi_c,\theta;t)}ds \rho(s,\theta;t)s^2, \label{eq:mmass}
\ea
where $\rho=\rho(r,\theta;t)$ is the density. Solving Eq.~(\ref{eq:mmass}) for $\Phi_c$, we obtain
\be
    \Phi_c = \Phi_c(m;t). \label{eq:phim}
\ee
Substituting Eq.~(\ref{eq:phim}) into (\ref{eq:rr}), we obtain 
the coordinate transformation relation from $(r,\theta;t)$ to $(m,\theta;t)$:
\be
r=R(\Phi_c(m;t),\theta;t)=r(m,\theta;t), \hspace{5mm} \theta=\theta, 
\hspace{5mm} t=t. \label{eq:r2m}
\ee
For any dependent variable $X$, 
for example, $P$, $T$, $F_r$, or $\rho$, we have
\be
 \pa{X}{\theta} = \left(\p{X}{\theta}\right)_r + \left(\p{X}{r}\right)_\theta \pa{r}{\theta}.
\ee

In order to achieve a high precision that is comparable to the one-dimensional 
solar model in the two-dimensional case, using limited computational resources, 
we cannot directly numerically solve those conservation equations and the Poisson 
equation. For example, even in the hydrostatic case, we have five dependent 
variables such as  pressure ($P$), temperature ($T$), radius ($r$), gravitational potential ($\Phi$), and flux 
($F_r$ or $L=4\pi r^2 F_r$). The coefficient matrix of the linearized 
difference equations with grids $M\times N$ has $\h{N}=5MN\times5MN$ 
elements, where $M$ ($N$) is the number of grid points for the mass 
(co-latitude) coordinate. The one-dimensional solar model has $M\ge 2000$. 
If we take $N=20$, we obtain $\h{N}\ge 4\times 10^{10}$. Since $2^{32}=4\times 1024^3$, 
a 32-bit computer can handle only $2\times 1024^3\sim 2\times 10^9$ elements, noting that 
1 bit is used to represent the sign of a number. Of course, a 64-bit computer does not 
impose such constraint, but the computation speed will become an obstacle.

Analytical solutions are accurate, but such solutions are hard to obtain in the general case.
The 1-D case is accurate because we do not need to numerically solve the second-order 
Poisson equation for the gravitational potential $\Phi_0$. It is well-known 
that the gravitational acceleration in the spherically symmetric case is 
\be
g=d\Phi_0/dr=Gm/r^2. \label{eq:g}
\ee
In order to take similar advantage in the 2-D case, we show in the paper that 
Eq.~(\ref{eq:g}) can be generalized as follows
\be
\p{\Phi}{r} = \frac{Gm}{r^2} + 2\pi G r(\rho-\rho_m)-\frac{\cot\theta}{2r}\p{\Phi}{\theta} +O(2), \label{eq:gg}
\ee
where $O(2)$ represents the much smaller correction than the retained, and $\rho_m$ is defined by
\be
  \rho_m(m,\theta;t) \equiv \frac{1}{2r^2}\int_0^\pi d\theta R^2(\Phi_c,\theta;t)
\rho(R(\Phi_c,\theta;t),\theta;t)\sin\theta.
\ee
Like Eq.~(\ref{eq:g}) in 1-D case, Eq.~(\ref{eq:gg}) substantially simplifies 
the 2-D stellar structure equations.

In the two-dimensional case, the radial component of the energy flux 
vector $\vb{F}$, $F_r$, and the $\theta$-dependent luminosity, 
$L\equiv 4\pi r^2 F_r(r,\theta;t)$, are equivalent to each other, but 
the actual luminosity $L^*$ is different than the 
$\theta$-dependent luminosity $L$ because
\be
  L^*\equiv 2\pi \int_0^\pi r^2 F_r(r,\theta;t) \sin\theta d\theta.
\ee

The basic equations are described in section \ref{sec:be}, and then the 
coordinate transformation from the radial coordinate to the mass 
coordinate is performed in section \ref{sec:ct}. Various possible 
magnetic field configurations are converted into suitable expressions 
that appear in the stellar structure equations in section \ref{sec:mf}. Boundary 
conditions are equally important. So we use a whole section 
(section \ref{sec:bc}) to elaborate them. The method of solution is 
detailed in section \ref{sec:ms}. The coefficient matrix and input physics 
used in section \ref{sec:ms} are presented in Appendix A and B, 
respectively. The evolution sequences without any magnetic 
field and with a shell-like magnetic field are presented in 
section \ref{sec:ssm} and section \ref{sec:smf} to test the method.

\section{Basic equations}\label{sec:be}

The basic equations consist of the time-dependent conservation 
laws of mass, momentum, energy, and composition, the Poisson 
equation (Deupree 1990), as well as the radiative transfer 
equation (Unno \&\ Spiegel 1966):
\begin{mathletters}
\ba
 &&  \p{\rho}{t}+\nabla\cdot(\rho\vb{v}) = 0, \label{eq:mass} \\
 &&  \rho\od{\vb{v}}{t} = -\nabla P
     -\rho\nabla\Phi+\frac{1}{4\pi}(\nabla\times\vb{B})\times\vb{B},  \label{eq:momentum} \\
 &&  \rho T\od{S_T}{t} = \rho\epsilon - \nabla\cdot \vb{F}_{\s{rad}},    \label{eq:energy} \\
 &&  \od{\rho_i}{t} = Q_i, \label{eq:composition} \\
 &&  \nabla^2\Phi=4\pi G\rho, \label{eq:poisson} \\
 &&  \nabla\cdot\vb{F}_{\s{rad}} = -4\kappa\rho(J-B), \label{eq:transfer}
\ea
\end{mathletters}
where
$\vb{v}$ is the velocity of a fluid element, $\vb{B}$ the magnetic field, 
$\epsilon$ the nuclear energy generation rate per unit mass, $\vb{F}_{\s{rad}}$ 
the radiative energy flux, $\rho_i$ the density of species $i$, $Q_i$ the 
creation rate of species $i$, $G$ the universal gravitational constant, 
$J$ the mean radiative intensity, $\kappa$ the absorption coefficient, and $B$ the 
Kirchhoff-Planck function. The total derivative is 
defined by $d/dt\equiv\partial/\partial t+\vb{v}\cdot\nabla$.

The specific entropy $S_T$ includes both nonmagnetic and magnetic 
components, as shown in the first law of thermodynamics (Callen 1996; 
Lydon \&\ Sofia 1995):
\be
   TdS_T = dU + PdV - d\chi,
\ee
where $U$ is the nonmagnetic specific internal energy, $V=1/\rho$ 
is the specific volume, $\chi=|\vb{B}|^2/8\pi\rho$ is the specific 
magnetic energy, $P$ is the nonmagnetic pressure. Since the 
magnetic work $d\chi$ is taken from the nonmagnetic internal 
energy, the total internal $U_T$ energy decreases:
\be
   U_T = U-\chi.
\ee
The isotropic magnetic pressure component $P_m$ can be 
expressed by $\chi$ and $\rho$:
\be
   P_m = \chi\rho.
\ee
The total isotropic pressure component $P_T$ can thus be 
defined as follows:
\be
  P_T = P + P_m.
\ee
Using $P_T$, $T$ and $\chi$ as independent thermodynamic 
variables, the equation of state and the first law of 
thermodynamics read (Lydon \&\ Sofia 1995):
\begin{mathletters}
\ba
   d\rho/\rho &=& \alpha dP_T/P_T -\delta dT/T - \nu d\chi/\chi, \label{eq:state} \\
   TdS_T &=& C_P dT -(\delta/\rho)dP_T + (P_T\delta\nu/P_m\alpha)d\chi, \label{eq:1stlaw}
\ea
\end{mathletters}
where
\begin{mathletters}
\ba
&&  \alpha\equiv (\partial\ln\rho/\partial\ln P_T)_{T,\chi;t}, 
  \hspace{5mm} \delta\equiv -(\partial\ln\rho/\partial\ln T)_{P_T,\chi;t}, \\
&& \nu \equiv (\partial\ln\rho/\partial\ln\chi)_{P_T,T;t},
  \hspace{5mm} C_P \equiv (\partial U_T/\partial T)_{P_T,\chi;t}.
\ea
\end{mathletters}
From the first law of thermodynamics (Eq.~\ref{eq:1stlaw}), 
we can define two adiabatic gradients, one fixes the specific 
magnetic energy:
\be
  \nabla_{\s{ad}} \equiv \left(\p{\ln T}{\ln P_T}\right)_{S_T,\chi} = \frac{P_T\delta}{\rho C_P T}, \\
\ee
another does not fix the specific magnetic energy:
\be
  \nabla'_{\s{ad}} \equiv \left(\p{\ln T}{\ln P_T}\right)_{S_T} = \nabla_{\s{ad}}(1-\nu \nabla_{\chi}/\alpha),
\ee
where the magnetic energy gradient $\nabla_\chi$ are defined as follows
\be
  \nabla_\chi \equiv \p{\ln\chi}{\ln P_T}.
\ee

In order to close the radiative transfer equation (Eq.~\ref{eq:transfer}), we use the 
Eddington approximation (Unno \&\ Spiegel 1966):
\be
   \vb{F}_{\s{rad}} = -\frac{4\pi}{3\kappa\rho}\nabla J. \label{eq:eddington}
\ee

Unlike Deupree (1990), we do not directly solve these equations. We first perform some 
analytic work to make some approximations in advance.

\subsection{Mass conservation equation}

Deupree (1990) uses the constancy of the total mass during the model 
evolution to determine the radius at the equator. In contrast, we want to determine 
the equipotential surface $S_\Phi$: $r=R(\Phi_c,\theta;t)=r(m,\theta;t)$, 
as in the one-dimensional case.

Mass conservation can be expressed by either Eq.~(\ref{eq:mmass}) or its differential form:
\be
   \p{m}{r}=\p{m}{R}=4\pi r^2(m,\theta;t) \rho_m(m,\theta;t), \label{eq:mass1}
\ee
where
\be
  r^2(m,\theta;t)\rho_m(m,\theta;t) \equiv \frac{1}{2}\int_0^\pi d\theta R^2(\Phi_c,\theta;t)
\rho(R(\Phi_c,\theta;t),\theta;t)\sin\theta=f(m;t).
\ee
It should be pointed out that in general,
\be
  \rho_m(m,\theta;t) \ne \rho(R(\Phi_c,\theta;t),\theta;t).
\ee
Nevertheless, in the spherically-symmetric case, $\rho_m(m;t)$ is indeed equal to $\rho(R(\Phi_c);t)$.
Since $f(m;t)$ is an integral, the two-dimensional case is much more complicated (i.e., nonlocal) than its 
one-dimensional counterpart (local). This complexity may be the price we have to pay from one dimension 
to two dimensions.

\subsection{Gravitational acceleration}

We want to show here that the last two terms [excluding $O(2)$] 
on the right hand side of Eq.~(\ref{eq:gg}) are due to the 2-D 
corrections to the gravitational acceleration. To this end we 
should start from the Poisson equation, Eq.~(\ref{eq:poisson}), 
which can be expanded as follows in the spherical polar coordinate system:
\be
\frac{1}{r^2}\p{}{r}\left(r^2\p{\Phi}{r}\right) 
+ \frac{1}{r^2\sin\theta}\p{}{\theta}\left(\sin\theta\p{\Phi}{\theta}\right) 
= 4\pi G \rho, \label{eq:poisson1}
\ee 
where we have assumed that $\Phi=\Phi(r,\theta;t)$ does not vary with 
the $\phi$ coordinate. We expand $\Phi$ around its spherically-symmetric state:
\be
  \Phi(r,\theta;t) = \Phi_0(r;t) + \delta\Phi(r,\theta;t), \label{eq:phi}
\ee
where $\delta\Phi$ is a small correction, and
\be
  \p{\Phi_0}{r} = \frac{Gm}{r^2}. \label{eq:phi0}
\ee
Substituting Eqs.~(\ref{eq:phi}-\ref{eq:phi0}) into Eq.~(\ref{eq:poisson1}), we obtain
\be
  \p{\Phi}{r} = \frac{Gm}{r^2} +2\pi G r (\rho-\rho_m) -\frac{\cot\theta}{2r}\p{\Phi}{\theta} +O(2), \label{eq:phi2} 
\ee
where
\be
  O(2) = -\frac{r}{2}\left(\pp{\delta \Phi}{r} + \frac{1}{r^2}\pp{\Phi}{\theta}\right).
\ee

\subsection{Momentum conservation equation}

Generally, we can decompose the total velocity $\vb{v}$ in the basic equations into three components:
\be
  \vb{v}=\vb{V}_0 + \vb{V}_{\s{rot}} + \vb{v}', 
\ee
where $\vb{V}_0$ is a secular evolution velocity, $\vb{V}_{\s{rot}}$ the rotation velocity, 
and $\vb{v}'$ the turbulent convection velocity. We neglect the secular 
expansion and rotation velocity components in the momentum conservation, i.e., we assume
\be
  \vb{v} = \vb{v}' \label{eq:turb}
\ee
in Eq.~(\ref{eq:momentum}). We checked in the one-dimensional case that the term $dV_0/dt$ in 
the momentum equation is negligible. Substituting Eq.~(\ref{eq:turb}) into (\ref{eq:momentum}) 
and averaging the resultant equation over the time $t$ and azimuthal angle $\phi$, we obtain
\be
 \rho\nabla\overline{{v'}^2}  = -\nabla P
     -\rho\nabla\Phi+\frac{1}{4\pi}(\nabla\times\vb{B})\times\vb{B}, \label{eq:momentum1}
\ee
where $\overline{{v'}^2}=\overline{{v'_x}^2} + \overline{{v'_y}^2}
+ \overline{{v'_z}^2}$ is computed by solving the basic equations in the 
three-dimensional convection simulations of the outer layers of the Sun 
(Robinson et al. 2003), in which the average is taken over the time $t$ and 
the horizontal coordinates $x$ and $y$ in a sample box. We have shown how to 
include turbulence in the one-dimensional case (Li et al. 2002). We neglect the 
turbulent contribution to the momentum equation here so as to stress 
the two-dimensional effects due to magnetic fields, i.e., we simply 
set
\be
\overline{{v'}^2}=0  \label{eq:assumption1}
\ee
in this paper.

We assume that the system is azimuthally symmetric. Under this assumption, the vector 
equation~(\ref{eq:momentum1}) is equivalent to the following two scalar equations:
\begin{mathletters}
\ba
   \p{P_T}{r} &=& -\rho \p{\Phi}{r} + \h{H}_r, \label{eq:pr}\\
   \frac{1}{r}\p{P_T}{\theta} &=& -\frac{\rho}{r} \p{\Phi}{\theta} + \h{H}_\theta,\label{eq:pa}
\ea
\end{mathletters}
where $P_T=P+P_m$ is the total pressure, including the magnetic pressure $P_m=B^2/8\pi$, and
\be
  \h{H} \equiv \frac{1}{4\pi} (\vb{B}\cdot\nabla)\vb{B}. \label{eq:hh}
\ee
Noticing that
\be
\frac{1}{4\pi}(\nabla\times\vb{B})\times\vb{B} = -\nabla\left(\frac{B^2}{8\pi}\right) 
   + \frac{1}{4\pi}(\vb{B}\cdot\nabla)\vb{B}.
\ee

In the one-dimensional case, we have only a single scalar equation 
to describe the momentum conservation, i.e., Eq.~(\ref{eq:pr}). 
In contrast, we need three scalar equations for the momentum conservation in the 
two-dimensional case, i.e., Eqs.~(\ref{eq:phi2}), (\ref{eq:pr}) 
and (\ref{eq:pa}). It would be much better if we can combine 
these three equations into a single scalar equation. Fortunately, we can. For this end, 
solving Eq.~(\ref{eq:pa}) for $\partial\Phi/\partial\theta$, we obtain
\be
  \p{\Phi}{\theta} = - \frac{1}{\rho}\p{P_T}{\theta} +\frac{r}{\rho}\h{H}_\theta. \label{eq:phia}
\ee
Then substituting this into Eq.~(\ref{eq:phi2}), we obtain
\be
  \p{\Phi}{r} = \frac{Gm}{r^2} + 2\pi G r(\rho-\rho_m) 
   +\frac{\cot\theta}{2r\rho}\p{P_T}{\theta} -\frac{\cot\theta}{2\rho}\h{H}_\theta +O(2). \label{eq:phir}
\ee
Substituting Eq.~(\ref{eq:phir}) into Eq.~(\ref{eq:pr}), we finally obtain
\be
\p{P_T}{r} = -\frac{Gm\rho}{r^2} +\h{H}_r -2\pi G r\rho(\rho-\rho_m) 
   -\frac{\cot\theta}{2r}\p{P_T}{\theta} +\frac{1}{2}\h{H}_\theta\cot\theta +O(2). \label{eq:pt}
\ee
This is our momentum conservation equation. The last three r.h.s. 
terms represent the two-dimensional effects.

\subsection{Energy conservation equation}

The energy conservation equation (Eq.~\ref{eq:energy}) depends 
on the velocity in the total derivative:
\be
   \od{S_T}{t}= \p{S_T}{t} +(\vb{V}_0+\vb{v}')\cdot\nabla S_T.
\ee
The secular expansion velocity $\vb{V}_0$ cannot be neglected, 
and from now on we define
\be
  \od{S_T}{t}\equiv \p{S_T}{t} + \vb{V}_0\cdot\nabla S_T.
\ee
The statistical average of $\rho T\vb{v}'\cdot\nabla S_T$, 
namely $<\rho T\vb{v}'\cdot\nabla S_T>$, will determine the 
divergence of the convective flux $\vb{F}_{\s{conv}}$:
\be
   \nabla\cdot\vb{F}_{\s{conv}} \equiv <\rho T \vb{v}'\cdot\nabla S_T>. \label{eq:fconv}
\ee
By defining the total energy flux to be the sum of 
both the convective and radiative flux, 
$\vb{F}=\vb{F}_{\s{rad}}+\vb{F}_{\s{conv}}$, Eq.~(\ref{eq:energy}) becomes
\be
   \nabla\cdot\vb{F} = \rho\left(\epsilon-T\od{S_T}{t}\right), \label{eq:energy1}
\ee
where
\be
    T\od{S_T}{t} = C_PT \left[\od{\ln T}{t} 
      - \nabla_{\s{ad}}\left(1-\frac{\nu\nabla_{\chi}}{\alpha}\right)\od{\ln P_T}{t}\right].
\ee

In the azimuthal case, Eq.~(\ref{eq:energy1}) is equivalent to the following equation
\be
  \frac{1}{r^2}\p{(r^2F_r)}{r} = \rho\left(\epsilon-T\od{S}{t}\right) 
    -\frac{1}{r\sin\theta}\p{(\sin\theta F_\theta)}{\theta}. \label{eq:energy2}
\ee
We will work out both the radial flux component $F_r$ and polar 
flux component $F_\theta$ in the next subsections.

\subsection{Energy transport by radiation}

The radiative flux is given by Eq.~(\ref{eq:eddington}), 
in which the mean radiative intensity $J$ is governed by 
the radiative transfer equation (Eq.~\ref{eq:transfer}). 
The Plank function $B$ is known:
\be
  B = \frac{ac}{4\pi} T^4, \label{eq:plank}
\ee
where $a$ is the radiative constant, and $c$ 
the speed of light in vacuum. In stellar interior, 
local thermodynamic equilibrium is a good 
approximation, which leads to
\be
J\approx B = \frac{ac}{4\pi} T^4. \label{eq:local}
\ee

The more accurate solution of Eqs.~(\ref{eq:transfer}) and (\ref{eq:eddington}) is (see Unno \&\ Spiegel 1966):
\be
  J= B+\frac{l_p^2}{3}\nabla^2 B+\frac{l_p^4}{5}\nabla^4B+\cdots, \label{eq:jj}
\ee
where $l_p=1/\kappa\rho$ is the mean free path of photons.
Since
\be
  \nabla^2 B = \frac{1}{r^2}\p{}{r}\left(r^2\p{B}{r}\right), \label{eq:nb}
\ee
using Eq.~(\ref{eq:plank}) in Eq.~(\ref{eq:nb}), we obtain
\be
  \nabla^2 B = 4\nabla_s\left(4\nabla_s-1+\alpha-\delta\nabla_s
   +\p{\ln\nabla_s}{\ln P_T}-\frac{4\pi r^2\ H_P\rho_m}{m}\right)\frac{B}{H_P^2},
   \label{eq:nb2}
\ee
where $H_P\equiv -dr/d\ln P_T=P/\rho g$ is the pressure scale height 
and $\nabla_s$ is the actual temperature gradient. 
Substituting Eq.~(\ref{eq:nb2}) 
into Eq.~(\ref{eq:jj}), we obtain the mean radiative intensity 
that goes beyond the local thermodynamic equilibrium approximation with one more term correction:
\be
  J = \left(1+\lambda_0\frac{l_p^2}{H_P^2}\right)B,
\ee
where
\be
  \lambda_0 = \frac{4}{3}\nabla_s\left(4\nabla_s-1+\alpha-\delta\nabla_s 
    +\p{\ln\nabla_s}{\ln P_T}\right).
\ee
We want to note that the term $(4\pi r^2 H_P\rho_m/m)(l_p^2/H_P^2)\ll 1$ 
is negligible in the whole star.
Using this solution in Eq.~(\ref{eq:eddington}), we obtain 
\be
   \vb{F}_{\s{rad}} = - \frac{4acT^3}{3\kappa\rho}(1+\lambda)\nabla T, \label{eq:frad}
\ee
where
\be
 \lambda\equiv \lambda_0  \left\{1-\frac{1}{2}\left(\p{\ln\kappa}{\ln T}\right)_{P_T}-\frac{1}{2}\frac{1}{\nabla_s}\left[1+\left(\p{\ln\kappa}{\ln P_T}\right)_T-\frac{2H_P}{r}\right]+\frac{1}{4}\p{\ln\lambda_0}{\ln T}\right\}\frac{l_p^2}{H_P^2}.
 \label{eq:ll0}
\ee
Since $l_p$ is much smaller than $H_P$ 
in the optically-thick region, we know $\lambda\approx 0$ so that 
Eq.~(\ref{eq:frad}) reduces to the widely-used approximation expression 
without $\lambda$. However, $l_p$ can be comparable to or larger than $H_P$ near the surface, 
the correction factor $\lambda$ cannot be neglected.

\subsection{Energy transport by convection}

Without solving the turbulent convection problem, Eq.~(\ref{eq:fconv}) only tells 
us that the convective flux may depend on the convective velocity $v_{\s{conv}}$ 
and the entropy $S_T$, where the convective velocity $v_{\s{conv}}$ has only the 
statistical meaning. We use the mixing-length theory to obtain an analytic 
expression for $\vb{F}_{\s{conv}}$ in terms of $v_{\s{conv}}$ and $S_T$ 
(e.g., Stix 1989; Lydon \&\ Sofia 1995). Since the convective velocity has 
only the statistical meaning we assume that the turbulent 
convection is isotropic so that $\vb{F}_{\s{conv}}$ 
depends on the amplitude of the convective velocity, $v_{\s{conv}}$:
\be
   \vb{F}_{\s{conv}} = -\frac{1}{2}\rho T l_m f(v_{\s{conv}})\nabla S_T, \label{eq:vfconv}
\ee
where $f(v)$ will be determined by the mixing-length theory, $l_m$ is 
the mixing length. It is well-known that $f(v)=v$ when the radiative loss of 
the convective element and the magnetic fields are neglected (e.g., Stix 1989).

The starting point of the mixing-length theory (MLT) is to calculate the excess heat flux
in the radial direction:
\ba
  F^r_{\s{conv}} &=& \rho v_{\s{conv}} DQ=\rho v_{\s{conv}}(Q_e-Q_s) \nonumber \\
 &=& \rho v_{\s{conv}}[C_P(T_e-T_s)-(\delta/\rho)(P_{Te}-P_{Ts}) 
+ (P_T\delta\nu/P_m\alpha)(\chi_e-\chi_s)],
\ea
where we have used the first law of thermodynamics $DQ=TDS_T$. 
The subscriptions ``e'' and ``s'' stand for a convective eddy and its surroundings. 
If the eddy is always assumed to be in pressure equilibrium ($DP_T=P_{Te}=P_{Ts}=0$) and magnetic equilibrium 
($D\chi=\chi_e-\chi_s=0$) with its surroundings, we have
\be
    F^r_{\s{conv}} = \rho v_{\s{conv}}C_P(T_e-T_s)=\frac{l_m}{2H_P}\rho v_{\s{conv}} C_P T(\nabla_s-\nabla_e), \label{eq:frc}
\ee
Where the mixing-length approximation in MLT is used to calculate the temperature (or density) difference:
\be
   T_e-T_s = \frac{l_m}{2}\left(\p{T_e}{r}-\p{T_s}{r}\right) = \frac{l_mT}{2H_P}(\nabla_s-\nabla_e).
\ee
We have also defined the eddy and surrounding temperature gradients and the pressure scale height:
\be
   \nabla_e \equiv \left(\p{\ln T}{\ln P_T}\right)_e, \hspace{5mm}
   \nabla_s \equiv \left(\p{\ln T}{\ln P_T}\right)_s, \hspace{5mm}
   H_P \equiv -\p{r}{\ln P_T}. 
\ee

The convective velocity $v_{\s{conv}}$ is generated by the radial buoyancy. The radial buoyancy acceleration 
is related to the density difference:
\be
  \dd{r}{t} = -g(D\rho/\rho), \label{eq:buoyancy}
\ee
where $g$ is the gravitational acceleration. For standard MLT, the density difference is related to the temperature difference via the equation of state with $DP_T=0$ and $D\chi=0$ (see Eq.~\ref{eq:state}):
\be
  D\rho/\rho = -(DT/T)\delta= \frac{l_m\delta}{2H_P}(\nabla_e-\nabla_s).  \label{eq:drho}
\ee
We also use the mixing-length approximation to calculate buoyancy acceleration
\be
  \dd{r}{t} = \frac{1}{2}\p{}{r}\left(\od{r}{t}\right)^2 = \frac{1}{2} \left(\od{r}{t}\right)^2_{\s{max}} \frac{2}{l_m}
    = 4v^2_{\s{conv}}/l_m, \label{eq:ddr}
\ee
where we have assumed that the convective velocity $v_{\s{conv}}$ equals half of the maximum velocity $(dr/dt)_{\s{max}}$.
Substituting Eqs.~(\ref{eq:drho}-\ref{eq:ddr}) into (\ref{eq:buoyancy}), we obtain
\be
  v^2_{\s{conv}} = g\delta(\nabla_s-\nabla_e)(l_m^2/8H_P). \label{eq:vconv}
\ee
This gives
\be
  \nabla_s-\nabla_e = \frac{8H_P}{gl_m^2\delta}v^2_{\s{conv}} \label{eq:nabla2v}
\ee
Substituting this into Eq.~(\ref{eq:frc}), we obtain
\be
   F^r_{\s{conv}} = (4\rho C_P T/gl_m\delta) v^3_{\s{conv}}. \label{eq:frc1}
\ee 

Eq.~(\ref{eq:frad}) yields
\be
   F^r_{\s{rad}}  = \frac{4acT^4}{3\kappa\rho H_P}(1+\lambda)\nabla_s.
\ee
Defining a ``radiative'' gradient
\be
  \nabla_{\s{rad}} = \frac{3\kappa\rho H_P F_r}{4acT^4},
\ee
we obtain 
\be
   F_r = \frac{4acT^4}{3\kappa\rho H_P}(1+\lambda) \nabla_{\s{rad}}.
\ee
We use the energy flux conservation law 
$
   F^r_{\s{conv}} + F^r_{\s{rad}} = F_r
$
to constrain the convective velocity:
\be
  \frac{1}{1+\lambda} \frac{4\rho C_PT}{gl_m\delta}\frac{3\kappa\rho H_P}{4acT^4}v^3_{\s{conv}}
    +\nabla_s  =\nabla_{\s{rad}}. \label{eq:flux}
\ee

\subsubsection{Nonmagnetic adiabatic approximation}

When the convective eddy is adiabatic, its temperature gradient equals the adiabatic gradient. The nonmagnetic approximation implies $\chi=0$. Therefore, the temperature gradient in a nonmagnetic adiabatic eddy is determined by
\be
  \nabla_e=\nabla'_{\s{ad}}=\nabla_{\s{ad}}.
   \label{eq:adiabatic}
\ee
Eq.~(\ref{eq:frc}) thus becomes
\be
  F^r_{\s{conv}} = - \frac{1}{2} \rho T l_m v_{\s{conv}} \left(\p{S}{r}\right)_s, \label{eq:frconvs}
\ee
where we have used the following equality:
\be
  \left(\p{S}{r}\right)_s = -\frac{C_P}{H_P}(\nabla_s-\nabla_{\s{ad}}). \label{eq:nabla2s}
\ee
Comparing Eq.~(\ref{eq:frconvs}) with the radial component of Eq.~(\ref{eq:vfconv}), we find
\be
   f(v) = v,
\ee
as stated above.

Using Eqs.~(\ref{eq:nabla2v}) and (\ref{eq:adiabatic}) in Eq.~(\ref{eq:flux}), we obtain the cubic equation of the convective velocity:
\be
   \frac{1}{1+\lambda}\frac{4\rho C_PT}{gl_m\delta}\frac{3\kappa\rho H_P}{4acT^4}v^3_{\s{conv}}
  + \frac{8H_P}{gl_m^2\delta}v^2_{\s{conv}}  =\nabla_{\s{rad}}-\nabla_{\s{ad}}.
\ee
The convective instability condition in the adiabatic approximation is
\be
  \nabla_{\s{rad}}\ge \nabla_s >\nabla_e=\nabla_{\s{ad}},
\ee
according to Eq.~(\ref{eq:vconv}).

\subsubsection{Nonmagnetic nonadiabatic approximation}\label{sect:nonmag}

During its rise the eddy radiates energy into its environment. For this reason the eddy 
gradient $\nabla_e$ differs from the adiabatic gradient, $\nabla_{\s{ad}}$. We decompose the convective flux (Eq.~\ref{eq:frc}) into the adiabatic (the first r.h.s. term) and nonadiabatic (the second r.h.s term) fluxes:
\ba
  F^r_{\s{conv}} &=& \frac{1}{2} \frac{l_m v_{\s{conv}}}{H_P} \rho T C_P (\nabla_s-\nabla_e) \nonumber\\
   &=& \frac{1}{2} \frac{l_m v_{\s{conv}}}{H_P} \rho T C_P (\nabla_s-\nabla_{\s{ad}}) 
    + \frac{1}{2} \frac{l_m v_{\s{conv}}}{H_P} \rho T C_P (\nabla_{\s{ad}}-\nabla_e). \label{eq:frconv}
\ea
If the effective cross section of the convective eddy is $q$, the heat energy loss rate of the eddy 
due to radiation can be expressed by
\be
   \od{Q_r}{t} = \frac{1}{2} \frac{l_m v_{\s{conv}}}{H_P} \rho T C_P (\nabla_{\s{ad}}-\nabla_e)q. \label{eq:dqtc}
\ee
We can also use Eq.~(\ref{eq:frad}) to calculate the radiative loss:
\be
  \od{Q_r}{t} = -\frac{4ac T^3}{3\kappa\rho}\frac{T_e-T_s}{d}\Sigma= -\frac{1}{2}\frac{l_m\Sigma}{H_Pd}
   \frac{4ac T^4}{3\kappa\rho}(\nabla_s-\nabla_e), \label{eq:dqtr}
\ee
where $d$ is the effective radius of the eddy, $\Sigma$ is the eddy surface. Comparing Eq.~(\ref{eq:dqtr}) with (\ref{eq:dqtc}), we obtain
\be
  \nabla_e-\nabla_{\s{ad}} = (v_0/v_{\s{conv}}) (\nabla_s-\nabla_e), \label{eq:ade}
\ee
where
\be
  v_0=\frac{l_m\Sigma}{qd} \frac{4ac T^3}{3\rho C_P}\frac{1}{l_m\kappa\rho}.
\ee
Substituting Eq.~(\ref{eq:ade}) into (\ref{eq:frconv}), we can express $\nabla_s-\nabla_e$ by $\nabla_s-\nabla_{\s{ad}}$:
\be
   \nabla_s-\nabla_e = \frac{1}{1+v_0/v_{\s{conv}}}(\nabla_s-\nabla_{\s{ad}})
    =-\frac{1}{1+v_0/v_{\s{conv}}}\frac{H_P}{C_P}\left(\p{S_T}{r}\right)_s, \label{eq:s2nabla}
\ee
where we have used Eq.~(\ref{eq:nabla2s}). Finally, using Eq.~(\ref{eq:s2nabla}) in (\ref{eq:frconv}) we obtain
\be
  F^r_{\s{conv}} = - \frac{1}{2} \rho T l_m \frac{v_{\s{conv}}}{1+v_0/v_{\s{conv}}} \left(\p{S}{r}\right)_s.
\ee
This shows
\be
  f(v) = \frac{v}{1+v_0/v}. \label{eq:fv}
\ee

Using Eq.~(\ref{eq:nabla2v}) in (\ref{eq:s2nabla}), we obtain
\be
   \nabla_s-\nabla_{\s{ad}} = \frac{8H_P}{gl_m^2\delta}v^2_{\s{conv}} (1+v_0/v_{\s{conv}}).
\ee
Substituting into Eq.~(\ref{eq:flux}), we obtain the cubic equation of the convective velocity:
\be
   \frac{4\rho C_PT}{gl_m\delta}\frac{3\kappa\rho H_P}{4acT^4}v^3_{\s{conv}}
+ \frac{8H_P}{gl_m^2\delta}v^2_{\s{conv}}(1+v_0/v_{\s{conv}})  =\nabla_{\s{rad}}-\nabla_{\s{ad}}.
\ee
The convective instability condition in the nonmagnetic nonadiabatic approximation is
\be
  \nabla_{\s{rad}}\ge \nabla_s >\nabla_e>\nabla_{\s{ad}},
\ee
according to Eq.~(\ref{eq:vconv}).

\subsubsection{General case}

When magnetic fields are present, we have
\be
  \left(\p{S_T}{r}\right)_s = -\frac{C_P}{H_P}(\nabla_s-\nabla'_{\s{ad}}). \label{eq:nabla2sp}
\ee
We decompose the convective flux (Eq.~\ref{eq:frc}) into the adiabatic (the first r.h.s. term) and nonadiabatic (the second r.h.s term) fluxes:
\ba
  F^r_{\s{conv}} &=& \frac{1}{2} \frac{l_m v_{\s{conv}}}{H_P} \rho T C_P (\nabla_s-\nabla_e) \nonumber\\
   &=& \frac{1}{2} \frac{l_m v_{\s{conv}}}{H_P} \rho T C_P (\nabla_s-\nabla'_{\s{ad}}) 
    + \frac{1}{2} \frac{l_m v_{\s{conv}}}{H_P} \rho T C_P ( \nabla'_{\s{ad}}-\nabla_e). \label{eq:frconvm}
\ea
The heat energy loss rate of the eddy due to radiation now can be expressed by
\be
   \od{Q_r}{t} = \frac{1}{2} \frac{l_m v_{\s{conv}}}{H_P} \rho T C_P (\nabla'_{\s{ad}}-\nabla_e)q. \label{eq:dqtcm}
\ee
The radiation loss rate calculated by Eq.~(\ref{eq:frad}) is the same as given in Eq.~(\ref{eq:dqtr}).
Comparing Eq.~(\ref{eq:dqtr}) with (\ref{eq:dqtcm}), we obtain
\be
  \nabla_e-\nabla'_{\s{ad}} = (v_0/v_{\s{conv}}) (\nabla_s-\nabla_e). \label{eq:adem}
\ee
Substituting Eq.~(\ref{eq:adem}) into (\ref{eq:frconvm}), we can express $\nabla_s-\nabla_e$ by 
$\nabla_s-\nabla'_{\s{ad}}$:
\be
   \nabla_s-\nabla_e = \frac{1}{1+v_0/v_{\s{conv}}}(\nabla_s-\nabla'_{\s{ad}})=-\frac{1}{1+v_0/v_{\s{conv}}}\frac{H_P}{C_P}\left(\p{S_T}{r}\right)_s, 
     \label{eq:s2nablam}
\ee
where we have used Eq.~(\ref{eq:nabla2sp}). 
Finally, substituting Eq.~(\ref{eq:s2nablam}) into (\ref{eq:frconvm}) we obtain
\be
  F^r_{\s{conv}} = - \frac{1}{2} \rho T l_m \frac{v_{\s{conv}}}{1+v_0/v_{\s{conv}}} \left(\p{S_T}{r}\right)_s,
\ee
which leads up to Eq.~(\ref{eq:fv}).

Using Eq.~(\ref{eq:nabla2v}) in (\ref{eq:s2nablam}), we obtain
\be
   \nabla_s-\nabla'_{\s{ad}} = \frac{8H_P}{gl_m^2\delta}v^2_{\s{conv}} \left(1+\frac{v_0}{v_{\s{conv}}}\right). \label{eq:nablas}
\ee
Substituting into Eq.~(\ref{eq:flux}), we obtain the cubic equation of the convective velocity in a magnetic system:
\be
   \frac{4\rho C_PT}{gl_m\delta}\frac{3\kappa\rho H_P}{4acT^4}v^3_{\s{conv}}
+ \frac{8H_P}{gl_m^2\delta}v^2_{\s{conv}}\left(1+\frac{v_0}{v_{\s{conv}}}\right)  =\nabla_{\s{rad}}-\nabla'_{\s{ad}}. \label{eq:cubicm}
\ee
The convective instability condition in the magnetic nonadiabatic case is
\be
  \nabla_{\s{rad}}\ge \nabla_s >\nabla_e>\nabla_{\s{ad}},
\ee
according to Eq.~(\ref{eq:vconv}).

Eq.~(\ref{eq:cubicm}) can be rewritten as follows
\be
   2A_0 y^3+Vy^2+V^2 y - V =0,
\ee
where we have defined the dimensionless variable
\be
   y=Vv_{\s{conv}}/v_0
\ee
and the following dimensionless parameters
\Ba
 v_0 &=& 6acT^3/\rho C_P l_m \kappa\rho,\\
 C &=& \frac{gl_m^2 \delta}{8 H_p}, \\
 V &=& v_0/[C^{1/2}(\nabla_{\s{rad}}-\nabla'_{\s{ad}})^{1/2}], \\
 A_0 &=& \frac{9}{8}\frac{1}{1+\lambda}.
\Ea
We choose $l_m\Sigma/qd=9/2$ for spherical eddies and $d/l_m=8/9$.
The convective gradient can be expressed by $y$:
\be
  \nabla_{\s{conv}}=\nabla_s = \nabla'_{\s{ad}} + (\nabla_{\s{rad}}-\nabla'_{\s{ad}})y(y+V),
\ee
according to Eq.~(\ref{eq:nablas}).

When magnetic fields are neglected, $\nabla'_{\s{ad}}=\nabla_{\s{ad}}$, all formulas automatically reduce to their counterparts in section \ref{sect:nonmag}. 

\subsection{Energy flux vector}

In the convective zone, the total energy flux vector equals the radiative flux (Eq.~\ref{eq:frad}):
\be
   \vb{F} = -\frac{4ac T^3}{3\kappa\rho}(1+\lambda)\nabla T.
\ee
In the convective zone, the total energy flux vector equals the sum of the radiative (Eq.~\ref{eq:frad}) and convective (Eq.~\ref{eq:vfconv}) fluxes:
\ba
\vb{F} &=& - \frac{4ac T^3}{3\kappa\rho}(1+\lambda) \nabla T 
     - \frac{1}{2} \frac{\rho T l_m v_{\s{conv}}}{1+v_{\s{conv}}/v_0} \nabla S_T \nonumber \\
&=& -\left[\frac{4ac T^3}{3\kappa\rho}(1+\lambda) 
     + \frac{1}{2} \frac{\rho C_P l_m v_{\s{conv}}}{1+v_{\s{conv}}/v_0}\right]  \nabla T 
  + \frac{1}{2} \frac{\rho C_P T \nabla'_{\s{ad}} l_m v_{\s{conv}}}{1+v_{\s{conv}}/v_0}\frac{1}{P_T}\nabla P_T,
  \label{eq:flux1}
\ea
where we have used the following formula
\be
   \nabla S_T = (C_P/T)\nabla T - (C_P\nabla'_{\s{ad}}/P_T)\nabla P_T.
\ee

\subsection{Composition conservation}

Eq.~(\ref{eq:composition}) describes the composition conservation law, which can be rewritten as follows:
\be
  \rho\p{X_i}{t} + X_i\p{\rho}{t} +\nabla\cdot (\rho X_i\vb{v}') = Q_i, \label{eq:cc1}
\ee
where we have used Eq.~(\ref{eq:turb}) and have used the mass fraction $X_i\equiv \rho_i/\rho$ 
to replace density $\rho_i$. We have also assumed $\nabla\cdot\vb{v}'=0$. 
Eq.~(\ref{eq:cc1}) involves two timescales: one is the thermonuclear reaction 
timescale $\tau_{\s{nucl}}$, which determines $Q_i$ and is quite long, 
another is the convection timescale $\tau_{\s{conv}}$, which determines the 
convection mixing and is much shorter than the former.

As before, taking the statistically average over Eq.~(\ref{eq:cc1}), we obtain
\be
  \rho\p{X_i}{t} +\frac{1}{\rho}\nabla \cdot <\rho X_i\vb{v}'> = q_i, \label{eq:cc2}
\ee
where we have used the assumption $<\partial \rho/\partial t>=0$ and defined $q_i\equiv Q_i/\rho$.
Using the mixing length theory, we can express the mass flux $\vb{F}_i\equiv <\rho X_i\vb{v}'>$ as follows:
\be
  \vb{F}_i = -\frac{1}{2}\rho v_{\s{conv}}l_m\nabla X_i. \label{eq:fi}
\ee
Substituting Eq.~(\ref{eq:fi}) into Eq.~(\ref{eq:cc2}), we obtain 
\be
  \p{X_i}{t} = q_i + \frac{1}{2\rho}\nabla\cdot(\rho v_{\s{conv}}l_m\nabla X_i). \label{eq:cc3}
\ee

In the radiative zone, the element diffusion velocity $w_i$ 
(e.g., Thoul et al. 1994) changes the local composition 
in addition to the thermonuclear reactions. Element diffusion 
in stars is driven by pressure gradients (or gravity), temperature 
gradients, composition gradients, and radiation pressure. Gravity 
tends to concentrate the heavier elements toward the center of the star. 
Temperature gradients lead to thermal diffusion, which tends to concentrate 
more highly charged and more massive species toward the hottest region of 
the star, its center. Concentration gradients oppose the above two processes. 
Radiation pressure causes negligible diffusion in the solar core. Element 
diffusion affects the element abundances, the mean molecular weight, and 
the radiative opacity in the radiative zone, and therefore affects 
the calculated neutrino fluxes and oscillation frequencies, on which 
observations impose strict constraints on the solar model.

The characteristic time for elements to diffuse a solar radius under 
solar conditions is of the order of $6\times10^{13}$ yr, much larger 
than the age of the Sun. Element diffusion therefore introduces only a small 
correction. Many authors have studied this topic carefully (see Thoul et al. 
1994 and references therein), and both portable subroutine and analytic formulae 
for element diffusion calculations are available. In particular, the formulae 
for the element diffusion velocity fits our theoretical framework developed 
in this paper. We use the formula given by Thoul et al. (1994) with $q_i$ 
included:
\be
\p{X_i}{t} = q_i  - \frac{1}{r^2\rho}\p{}{r}(r^2\rho X_i w_i),
\ee
where
\be
w_i(r) = \frac{T^{5/2}}{\rho}\left( A^i_p\p{\ln P_T}{r} + A^i_T\p{\ln T}{r} 
  + A^i_H \p{\ln C_H}{r}\right).
\ee
See Thoul et al. (1994) for the expansion coefficients, which are actually 
computed by numerically solving the multi-fluid equations for all species. 
These formulae just give readers the main idea. We use the portable subroutine 
provided by the authors to compute the element diffusion correction. 
Diffusion in the polar direction is negligible.

\section{Coordinate transformation from $r$ to $m$}\label{sec:ct}

So far, all derivatives with respect to $\theta$ assume $r$ to be constant. What 
we need is to obtain the corresponding derivatives at the constant $m$. This 
can be done by using the so-call implicit-function rule, that is
\be
\pa{}{\theta} = \pb{}{\theta} + \pa{r}{\theta}\p{}{r} = \pb{}{\theta} 
    + \pa{\ln r}{\theta}\p{}{\ln r}. \label{eq:implicit}
\ee
From now on, we use the following shortcuts to save writings:
\be
r'=\ln r, \hspace{5mm} \rho'=\ln \rho, \hspace{5mm} P'=\ln P_T, 
 \hspace{5mm} T'=\ln T, \mbox{ and } s=\ln m.
\ee
We note that $\ln$ is the natural logarithm.

Another formulas we need for this purpose is the mass conservation equation, Eq.~(\ref{eq:mass1}), which can be rewritten as follows:
\be
   \p{r'}{s} = \frac{m}{4\pi r^3\rho_m}. \label{eq:mass2}
\ee

\subsection{Momentum conservation equation}

We perform the necessary coordinate 
transformation from $r$ to $m$ in Eq.~(\ref{eq:pt}). The only term that needs to be transformed is the term that contains $\partial P_T/\partial \theta$, which is equivalent to $(\partial P_T/\partial \theta)_r$. Using Eq.~(\ref{eq:implicit}), we obtain
\be
  \p{P_T}{\theta} = \pa{P_T}{\theta} -\p{P_T}{r'} \pa{r'}{\theta}. \label{eq:pp1}
\ee
Consequently, Eq.~(\ref{eq:pt}) becomes
\ba
\p{P_T}{r} &=&  \left[1-\frac{\cot\theta}{2}\pa{r'}{\theta}\right]^{-1} \left[-\frac{Gm\rho}{r^2} -2\pi G r\rho(\rho-\rho_m) -\frac{\cot\theta}{2r}\pa{P_T}{\theta}\nonumber \right.\\
&& \left. +\h{H}_r +\frac{1}{2}\h{H}_\theta\cot\theta\right] +O(2). \label{eq:ptpp}
\ea
The first factor on r.h.s. is caused by the coordinate transformation from $r$ to $m$.

Since $\partial P_T/\partial r=(P_T/r)(\partial s/\partial r')(\partial P'/\partial s)$, using Eq.~(\ref{eq:mass2}), we can rewritten Eq.~(\ref{eq:ptpp}) as follows:
\ba
\p{P'}{s}&=& -\frac{Gm^2}{4\pi r^4P_T}\frac{\rho}{\rho_m} + \Theta + \h{M} +O(2), 
\ea
where
\begin{mathletters}
\Ba
\Theta &=&  -\frac{Gm(\rho-\rho_m)}{2rP_T}\frac{\rho}{\rho_m}
\left[1-\frac{\cot\theta}{2}\pa{r'}{\cot\theta}\right]^{-1} \\
&&  -\frac{m}{4\pi r^3\rho_m}\frac{\cot\theta}{2}\pa{P'}{\theta}
  \left[1-\frac{\cot\theta}{2}\pa{r'}{\theta}\right]^{-1}   \\
&&-\frac{Gm^2}{4\pi r^4 P_T}\frac{\rho}{\rho_m}\frac{\cot\theta}{2}\pa{r'}{\theta} 
 \left[1-\frac{\cot\theta}{2}\pa{r'}{\theta}\right]^{-1},    \\
\h{M} &=& \frac{m}{4\pi r^2\rho_m P_T}\left[1-\frac{\cot\theta}{2}\pa{r'}{\theta}\right]^{-1}(\h{H}_r+\frac{1}{2}\h{H}_\theta\cot\theta). 
\Ea
\end{mathletters}

\subsection{Energy conservation equation}

The starting equation is Eq.~(\ref{eq:energy2}). The only term that needs 
to be transformed is the term that contains the derivative of 
$(\sin\theta F_\theta)$ with respect to $\theta$. This term is 
a small 2-D correction to the Energy conservation equation 
since $F_\theta$, which is given in Eq.~(\ref{eq:flux1}), 
is already a combination of the first-order derivatives of $T$ and $P_T$,
\be
F_\theta = -\left[\frac{4ac T^4}{3\kappa\rho}(1+\lambda) 
  + \frac{1}{2} \frac{\rho C_PT l_m v_{\s{conv}}}{1+v_{\s{conv}}/v_0}\right]  \frac{1}{r}\p{T'}{\theta} 
  + \frac{1}{2} \frac{\rho C_P T l_m v_{\s{conv}}}{1+v_{\s{conv}}/v_0}\frac{\nabla'_{\s{ad}}}{r}\p{P'}{\theta}.
  \label{eq:ftt}
\ee
Therefore, after neglecting the higher-order corrections 
as we did above, the energy conservation equation becomes
\be
  \frac{1}{r^2}\p{(r^2F_r)}{r} = \rho\left(\epsilon-T\od{S_T}{t}\right) 
    -\frac{F_\theta\cot\theta}{r} +O(2). \label{eq:energy3}
\ee
This shows that we only need to transform $F_\theta$ from $r$ to $m$. 
Applying Eq.~(\ref{eq:implicit}) to $(\partial T'/\partial\theta)_r$ and 
$(\partial P'/\partial\theta)_r$ in Eq.~(\ref{eq:ftt}), we obtain
\ba
F_\theta &=& -\left[\frac{4ac T^4}{3\kappa\rho}(1+\lambda)
  + \frac{1}{2} \frac{\rho C_PT l_m v_{\s{conv}}}{1+v_{\s{conv}}/v_0}\right]  \frac{1}{r}
  \left[\pa{T'}{\theta} -\p{T'}{r'}\pa{r'}{\theta}\right] \nonumber \\
&&  + \frac{1}{2} \frac{\rho C_P T l_m v_{\s{conv}}}{1+v_{\s{conv}}/v_0}
  \frac{\nabla'_{\s{ad}}}{r}\left[ \pa{P'}{\theta}-\p{P'}{r'}\pa{r'}{\theta}\right] \nonumber \\
&=&  -\left[\frac{4ac T^4}{3\kappa\rho}(1+\lambda)
  + \frac{1}{2} \frac{\rho C_PT l_m v_{\s{conv}}}{1+v_{\s{conv}}/v_0}\right]  \frac{1}{r}
  \left[\pa{T'}{\theta} + \frac{Gm\rho}{rP_T}\nabla\pa{r'}{\theta}\right] \nonumber \\
&&  + \frac{1}{2} \frac{\rho C_P T l_m v_{\s{conv}}}{1+v_{\s{conv}}/v_0}
  \frac{\nabla'_{\s{ad}}}{r}\left[ \pa{P'}{\theta}+ \frac{Gm\rho}{rP_T}\pa{r'}{\theta}\right],
\ea
where $\nabla$ is the temperature gradient.

The second step is to use Eq.~(\ref{eq:mass2}) to replace $\partial r$ by $\partial s$ in 
Eq.~(\ref{eq:energy2}). Unlike $r'$, $P'$, and $T'$, which are the natural logarithms, we define
\be
  L' \equiv 4\pi r^2 F_r/L_\sun,
\ee
which is not a logarithm at all. The resultant equation is
\be
  \p{L'}{s} = \frac{1}{L_\sun}m \left(\epsilon-T\od{S_T}{t}\right)\frac{\rho}{\rho_m}
     -\frac{1}{L_\sun}\frac{mF_\theta\cot\theta}{r\rho_m} +O(2).
\ee

\subsection{Composition conservation}

Eq.~(\ref{eq:cc3}) involves the derivatives with respect to $\theta$ at constant $r$. Since what we need are the corresponding derivatives at constant $m$, this equation needs a coordinate transformation from $r$ to $m$. 
To this end, we first expand it as follows:
\be
  \p{X_i}{t} = q_i + \frac{1}{2r^2\rho}\p{}{r}\left(r^2\rho v_{\s{conv}}l_m\p{X_i}{r}\right)
   + \frac{1}{2r^2\rho\sin\theta}\p{}{\theta}\left(\sin\theta\rho v_{\s{conv}}l_m\p{X_i}{\theta}\right).
\ee
The last rhs term needs the transformation. We  retain its most important part. The resultant equation is
\be
   \p{X_i}{t} = q_i + \left(1+\frac{1}{2}\p{\rho'}{r'}\right) v_{\s{conv}}l_m\p{X_i}{r}
   + \frac{\cot\theta v_{\s{conv}}l_m}{2r^2}\pa{X_i}{\theta}+O(2),
\ee
where
\be
  O(2) = \frac{1}{2}\pp{X_i}{r} 
   + \frac{v_{\s{conv}}l_m}{2r^2}\p{\rho'}{\theta} \p{X_i}{\theta}
+ \frac{v_{\s{conv}}l_m}{2r^2}\pp{X_i}{\theta}
- \frac{\cot\theta v_{\s{conv}}l_m}{2r^2}\p{X_i}{r'}\pa{r'}{\theta}.
\ee
We have takan advantage of the fact that $v_{\s{conv}}l_m\approx \mbox{constant}$ in stars.

Since the convection timescale is much shorter than the evolution timescale, 
the convection zone is well-mixed on the evolution timescale. As a result, 
the detailed expression for the composition conservation equation in the 
convection zone does not matter much. We do it here just for the sake of completeness.

\subsection{Two-dimensional stellar structure equations}

In summary, we obtain the two-dimensional stellar structure equations 
with $(m,\theta)$ as independent variables:
\begin{mathletters}
\ba
\p{r'}{s}  &=& \frac{m}{4\pi r^3\rho}\left\{\frac{\rho}{\rho_m}\right\}, \label{eq:st1} \\
\p{P'}{s}&=& -\frac{Gm^2}{4\pi r^4P_T}\left\{\frac{\rho}{\rho_m}\right\} 
    + \left\{\Theta\right\}  + \left\{\h{M}\right\} +O(2), \label{eq:st2} \\
  \p{T'}{s} &=& \p{P'}{s} \left\{ \begin{array}{ll}
         \nabla_{\s{rad}} & \mbox{ radiative} \\
         \nabla_{\s{c}} & \mbox{ convective} \\
    \end{array} \right. \label{eq:st3} \\
  \p{L'}{s} &=& \frac{1}{L_\sun}m \left(\epsilon-T\od{S_T}{t}\right)\left\{\frac{\rho}{\rho_m}\right\}
     -\left\{\frac{1}{L_\sun}\frac{mF_\theta\cot\theta}{r\rho_m}\right\}+O(2), \label{eq:st4} \\
  \p{X_i}{t} &=& q_i + \left\{ \begin{array}{ll} -\frac{1}{r^2\rho}\p{}{r}(r^2\rho X_iw_i) & \mbox{ (radiative)} \\
          \left(1+\frac{1}{2}\p{\rho'}{r'}\right) v_{\s{conv}}l_m\p{X_i}{r}
   + \left\{\frac{\cot\theta }{2r^2}v_{\s{conv}}l_m\pa{X_i}{\theta}\right\}+O(2). & \mbox{ (convective)} 
  \end{array} \right. \label{eq:st5}
\ea
\end{mathletters}
Those terms or factors associated with 2-D effects are indicated by $\{\}$ in the equations. 
The symbols used above are defined as follows:
\begin{mathletters}
\ba
\Theta &=&    -\frac{m}{4\pi r^3\rho_m} \frac{\cot\theta}{2} \pa{P'}{\theta}
  \left[1-\frac{\cot\theta}{2}\pa{r'}{\theta}\right]^{-1}   \nonumber \\
&& -\frac{Gm(\rho-\rho_m)}{2rP_T}\frac{\rho}{\rho_m}
   \left[1-\frac{\cot\theta}{2}\pa{r'}{\theta}\right]^{-1} \nob
&&-\frac{Gm^2}{4\pi r^4 P_T}\frac{\rho}{\rho_m}\frac{\cot\theta}{2}\pa{r'}{\theta} \left[1-\frac{\cot\theta}{2}\pa{r'}{\theta}\right]^{-1},   \label{eq:q1} \\
\h{M} &=& \frac{m}{4\pi r^2\rho_m P_T}\left[1-\frac{\cot\theta}{2}\pa{r'}{\theta}\right]^{-1}(\h{H}_r+\frac{1}{2}\h{H}_\theta\cot\theta), \label{eq:q2a} \\
F_\theta &=&  -\left[\frac{4ac T^4}{3\kappa\rho}(1+\lambda)
  + \frac{1}{2} \frac{\rho C_PT l_m v_{\s{conv}}}{1+v_{\s{conv}}/v_0}\right]  \frac{1}{r}
  \left[\pa{T'}{\theta} + \frac{Gm\rho}{rP_T}\nabla\pa{r'}{\theta}\right] \nonumber \\
&&  + \frac{1}{2} \frac{\rho C_P T l_m v_{\s{conv}}}{1+v_{\s{conv}}/v_0}
  \frac{\nabla'_{\s{ad}}}{r}\left[ \pa{P'}{\theta}+ \frac{Gm\rho}{rP_T}\pa{r'}{\theta}\right]. \label{eq:q3}
\ea
\end{mathletters}

From now on, we'll drop the subscript $m$ in the derivatives such as $(\partial r'/\partial\theta)_m$,
\be
   \pa{r'}{\theta} \Longrightarrow \p{r'}{\theta}.
\ee
Wherever needed, we'll specify the subscript $m$ or $r$ to avoid confusion.

\section{Magnetic fields}\label{sec:mf}

Our strategy is to take advantage of analytical results as much as possible. For this 
purpose, in this section we work out the explicit expressions for the terms associated with magnetic
fields.

Generally, a magnetic field has three components. Using the spherical coordinate system, 
it can be expressed by
\be
  \vb{B} = (B_r,B_\theta,B_\phi).
\ee
All three components are functions of $m$ and $\theta$ in the azimuthally-symmetric case 
treated in this paper. The $\vb{B}$-related terms are expressed by $\h{H}$, see Eq.~(\ref{eq:hh}), which can 
be expanded as follows
\ba
 4\pi r\h{H} &=& \vb{e}_r\left( r\vb{B}\cdot\nabla B_r - B_\theta B_\theta - B_\phi B_\phi \right)\nonumber \\
   && +\vb{e}_\theta\left( r\vb{B}\cdot\nabla B_\theta - B_\phi B_\phi\cot\theta + B_\theta B_r \right) \nonumber \\
   && +\vb{e}_\phi\left( r\vb{B}\cdot\nabla B_\phi + B_\phi B_r - B_\phi B_\theta\cot\theta \right). \label{eq:h}
\ea
Consequently, we see
\begin{mathletters}
\ba
  4\pi r \h{H}_r &=& r\vb{B}\cdot\nabla B_r - B_\theta B_\theta - B_\phi B_\phi, \label{eq:hr} \\
  4\pi r \h{H}_\theta &=&  r\vb{B}\cdot\nabla B_\theta - B_\phi B_\phi\cot\theta + B_\theta B_r, \label{eq:ht} \\ 
  4\pi r \h{H}_\phi &=&  r\vb{B}\cdot\nabla B_\phi + B_\phi B_r - B_\phi B_\theta\cot\theta.
\ea
\end{mathletters}

We use $\vb{B}$ to define three stellar magnetic parameters, in addition to the 
conventional stellar parameters such as pressure, temperature, radius and 
luminosity. The first magnetic parameter is the magnetic kinetic energy 
per unit mass, $\chi$,
\be
   \chi = B^2/(8\pi\rho).
\ee
The second is the heat index due to the magnetic field, or the ratio of the magnetic 
pressure in the radial direction to the magnetic energy density, $\gamma-1$,
\be
  \gamma = 1 + (B_\theta^2+B_\phi^2)/B^2.
\ee
Lydon and Sofia (1995) introduced the first two in the one-dimensional case. 
Here we introduce the third one, the ratio of the magnetic pressure in the 
co-latitude direction to the magnetic energy density, $\vartheta-1$,

\be
 \vartheta=1+(B_\phi^2+B_r^2)/B^2.
\ee
We can use these three magnetic parameters to express 
three components of a magnetic field as follows:
\begin{mathletters}
\ba
   B_r &=& [8\pi(2-\gamma) \chi\rho]^{1/2}, \label{eq:br0} \\
   B_\theta &=& [8\pi(2-\vartheta)\chi\rho]^{1/2}, \label{eq:bt0} \\
   B_\phi   &=& [8\pi(\gamma+\vartheta-3)\chi\rho]^{1/2}. \label{eq:bp0}
\ea
\end{mathletters} 

We discuss below various possible cases. Note that any case 
should satisfy the restriction:
\be
  \nabla\cdot\vb{B}=0. \label{eq:maxwell}
\ee

\subsection{$\vb{B}=(0,0,0)$}

In this case,
\be
   \chi = 0, \hspace{5mm} \vartheta=1, \hspace{5mm} \gamma = 1, \hspace{5mm} \h{H}=0.
\ee
Consequently, the term associated with magnetic fields vanishes, namely,
\be
 \h{M} = 0. \\
\ee
Defining
\begin{mathletters}
\ba
\h{B}^1 &=&  -\frac{Gm^2}{4\pi r^4 P_T} \frac{\rho}{\rho_m} , \\
\h{B}^2 &=& -\frac{Gm(\rho-\rho_m)}{2rP_T}\frac{\rho}{\rho_m}, \\
\h{B}^3 &=& -\frac{m}{4\pi r^3 \rho_m},
\ea
\end{mathletters}
we can rewrite $\Theta$ as follows
\ba
\Theta &=& \h{B}^1\frac{\cot\theta}{2}\p{r'}{\theta} \left(1-\frac{\cot\theta}{2}\p{r'}{\theta}\right)^{-1} \nob
&&+\h{B}^2 \left(1-\frac{\cot\theta}{2}\p{r'}{\theta}\right)^{-1} \nob
&&+\h{B}^3 \frac{\cot\theta}{2}\p{P'}{\theta} \left(1-\frac{\cot\theta}{2}\p{r'}{\theta}\right)^{-1}.
\ea

Solving this case will provide us with a standard 2-D stellar model.

\subsection{$\vb{B}=(0,0,B_\phi)$}

Since $B_\phi$ is assumed to depend upon only $r$ and $\theta$, 
Eq.~(\ref{eq:maxwell}) is satisfied for any arbitrary function 
$B_\phi=B_\phi(r,\theta)$. In this case, since
\Ba
   \chi &=& B^2_\phi/(8\pi\rho), \\
   \vartheta &=& 2, \\
   \gamma &=& 2,
\Ea
we have
\begin{mathletters}
\ba
   B_r &=& 0, \label{eq:br1} \\
   B_\theta &=& 0, \label{eq:bt1} \\
   B_\phi &=& (8\pi\chi\rho)^{1/2}. \label{eq:bp1}
\ea
\end{mathletters}
Substituting them into Eqs.~(\ref{eq:hr}-\ref{eq:ht}), we obtain
\begin{mathletters}
\ba
   H_r &=& -2\chi\rho/r, \\
   H_\theta &=& H_r\cot\theta.
\ea
\end{mathletters}
Substituting them into Eq.~(\ref{eq:q2a}), we obtain
\ba
\h{M} =  -\h{B}^4 (1+\frac{1}{2}\cot^2\theta) \left(1-\frac{\cot\theta}{2}\p{r'}{\theta}\right)^{-1},
\ea
where
\be
 \h{B}^4 = \frac{m}{2\pi r^3\rho_m}\frac{\chi\rho}{P_T}.
\ee

\subsection{$\vb{B}=(0,B_\theta,0)$}

Eq.~(\ref{eq:maxwell}) requires
\be
   \p{(\sin\theta B_\theta)}{\theta}=0.
\ee
This leads to
\be
   B_\theta = B(r)/\sin\theta, \label{eq:btheta}
\ee
where $B(r)$ is an arbitrary function of $r$.

In this case, since
\Ba
   \chi &=& B^2_\theta/(8\pi\rho), \\
   \vartheta &=& 1, \\
   \gamma &=& 2,
\Ea
we have
\begin{mathletters}
\ba
   B_r &=& 0, \label{eq:br2}\\
   B_\theta &=& (8\pi\chi\rho)^{1/2}, \label{eq:bt2} \\
   B_\phi &=& 0. \label{eq:bp2}
\ea
\end{mathletters}
Eq.~(\ref{eq:btheta}) requires that $B=(8\pi\chi\rho)^{1/2}\sin\theta$ does not depend upon $\theta$.

In order to calculate $\h{M}$, we have to calculate $\h{H}_r$ and $\h{H}_\theta$ first. Substituting Eqs.~(\ref{eq:br2}-\ref{eq:bp2}) into Eqs.~(\ref{eq:hr}-\ref{eq:ht}), we obtain
\begin{mathletters}
\ba
H_r &=& -2\chi\rho/r, \\
H_\theta &=& \frac{1}{4\pi r} \frac{1}{2} \left(\p{B_\theta^2}{\theta}\right)_r \\
&=& \frac{\chi\rho}{r} \left[ \left(\p{\rho'}{\theta}+\p{\chi'}{\theta}\right)_m
  - \frac{4\pi r^3\rho_m}{m} \left(\p{\rho'}{s}+\p{\chi'}{s}\right)\pa{r'}{\theta}\right].
\ea
\end{mathletters}
Substituting them into Eq.~(\ref{eq:q2a}), we obtain $\h{M}$:
\ba
\h{M} &=& -\h{B}^4\left[1-\frac{\cot\theta}{4}\left(\p{\rho'}{\theta}+\p{\chi'}{\theta}\right)\right]\left(1-\frac{\cot\theta}{2}\p{r'}{\theta}\right)^{-1}  \nob
  && -\h{B}^5\left(\p{\rho'}{s}+\p{\chi'}{s}\right)\frac{\cot\theta}{2}\p{r'}{\theta}\left(1-\frac{\cot\theta}{2}\p{r'}{\theta}\right)^{-1},
\ea
where
\be
\h{B}^5 = \chi\rho/P_T.
\ee

\subsection{$\vb{B}=(B_r,0,0)$}

In this case, since
\Ba
   \chi &=& B^2_r/(8\pi\rho), \\
   \vartheta &=& 2, \\
   \gamma &=& 1,
\Ea
we have
\begin{mathletters}
\ba
   B_r &=& (8\pi\chi\rho)^{1/2}, \label{eq:br3} \\
   B_\theta &=& 0, \label{eq:bt3} \\
   B_\phi &=& 0. \label{eq:bp3}
\ea
\end{mathletters}

Eq.~(\ref{eq:maxwell}) requires
\be
   \p{(r^2 B_r)}{r}=0.
\ee
This leads to
\be
   B_r = B(\theta)/r^2,
\ee
where $B(\theta)$ is an arbitrary function of $\theta$. Therefore, we know
\be
   B=(8\pi\chi\rho)^{1/2}r^2 
\ee
varies with only $\theta$. 

Substituting Eqs.~(\ref{eq:br3}-\ref{eq:bp3}) into Eqs.~(\ref{eq:hr}-\ref{eq:ht}), we obtain
\Ba
   H_r &=& \frac{4\pi r^3\rho_m}{m}\frac{\chi\rho}{r} \left(\p{\rho'}{s}+\p{\chi'}{s}\right), \\
   H_\theta &=& 0.
\Ea
Substituting them into Eq.~(\ref{eq:q2a}), we obtain $\h{M}$:
\be
  \h{M} = \h{B}^5\left(\p{\rho'}{s}+\p{\chi'}{s}\right)\left(1-\frac{\cot\theta}{2}\p{r'}{\theta}\right)^{-1}. 
    \label{eq:q2b1}
\ee

\subsection{$\vb{B}=(0,B_\theta,B_\phi)$}

In this case, since
\Ba
   \chi &=& (B^2_\theta+B^2_\phi)/(8\pi\rho), \\
   \vartheta &=& 1+\frac{B_\phi^2}{B^2_\theta+B^2_\phi}, \\
   \gamma &=& 2,
\Ea
we have
\begin{mathletters}
\ba
   B_r &=& 0, \label{eq:br4} \\
   B_\theta &=& [8\pi(2-\vartheta)\chi\rho]^{1/2}, \label{eq:bt4} \\
   B_\phi &=& [8\pi(\vartheta-1)\chi\rho]^{1/2}. \label{eq:bp4}
\ea
\end{mathletters}

Eq.~(\ref{eq:maxwell}) requires
\be
   \p{(\sin\theta B_\theta)}{\theta}=0.
\ee
This leads to
\be
   B_\theta = B(r)/\sin\theta,
\ee
where $B(r)$ is an arbitrary function of $r$. Therefore, we have the constraint that
$[8\pi(2-\vartheta)\chi\rho]^{1/2}\sin\theta$ depend only upon $r$.

Substituting Eqs.~(\ref{eq:br4}-\ref{eq:bp4}) into Eqs.~(\ref{eq:hr}-\ref{eq:ht}), we obtain
\begin{mathletters}
\ba
H_r &=& -2\chi\rho/r, \\
H_\theta &=& -\frac{2\chi\rho}{r}(\vartheta-1)\cot\theta 
  + \frac{1}{r}\left\{\p{}{\theta} [(2-\vartheta)\chi\rho]\right\}_r \nob
&=& -\frac{2\chi\rho}{r}(\vartheta-1)\cot\theta + \frac{1}{r} \left\{\p{}{\theta} [(2-\vartheta)\chi\rho]\right\}_m \nob
&& - \frac{1}{r} \frac{4\pi r^3\rho_m}{m} \pa{r'}{\theta} \p{}{s} [(2-\vartheta)\chi\rho]. 
\ea
\end{mathletters}
Substituting these expressions into Eq.~(\ref{eq:q2a}), we obtain
\ba
\h{M} &=& -\h{B}^6\left(1-\frac{\cot\theta}{2}\p{r'}{\theta}\right)^{-1}   \nob
&&  +\h{B}^7 \frac{\cot\theta}{2} \left(\p{\chi'}{\theta}+\p{\rho'}{\theta}+\p{\vartheta''}{\theta}\right) \left(1-\frac{\cot\theta}{2}\p{r'}{\theta}\right)^{-1} \nob
&&  -\h{B}^8 \left(\p{\chi'}{s}+\p{\rho'}{s}+\p{\vartheta''}{s}\right) \frac{\cot\theta}{2}\p{r'}{\theta} \left(1-\frac{\cot\theta}{2}\p{r'}{\theta}\right)^{-1},   \label{eq:q2q23}
\ea
where
\Ba
  \h{B}^6 &=& \h{B}^4[1+\frac{1}{2}(\vartheta-1)\cot^2\theta], \\
  \h{B}^7 &=& \frac{1}{2}\h{B}^4(2-\vartheta), \\
  \h{B}^8 &=& \h{B}^5(2-\vartheta), \\
  \p{\vartheta''}{s} &=& \p{}{s}\log(2-\vartheta), \\
  \p{\vartheta''}{\theta} &=& \p{}{\theta}\log(2-\vartheta).
\Ea

\subsection{$\vb{B}=(B_r,B_\theta,0)$}

In this case, since
\Ba
   \chi &=& (B^2_r+B^2_\theta)/(8\pi\rho), \\
   \vartheta &=& 1+\frac{B^2_r}{B^2_r+B^2_\theta}, \\
   \gamma &=& 1+\frac{B^2_\theta}{B^2_r+B^2_\theta},
\Ea
we have
\begin{mathletters}
\ba
   B_r &=& [8\pi(2-\gamma)\chi\rho]^{1/2}, \label{eq:br5} \\
   B_\theta &=& [8\pi(2-\vartheta)\chi\rho]^{1/2}, \label{eq:bt5} \\
   B_\phi &=& 0. \label{eq:bp5}
\ea
\end{mathletters}
We have used the fact that
\be
  \gamma+\vartheta=3.
\ee

A meaningful magnetic field should satisfy Eq.~(\ref{eq:maxwell}). For example,
$r^2B_r$ does not vary with $r$ and $\sin\theta B_\theta$ does not vary with $\theta$.

Substituting Eqs.~(\ref{eq:br5}-\ref{eq:bp5}) into Eqs.~(\ref{eq:hr}-\ref{eq:ht}), we obtain
\begin{mathletters}
\Ba
H_r &=& -(2-\vartheta)\frac{2\chi\rho}{r}+ (2-\gamma)\frac{4\pi r^3\rho_m}{m} \frac{\chi\rho}{r}\left(\p{\chi'}{s}
  +\p{\rho'}{s}+\p{\gamma''}{s}\right) \nob
&& +[(2-\gamma)(2-\vartheta)]^{1/2}\frac{\chi\rho}{r}\left(\p{\chi'}{\theta}
  +\p{\rho'}{\theta}+\p{\gamma''}{\theta}\right) \nob
&& -[(2-\gamma)(2-\vartheta)]^{1/2}\frac{\chi\rho}{r}\frac{4\pi r^3\rho_m}{m}\p{r'}{\theta}\left(\p{\chi'}{s}
  +\p{\rho'}{s}+\p{\gamma''}{s}\right), \\
H_\theta &=& [(2-\gamma)(2-\vartheta)]^{1/2}\frac{2\chi\rho}{r} \nob
&&+ [(2-\gamma)(2-\vartheta)]^{1/2}\frac{4\pi r^3\rho_m}{m} \frac{\chi\rho}{r}\left(\p{\chi'}{s}
  +\p{\rho'}{s}+\p{\vartheta''}{s}\right) \nob
&&  + (2-\vartheta)\frac{\chi\rho}{r}\left(\p{\chi'}{\theta}+\p{\rho'}{\theta}+\p{\vartheta''}{\theta}\right) \nob
&&  - (2-\vartheta)\frac{\chi\rho}{r}\frac{4\pi r^3\rho_m}{m}\p{r'}{\theta}\left(\p{\chi'}{s}+\p{\rho'}{s}+\p{\vartheta''}{s}\right),
\Ea
\end{mathletters}
Substituting these expressions into Eq.~(\ref{eq:q2a}), we obtain
\ba
\h{M} &=& -\h{B}^9 \left(1-\frac{\cot\theta}{2}\p{r'}{\theta}\right)^{-1} +\h{B}^{10}\left(\p{\rho'}{s}
   +\p{\chi'}{s}+\p{\gamma''}{s}\right)\left(1-\frac{\cot\theta}{2}\p{r'}{\theta}\right)^{-1}  \nob
&& -\h{B}^{11}\p{r'}{\theta}\left(\p{\chi'}{s}+\p{\rho'}{s}+\p{\gamma''}{s}\right) \left(1-\frac{\cot\theta}{2}\p{r'}{\theta}\right)^{-1} \nob
&& +\h{B}^{12}\left(\p{\chi'}{\theta}+\p{\rho'}{\theta}+\p{\gamma''}{\theta}\right) \left(1-\frac{\cot\theta}{2}\p{r'}{\theta}\right)^{-1}  \nob
&& +\h{B}^{13}\left(\p{\chi'}{s}+\p{\rho'}{s}+\p{\vartheta''}{s}\right)\left(1-\frac{\cot\theta}{2}\p{r'}{\theta}\right)^{-1} \nob
&& -\h{B}^{14} \left(\p{\chi'}{s}+\p{\rho'}{s}+\p{\vartheta''}{s}\right)\frac{\cot\theta}{2}\p{r'}{\theta} \left(1-\frac{\cot\theta}{2}\p{r'}{\theta}\right)^{-1}   \nob
&& +\h{B}^{15}\frac{\cot\theta}{2}\left(\p{\chi'}{\theta}+\p{\rho'}{\theta}+\p{\vartheta''}{\theta}\right)\left(1-\frac{\cot\theta}{2}\p{r'}{\theta}\right)^{-1}, \label{eq:q2q12}
\ea
where
\Ba
  \h{B}^{9} &=& \h{B}^4\{(2-\vartheta)-\frac{\cot\theta}{2}[(2-\gamma)(2-\vartheta)]^{1/2}\},\\
  \h{B}^{10} &=& \h{B}^5(2-\gamma),\\
  \h{B}^{11} &=& \h{B}^5[(2-\vartheta)(2-\gamma)]^{1/2},\\
  \h{B}^{12} &=& \frac{1}{2}\h{B}^4[(2-\vartheta)(2-\gamma)]^{1/2}, \\
  \h{B}^{13} &=& \frac{1}{2}\h{B}^5\cot\theta[(2-\vartheta)(2-\gamma)]^{1/2},\\
  \h{B}^{14} &=& \h{B}^8, \\
  \h{B}^{15} &=& \h{B}^7, \\
  \p{\gamma''}{s} &=& \p{}{s}\log(2-\gamma), \\
  \p{\gamma''}{\theta} &=& \p{}{\theta}\log(2-\gamma).
\Ea

\subsection{$\vb{B}=(B_r,0,B_\phi)$}

In this case, since
\Ba
   \chi &=& (B^2_r+B^2_\theta)/(8\pi\rho), \\
   \vartheta &=& 2, \\
   \gamma &=& 1+\frac{B^2_\phi}{B^2_r+B^2_\phi},
\Ea
we have
\begin{mathletters}
\ba
   B_r &=& [8\pi(2-\gamma)\chi\rho]^{1/2}, \label{eq:br6} \\
   B_\theta &=& 0, \label{eq:bt6} \\
   B_\phi &=& [8\pi(\gamma-1)\chi\rho]^{1/2}. \label{eq:bp6}
\ea
\end{mathletters}
A meaningful magnetic field should satisfy Eq.~(\ref{eq:maxwell}), which 
requires that $r^2B_r$ not to vary with $r$.

Substituting Eqs.~(\ref{eq:br6}-\ref{eq:bp6}) into Eqs.~(\ref{eq:hr}-\ref{eq:ht}), we obtain
\Ba
   H_r &=& -(\gamma-1)\frac{2\chi\rho}{r} +\frac{4\pi r^3\rho_m}{m}\frac{\chi\rho}{r}(2
   -\gamma)\left(\p{\rho'}{s}+\p{\chi'}{s}+\p{\gamma''}{s}\right), \\
   H_\theta &=& -(\gamma-1)\frac{2\chi\rho}{r}\cot\theta.
\Ea
Substituting these expressions into Eq.~(\ref{eq:q2a}), we obtain
\be
  \h{M} =-\h{B}^{16}\left(1-\frac{\cot\theta}{2}\p{r'}{\theta}\right)^{-1} + \h{B}^{10}\left(\p{\rho'}{s}+\p{\chi'}{s}+\p{\gamma''}{s}\right) \left(1-\frac{\cot\theta}{2}\p{r'}{\theta}\right)^{-1}, \label{eq:q2q13}
\ee
where
\[
  \h{B}^{16} = \h{B}^4(\gamma-1)(1+\frac{1}{2}\cot^2\theta).
\]

\subsection{$\vb{B}=(B_r,B_\theta,B_\phi)$}

This is the general case, in which all magnetic field parameters $\chi$, 
$\vartheta$ and $\gamma$ are variables. Therefore, we use the general 
expressions for $B_r$, $B_\theta$ and $B_\phi$ given at the beginning 
of this section. Substituting Eqs.~(\ref{eq:br0}-\ref{eq:bp0}) into Eqs.~(\ref{eq:hr}-\ref{eq:ht}) to calculate $H_r$ and $H_\theta$, we obtain
\Ba
 H_r &=& -(\gamma-1)\frac{2\chi\rho}{r} 
   + (2-\gamma)\frac{4\pi r^3\rho_m}{m}\frac{\chi\rho}{r}\left(\p{\rho'}{s}+\p{\chi'}{s}+\p{\gamma''}{s}\right) \\
&& +[(2-\gamma)(2-\vartheta)]^{1/2}\frac{\chi\rho}{r}\left(\p{\chi'}{\theta}
  +\p{\rho'}{\theta}+\p{\gamma''}{\theta}\right) \\
&& - [(2-\gamma)(2-\vartheta)]^{1/2}\frac{4\pi r^3\rho_m}{m}\frac{\chi\rho}{r}\p{r'}{\theta}\left(\p{\chi'}{s}
  +\p{\rho'}{s}+\p{\gamma''}{s}\right), \\
 H_\theta &=& [(2-\gamma)(2-\vartheta)]^{1/2}\frac{2\chi\rho}{r} - (\gamma+\vartheta-3)\frac{2\chi\rho}{r}\cot\theta \nob
&&+ [(2-\gamma)(2-\vartheta)]^{1/2}\frac{4\pi r^3\rho_m}{m} \frac{\chi\rho}{r}\left(\p{\chi'}{s}
  +\p{\rho'}{s}+\p{\vartheta''}{s}\right) \nob
&&  +(2-\vartheta)\frac{\chi\rho}{r}\left(\p{\rho'}{\theta}+\p{\chi'}{\theta}+\p{\vartheta''}{\theta} \right) \\
&&  -(2-\vartheta)\frac{4\pi r^3\rho_m}{m}\frac{\chi\rho}{r}\p{r'}{\theta}\left(\p{\rho'}{s}+\p{\chi'}{s}+\p{\vartheta''}{s} \right).
\Ea
Substituting these expressions into Eq.~(\ref{eq:q2a}), we obtain
\ba
\h{M} &=& -\h{B}^{17} \left(1-\frac{\cot\theta}{2}\p{r'}{\theta}\right)^{-1} +\h{B}^{10}\left(\p{\rho'}{s}
   +\p{\chi'}{s}+\p{\gamma''}{s}\right)\left(1- \frac{\cot\theta}{2}\p{r'}{\theta}\right)^{-1}  \nob
&& -\h{B}^{11}\p{r'}{\theta}\left(\p{\chi'}{s}+\p{\rho'}{s}+\p{\gamma''}{s}\right) \left(1-\frac{\cot\theta}{2}\p{r'}{\theta}\right)^{-1} \nob
&& +\h{B}^{12}\left(\p{\chi'}{\theta}+\p{\rho'}{\theta}+\p{\gamma''}{\theta}\right) \left(1-\frac{\cot\theta}{2}\p{r'}{\theta}\right)^{-1}  \nob
&& +\h{B}^{13}\left(\p{\chi'}{s}+\p{\rho'}{s}+\p{\vartheta''}{s}\right)\left(1-\frac{\cot\theta}{2}\p{r'}{\theta}\right)^{-1} \nob
&& -\h{B}^{14} \left(\p{\chi'}{s}+\p{\rho'}{s}+\p{\vartheta''}{s}\right)\frac{\cot\theta}{2}\p{r'}{\theta} \left(1-\frac{\cot\theta}{2}\p{r'}{\theta}\right)^{-1}   \nob
&& +\h{B}^{15}\frac{\cot\theta}{2}\left(\p{\chi'}{\theta}+\p{\rho'}{\theta}+\p{\vartheta''}{\theta}\right)\left(1-\frac{\cot\theta}{2}\p{r'}{\theta}\right)^{-1}, \label{eq:q2q14}
\ea
where
\be
  \h{B}^{17} = \h{B}^{4}\{\gamma-1-\frac{1}{2}\cot\theta[(2-\gamma)(2-\vartheta)]^{1/2}+(\gamma+\vartheta-3)\frac{1}{2}\cot^2\theta\}.
\ee

Realistic magnetic fields in the stellar interior should satisfy the Maxwell equations. 
One of them is the divergence-free condition specified by Eq.~(\ref{eq:maxwell}). 
Using the coordinate $(m,\theta)$, this equation reads
\be
   \frac{4\pi r\rho_m\sin\theta}{m}\p{(r^2B_r)}{s} + \p{(\sin\theta B_\theta)}{\theta}=O(2).
\ee
Assuming $B_r=C(m)\cos\theta/r^2$, we obtain by solving this equation for $B_\theta(r,\theta)$,
\be
  B_\theta(r,\theta)=-\frac{4\pi r\rho_m}{m}\od{C(s)}{s}\sin\theta.
\ee

So far, we have finished the coordinate transformation from $(r,\theta)$ 
to $(m,\theta)$. This allows us to use the analytical formula, for instance, 
$\Theta$ and $\h{M}$, to describe the two-dimensional effects. This effort has at least the following rewards
\begin{description}
\item{a)} We can control the approximations by neglecting certain terms.
\item{b)} We can understand if certain factor(s) play an important role by including 
  or excluding the corresponding term(s) in the numerical calculations.
\item{c)} We may use the existing technique to numerically solve the two-dimensional 
  stellar structure equations.
\item{d)} We can use the analytical expressions to calculate the matrix element coefficients 
  for the linearization correction equations.
\end{description}
We'll make use of these advantages below.

\section{Boundary conditions}\label{sec:bc}

As usual in mathematical physics, the boundary conditions constitute 
a serious part of the whole problem, and their influence on the 
solutions is not easy to foresee. In the one-dimensional stellar 
model calculations, the boundary conditions cannot be specified 
at one end of the interval $[0,M_{\s{tot}}]$ only, but rather 
are split into some that are given at the center and some near 
the surface of the star. The central conditions are simple, 
whereas the surface conditions involve observable quantities. 
The boundaries in the angular direction are located at $\theta=0$ 
and either $\theta=\pi/2$ or $\theta=\pi$. We'll follow Deupree 
(1990) in using symmetry conditions to determine them. Otherwise, 
the treatment of the boundary conditions is as described in 
Prather (1976) and as implemented in YREC (Pinsonneault 1988).

\subsection{Central conditions}

Two boundary conditions can be specified for the center, defined by $m=0$:
\be
  m=0: \hspace{5mm} r=0, \hspace{4mm} L=0. \label{eq:cb}
\ee
Rewriting Eq.~(\ref{eq:st1}) as follows
\be
  dr^3 =\frac{3}{4\pi\rho_m}dm,
\ee
we can integrate it over a small mass interval $[0,m]$ in which 
$\rho_m=\rho_{m{\s{c}}}$ can be considered to be constant. The result
\be
  r=\left(\frac{3}{4\pi\rho_{m{\s{c}}}}\right)^{1/3}m^{1/3}
\ee
can be considered to be the first term in a series expansion of $r$ 
around $m=0$. Taking the logarithm, we obtain
\be
  r'=\frac{1}{3}[s-\log(4\pi\rho_m/3)]. \label{eq:cb1}
\ee
A corresponding integration of Eq.~(\ref{eq:st4}) yields
\be
  L' = \frac{m}{L_\sun}\left(\epsilon-T\od{S_T}{t}\right)\frac{\rho}{\rho_m}-\frac{m}{L_\sun}\frac{F_\theta}{r\rho_m}\cot\theta, \label{eq:cb2}
\ee
In both cases we have used the proper boundary conditions 
(\ref{eq:cb}) by taking the lower limit of integration to be zero.
 
Eqs.~(\ref{eq:cb1}-\ref{eq:cb2}) are two central boundary 
conditions that are equivalent to Eq.~(\ref{eq:cb}).

\subsection{Surface boundary conditions}

\subsubsection{1-D surface boundary conditions}

Nothing is a priori known about the central values of pressure 
$P_c$ and temperature $T_c$. So we need to define the surface 
and specify the surface values of pressure and temperature.

In principle, we can use a definition for the surface such as
\be
   m=M_{\s{tot}}. \label{def:sb1}
\ee
However, since near the surface $m$ does not change much, 
this definition is not accurate enough. The theory of 
stellar atmospheres suggests the use of the photosphere, 
from where the bulk of the radiation is emitted into space:
\be
  T=T_{\s{eff}},  \label{def:sb2}
\ee
where $T_{\s{eff}}$ is the effective temperature. The optical 
depth $\tau_s$ of the overlying layers,
\[
  \tau=\int_R^\infty\kappa\rho dr \label{def:sb3}
\]
is equal to $2/3$ for the Eddington approximation,
\be
  T^4=\frac{3}{4}T^4_{\s{eff}} \left(\tau+\frac{2}{3}\right).
\ee
where $R$ is the total stellar radius. In contrast, the optical 
depth $\tau_s=0.312155$ of the overlying layers is different than 
$2/3$ if the atmosphere is assumed to obey a scaled solar $T(\tau)$ 
relation given by Krishna-Swamy (1966)
\be
  T^4(\tau)= \frac{3}{4}T^4_{\s{eff}}[\tau+1.39-0.815\exp(-2.54\tau)-0.025\exp(-30.0\tau)].
\ee
Since $T_{\s{eff}}$ is the temperature of that black body that 
yields the same surface flux of energy as the star, then
\be
   m=M_{\s{tot}}: \hspace{5mm} L_s=4\pi R^2 \sigma T^4_{\s{eff}}, \label{eq:sb1a}
\ee
where $\sigma=ac/4$ is the Stefan-Boltzmann constant of radiation, 
$L_s$ is the total luminosity. This is one of two surface boundary 
conditions.

The second surface boundary condition is the hydrostatic 
equilibrium condition: the pressure at the surface is given 
by the weight of the matter above. We can well approximate 
the gravitational acceleration by the constant value 
$g_0=GM_{\s{tot}}/R^2$, since the bulk of the matter 
above the surface is anyway very close to the photosphere. 
We hence have
\be
  m=M_{\s{tot}}: \hspace{5mm} P_s = \int_R^{\infty}g\rho dr = \frac{GM_{\s{tot}}}{R^2} I, \label{eq:sb2a}
\ee
where the integration
\[
   I = \int_0^{\tau_s}\frac{1}{\kappa} d\tau
\]
is calculated in the following way: the starting values of 
($P_0$, $\tau_0$) are chosen by selecting a small density 
$\rho_0$ and then computing,
\[
  P_0 = (a/3)T^4_0+\rho_0\h{R}T_0,
\]
where $T_0\equiv T(\tau=0)$. Then ($P_0$, $\tau_0$) 
gives $\rho_1$ which gives $\kappa_0(\rho_1,T_0)$ 
which gives $\tau_1=\kappa_0P_0/g$ or $\delta\tau=\tau_1-\tau_0$. 
Thus we have $I_0=\delta\tau/\kappa_0$. Then redefining 
$\tau_0=\tau_1$ and $\rho_0=\rho_1$. This method could 
be iterated upon by redefining $T_0=T(\tau_0)$ and so forth:
\[
  I=I_0+ \frac{1}{2}\left(\frac{1}{\kappa_0}+\frac{1}{\kappa_1}\right)\delta\tau + \cdots
\]
Sufficient accuracy was achieved in the atmosphere integration 
by choosing a small enough $\rho_0$ (e.g., $\rho_0=10^{-10}$) 
such that $\tau<10^{-4}$. 

From the calculation description, we can see that 
$I=I(P_s,T_{\s{eff}})$. However, we do not know 
the explicit expression. Therefore, we cannot 
directly use Eqs.~(\ref{eq:sb1a}) and (\ref{eq:sb2a}) 
as our surface boundary conditions. Instead, we solve 
the following system (Kippenhahn 1967),
\be
\left(\begin{array}{lll}
  P'_1 & T'_1 & 1 \\
  P'_2 & T'_2 & 1 \\
  P'_3 & T'_3 & 1 \\
\end{array}\right) 
\left( \begin{array}{lll}
  a_1 & a_4 & a_7 \\
  a_2 & a_5 & a_8 \\
  a_3 & a_6 & a_9 \\
\end{array}\right) 
=
\left(\begin{array}{lll}
  R'_1 & \ln L_1 & T'_{\s{eff$_1$}} \\
  R'_2 & \ln L_2 & T'_{\s{eff$_2$}} \\
  R'_3 & \ln L_3 & T'_{\s{eff$_3$}} \\
\end{array}\right) 
\ee
for the $a_i's$ which are used for the surface boundary conditions,
\begin{mathletters}
\ba
  R' &=& a_1 P' + a_2 T' + a_3, \label{eq:sb1}\\
  \ln L &=& a_4 P' + a_5 T' + a_6, \label{eq:sb2}
\ea
\end{mathletters}
and for the calculation of the effective temperature,
\be
  T'_{\s{eff}} = a_7 P' + a_8 T' + a_9.
\ee
Here, the ($P'$, $T'$) refer to the values at the outermost mass 
point in the model. The last three equations can be considered 
to be the first term in the series expansions 
of Eqs.~(\ref{eq:sb1a}-\ref{eq:sb2a}).

The initial model with an estimated ($\ln L^*$, $T^{'*}_{\s{eff}}$) 
is triangulated in the ($\ln L$, $T'_{\s{eff}}$)-plane by 
constructing three atmospheres of the form
\[
\begin{array}{ll}
  A1: & (\ln L^*-\frac{1}{2}\Delta_L, T^{'*}_{\s{eff}}+\frac{1}{2}\Delta_T) \\
  A2: & (\ln L^*-\frac{1}{2}\Delta_L, T^{'*}_{\s{eff}}-\frac{1}{2}\Delta_T) \\
  A3: & (\ln L^*+\frac{1}{2}\Delta_L, T^{'*}_{\s{eff}}).
\end{array}
\]
If subsequent models or if the model itself during convergence moves 
significantly out of the triangle, the triangle is flipped until it 
once again constrains the model. The decision as to which point of 
the triangle should be flipped (if any) can be made by testing,
\[
  c_i=f[(\ln L_{i+1}-\ln L_{i+2})(T'_{\s{eff}}-T'_{\s{eff$_{i+1}$}})
        +( T'_{\s{eff$_{i+2}$}}- T'_{\s{eff$_{i+1}$}})(\ln L-\ln L_{i+1})],
\]
where $f=\pm 1$ is the orientation of the triangle (e.g., in the example
given $f=+1$) and $\{i,i+1,i+2\}$ is $\{123\}$, $\{231\}$ or $\{312\}$. 
The value of $c_i$ is tested against $\epsilon\Delta_L\Delta_T$ where 
setting $\epsilon=0$ gives exact triangulation and $\epsilon>0$ allows 
the point $(\ln L, T'_{\s{eff}})$ to be at most $\epsilon$ of a triangle 
outside. Begin testing with i = 1 to 3, if $c_i<-\epsilon\Delta_L\Delta_T$ 
then flip point i,
\Ba
  \ln L_i &\Leftarrow & \ln L_{i+1}+\ln L_{i+2} - \ln L_i, \\
  T'_{\s{eff$_i$}} &\Leftarrow & T'_{\s{eff$_{i+1}$}} + T'_{\s{eff$_{i+2}$}} - T'_{\s{eff$_{i}$}}, \\
  f &\Leftarrow & -f
\Ea
and repeat the testing again starting with i = 1 until $c_i$ passes 
for i = 1 to 3. The atmospheres that have been flipped are then 
recomputed as are all the coefficients $a_i$.

This treatment of the surface boundary conditions is the same 
as that in one-dimensional model calculation, except that we move the fitting 
point to the surface where $T=T_{\s{eff}}$. Therefore, we do not 
need an envelope integration. This has been tested for the one-dimensional model 
calculations and it turns out to be acceptable. This saves much 
computation time in the two-dimensional case.

Our surface boundary conditions are much more complicated 
than Deupree's (1990) because our applications to the Sun 
are very sensitive to the surface conditions.

\subsubsection{Deupree's 2-D surface boundary conditions (Deupree 1990)}\label{sec:bob}

In his two-dimensional rotational models, Deupree (1990) 
uses the following surface boundary conditions:
\[
   \rho = \rho_{\s{ref}}, \hspace{5mm}  T=T_{\s{ref}},
\]
where $\rho_{\s{ref}}$ and $T_{\s{ref}}$ are reference density and temperature, respectively.
The most difficult part of using these surface boundary 
condition is how to select the reference density and 
temperature at the surface.

\begin{figure}
\plotone{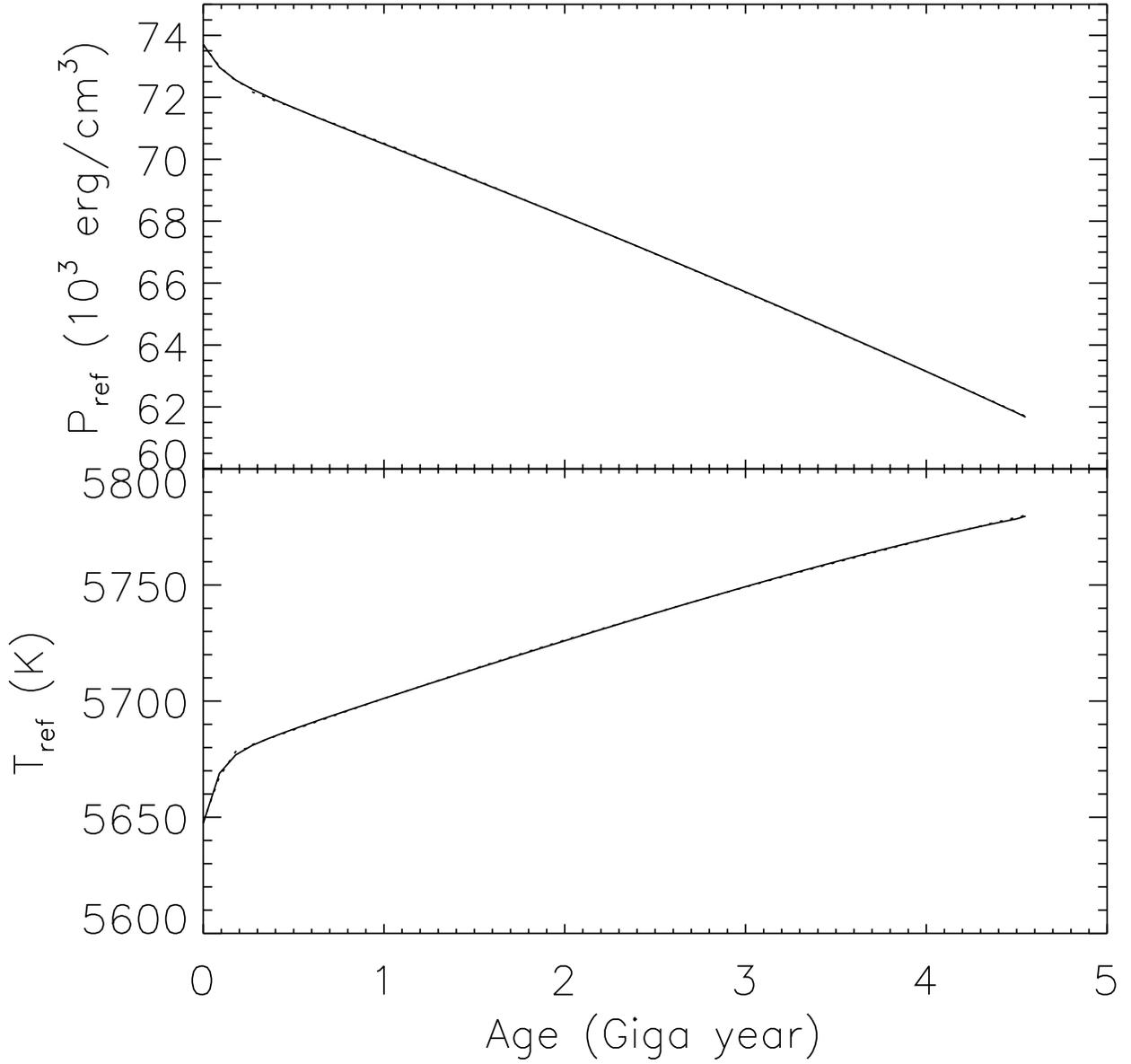}
\caption{Reference values of pressure and temperature at the surface as functions of age. 
The dotted lines are polynomial fits to the calculated model (solid lines) using the 
standard surface boundary condition.}\label{fig:ref}
\end{figure}

Unlike Deupree, we use $P_T$ and T as independent 
thermodynamical variables. Since $\rho=\rho(P_T,T)$, 
the equivalent surface boundary condition is
\be
  P_T = P_{\s{ref}}, \hspace{5mm} T=T_{\s{ref}}. \label{eq:deupree}
\ee
In order to compare this with the standard surface boundary 
condition given above, we use the surface values of $P_T$ and 
T obtained by using the standard surface boundary condition 
for the current Sun as the reference. Fig.~\ref{fig:ref} shows the reference values 
as functions of age and their polynomial fits. The fitting formulas are:
\begin{mathletters}
\ba
   P_{\s{ref}} &=& \left\{   \begin{array}{ll} 
     73695.514 -9004.5498\cdot t + 13898.511\cdot t^2 & 0\le t \le 0.27 \\
     72777.060 -2211.7088\cdot t - 49.075155\cdot t^2 & 0.27\le t \le 4.55 \\
  \end{array}\right.  \label{eq:bob1} \\
   T_{\s{ref}} &=& \left\{ \begin{array}{ll} 
     5647.8836 + 266.07365\cdot t -539.35360 \cdot t^2 & 0\le t \le 0.27 \\
     5673.6126 + 28.625469\cdot t -1.1516435 \cdot t^2 & 0.27\le t \le 4.55 \\
  \end{array} \right.   \label{eq:bob2}
\ea
\end{mathletters}
The age t is in giga year.

\subsection{Polar boundary conditions}

Eqs.~(\ref{eq:st2}-\ref{eq:st5}) are singular at the poles 
($\theta=0$ and $\pi$) because of $\Theta$, $\h{M}$, $\h{F}_\theta$, and $u_\theta$. 
However, if $\partial r'/\partial\theta=0$, the singularity 
due to $\Theta$ will disappear. In order to guarantee 
$\partial r'/\partial\theta=0$, we also need to zero 
the other derivatives. Therefore, we require
\be
  \p{r'}{\theta}=\p{L'}{\theta}=\p{P'}{\theta}=\p{T'}{\theta}=0 \label{eq:pole1}
\ee
at the poles. In order to remove the singularity due to $\h{M}$, 
we have to zero $\chi$ at the poles, namely,
\be
   \chi = 0 \label{eq:pole2}
\ee
at the poles. Eqs.~(\ref{eq:pole1}-\ref{eq:pole2}) are the 
polar boundary conditions. Eq.~(\ref{eq:pole1}) are similar 
to Deupree's (1990) polar boundary conditions, which are 
the symmetry conditions.

\subsection{Equatorial conditions}

Eqs.~(\ref{eq:st1}-\ref{eq:st4}) show that the two-dimensional 
stellar structure equations are not singular at the equator. 
Therefore, there are no special constraints there. If we neglect 
$O(2)$ in Eqs.~(\ref{eq:st2}, \ref{eq:st3}), the two-dimensional 
stellar structure equations are a set of first-order 
differential equations. Since we have already specified four 
boundary conditions at the north pole ($\theta=0$), we 
do not need extra boundary conditions at the equator. 
If we want to include those terms that contain those 
second-order derivatives in $O(2)$, we have to specify four equatorial 
boundary conditions or five polar boundary conditions 
at the south pole ($\theta=\pi$). We do not include those second-order 
derivatives in $O(2)$ in this paper for the following reasons:
\begin{description}
\item{1)} They are much smaller corrections than the retained;
\item{2)} They may cause a much bigger numerical error than the actual corrections;
\item{3)} They require a totally different method of solution (e.g., Deupree 1990).
\end{description}

\section{Method of solution}\label{sec:ms}

\subsection{Linearization of the two-dimensional stellar structure equations}

The dependent variables to be solved for are pressure $P_T$, 
temperature $T$, radius $r$, luminosity $L$ (hereafter we use
$L$ to replace $L'$, but remember that $L$ is in solar units); 
the independent variables are chosen to be mass $m$ 
(or $s=\ln m$) and angular coordinate $\theta$. The 
magnetic field variables $\chi$, $\vartheta$, and 
$\gamma$ are also dependent variables. However, 
since we do not introduce their governing 
equations (such as the dynamo equations), we 
consider them to be given. All units are in 
cgs, except for the luminosity that is in solar units.

The construction of a two-dimensional stellar model begins 
by dividing the star into $M$ mass shells and $N$ angular 
zones. The mass shells are assigned a value $s_i=\log m_i$, 
where $m_i$ is the interior mass at the midpoint of shell i. 
The angular zones are assigned a value $\theta_j$. A starting 
(or previous in evolutionary time) model is supplied with a 
run of ($P'_{ij}$, $T'_{ij}$, $r'_{ij}$, $L_{ij}$,
$\chi'_{ij}$, $\vartheta_{ij}$, $\gamma_{ij}$) for 
i=1 to $M$ and j=1 to $N$. 

Here we take the general case $\vb{B}=(B_r,B_\theta,B_\phi)$ as the example to 
show how to solve the two-dimensional stellar structure 
equations. In order to write down the linearization 
equations, we introduce the following notations:
\begin{mathletters}
\begin{eqnarray}
\h{P} &\equiv& -\frac{Gm^2}{4\pi r^4 P_T}\frac{\rho}{\rho_m}, \\
\h{R} &\equiv& \frac{m}{4\pi r^3\rho}\frac{\rho}{\rho_m}, \\
\h{T} &\equiv& \h{P}\nabla, \\ 
\h{L}&\equiv&  \frac{1}{L_\sun} m \left( \epsilon-T\od{S_T}{t}\right)\frac{\rho}{\rho_m}, \\
\h{T}^\ell &\equiv& \h{B}^i\nabla, \mbox{ $\ell=1,2,3,10, \cdots, 15,17$}\\
\h{D}^1 &\equiv& \frac{\cot\theta}{2}\p{r'}{\theta}\left(1
  -\frac{\cot\theta}{2}\p{r'}{\theta}\right)^{-1}, \\
\h{D}^2 &\equiv& \left(1-\frac{\cot\theta}{2}\p{r'}{\theta}\right)^{-1}, \\
\h{D}^3 &\equiv& \frac{\cot\theta}{2}\p{P'}{\theta}
  \left(1-\frac{\cot\theta}{2}\p{r'}{\theta}\right)^{-1}, \\
\h{D}^{10} &\equiv & \h{D}^2\left(\p{\chi'}{s}+\p{\rho'}{s}+\p{\gamma''}{s}\right),\\
\h{D}^{11} &\equiv & -\h{D}^{10}\p{r'}{\theta},\\
\h{D}^{12} &\equiv & \h{D}^2\left(\p{\chi'}{\theta}+\p{\rho'}{\theta}+\p{\gamma''}{\theta}\right), \\
\h{D}^{13} &\equiv & \h{D}^2\left(\p{\chi'}{s}+\p{\rho'}{s}+\p{\vartheta''}{s}\right), \\
\h{D}^{14} &\equiv & -\h{D}^1\left(\p{\chi'}{s}+\p{\rho'}{s}+\p{\vartheta''}{s}\right), \\
\h{D}^{15} &\equiv & \h{D}^2\frac{\cot\theta}{2}\left(\p{\chi'}{\theta}+\p{\rho'}{\theta}
+\p{\vartheta''}{\theta}\right), \\
\h{D}^{17} &\equiv & -\h{D}^2, \\
\h{F}^1 &\equiv & \frac{4ac}{3L_\sun}\frac{mT^4}{r^2\kappa\rho\rho_m}(1+\lambda), \\
\h{F}^2 &\equiv & \frac{1}{2L_\sun}\frac{m}{r^2\rho_m}
\frac{\rho C_p T l_m v_{\s{conv}}}{1+v_{\s{conv}}/v_0},\\
\h{F}^3 &\equiv & -\h{F}^2\nabla'_{\s{ad}}, \\
\h{F}^4 &\equiv & \h{F}^1\frac{Gm\rho\nabla}{rP_T}, \\
\h{F}^5 &\equiv & \h{F}^2\frac{Gm\rho\nabla}{rP_T}, \\
\h{F}^6 &\equiv & \h{F}^3\frac{Gm\rho}{rP_T}.
\end{eqnarray}
\end{mathletters}
Consequently, the stellar structure equations in the general
case can be rewritten as follows:
\begin{mathletters}
\ba
\p{P'}{s} &=& \h{P} + \sum_{i=1,2,3,10}^{15,17} \h{B}^i \h{D}^i +O(2), \label{eq:pp} \\
\p{T'}{s} &=& \h{T} + \sum_{i=1,2,3,10}^{15,17}\h{T}^i\h{D}^i +O(2), \label{eq:tp} \\
\p{r'}{s} &=& \h{R}, \label{eq:rp} \\
\p{L}{s} &=& \h{L} + \left(\sum_{\ell=1}^2\h{F}^\ell\p{T'}{\theta} 
  + \h{F}^3\p{P'}{\theta}  +\sum_{\ell=4}^6\h{F}^\ell\p{r'}{\theta}\right)\cot\theta
+O(2), \label{eq:lp}
\ea
\end{mathletters}
where $\nabla=\nabla_{\s{rad}}$ in the radiative zone, $\nabla=\nabla_c$ in the convective zone.

We calculate the derivatives of the dependent variables with respect 
to $s$ by the central difference scheme, e.g.,
\be
\frac{P'_{ij}-P'_{i-1j}}{s_i-s_{i-1}} = \frac{1}{2}\left[ \left( \p{P'}{s}\right)_{ij}
  + \left( \p{P'}{s}\right)_{i-1j} \right],
\ee
but we simply use the difference scheme
\be
  \frac{r'_{ij}-r'_{ij-1}}{\theta_j-\theta_{j-1}} = \left(\p{r'}{\theta}\right)_{ij} 
\ee
to calculate the derivatives with respect to $\theta$ because 
the first two of the two-dimensional stellar structure equations 
are singular at the poles. Thus, we can define a set of functions 
that should vanish at the solution of the stellar structure equations,
\begin{mathletters}
\ba
F_P^{ij} &\equiv& (P'_{ij}-P'_{i-1j}) -\frac{1}{2}\Delta s_i[ (\h{P}_{ij}+\h{P}_{i-1j}) 
       + \sum_{\ell=1,2,3,10}^{15,17}( \h{B}^\ell_{ij}+ \h{B}^\ell_{i-1j} )\h{D}^\ell], \label{eq:fp} \\
F_T^{ij} &\equiv & (T'_{ij}-T'_{i-1j}) -\frac{1}{2}\Delta s_i[ (\h{T}_{ij}+\h{T}_{i-1j})
       + \sum_{\ell=1,2,3,10}^{15,17}(\h{T}^\ell_{ij}+\h{T}^\ell_{i-1j})\h{D}^\ell], \label{eq:ft} \\
F_R^{ij} &\equiv & (r'_{ij}-r'_{i-1j}) -\frac{1}{2}\Delta s_i (\h{R}_{ij}+\h{R}_{i-1j}), \label{eq:fr} \\
F_L^{ij} &\equiv & (L_{ij}-L_{i-1j}) -\frac{1}{2}\Delta s_i \{(\h{L}_{ij}+\h{L}_{i-1j}) \nob
&&  +\sum_{\ell=1}^2 (\h{F}^\ell_{ij} + \h{F}^\ell_{i-1j}))
  \frac{\cot\theta_j}{\Delta\theta_j}(T'_{ij}-T'_{ij-1})  \nob
&&   + (\h{F}^3_{ij} + \h{F}^3_{i-1j}) \frac{\cot\theta_j}{\Delta\theta_j}(P'_{ij}-P'_{ij-1}) \nob
&&  + \sum_{\ell=4}^6(\h{F}^\ell_{ij} + \h{F}^\ell_{i-1j}) 
  \frac{\cot\theta_j}{\Delta\theta_j}(r'_{ij}-r'_{ij-1})  \}, \label{eq:fl}
\ea
\end{mathletters}
where $\Delta s_i\equiv (s_i-s_{i-1})$ and i=2 to $M$, j=2 
to $N$. $\h{D}^1_{ij}$, $\h{D}^{10}_{ij}, \cdots, \h{D}^{14}_{ij}$ 
are defined as follows:
\begin{mathletters}
\ba
\h{D}^1 &=& \frac{\cot\theta_j}{2\Delta\theta_j}(r'_{ij}-r'_{ij-1})\left[1
    -\frac{\cot\theta_j}{2\Delta\theta_j}(r'_{ij}-r'_{ij-1})\right]^{-1}, \\
\h{D}^2 &=& \left[1-\frac{\cot\theta_j}{2\Delta\theta_j}(r'_{ij}-r'_{ij-1})\right]^{-1}, \\
\h{D}^3 &=& \frac{\cot\theta_j}{2\Delta\theta_j}(P'_{ij}-P'_{ij-1})\left[1
    -\frac{\cot\theta_j}{2\Delta\theta_j}(r'_{ij}-r'_{ij-1})\right]^{-1}, \\
\h{D}^{10} &=& \frac{\h{D}^2}{\Delta s_i}[(\chi'_{ij}-\chi'_{i-1j}) + (\rho'_{ij}-\rho'_{i-1j}) 
  + (\gamma''_{ij}-\gamma''_{i-1j}) ], \\
\h{D}^{11} &=& -\frac{\h{D}^{10}}{\Delta\theta_j}(r'_{ij}-r'_{ij-1}), \\
\h{D}^{12} &=& \frac{\h{D}^2}{\Delta \theta_j}[(\chi'_{ij}-\chi'_{ij-1}) + (\rho'_{ij}-\rho'_{ij-1}) + (\gamma''_{ij}-\gamma''_{ij-1})],\\
\h{D}^{13} &=& \frac{\h{D}^2}{\Delta s_i}[(\chi'_{ij}-\chi'_{i-1j}) + (\rho'_{ij}-\rho'_{i-1j}) + (\vartheta''_{ij}-\vartheta''_{i-1j})],\\
\h{D}^{14} &=& -\frac{\h{D}^1}{\Delta s_i}[(\chi'_{ij}-\chi'_{i-1j}) + (\rho'_{ij}-\rho'_{i-1j}) + (\vartheta''_{ij}-\vartheta''_{i-1j})],\\
\h{D}^{15} &=& \h{D}^2\frac{\cot\theta_j}{2\Delta \theta_j}[(\chi'_{ij}-\chi'_{ij-1}) + (\rho'_{ij}-\rho'_{ij-1}) + (\vartheta''_{ij}-\vartheta''_{ij-1})],\\
\h{D}^{17} &=& -\h{D}^2.
\ea
\end{mathletters}
We want then to solve for the set of ($P'_{ij}$, 
$T'_{ij}$, $r'_{ij}$, $L_{ij}$) such that 
$F_P^{ij}=F_T^{ij}=F_R^{ij}=F_L^{ij}=0$ with $\chi'$, $\vartheta$, and 
$\gamma$ specified.

The linearization of Eqs.~(\ref{eq:fp}-\ref{eq:fl}) with respect 
to ($\delta P'_{ij}$, $\delta T'_{ij}$, $\delta r'_{ij}$, 
$\delta L_{ij}$) yields $4MN-4(N-1)-4M$ equations 
for the $4MN$ unknowns. The $2(N-1)$ additional 
equations are supplied by the boundary conditions at 
the center. From Eqs.~(\ref{eq:cb1}-\ref{eq:cb2}), we can define
\begin{mathletters}
\ba
  F^{1j}_R &\equiv & r'_{1j}-\frac{1}{3}[s_1-\log(4\pi\rho_{m1j}/3)], \\
  F^{1j}_L &\equiv & L_{1j}-\h{L}_{1j} 
 -\frac{\cot\theta_j}{\Delta\theta_j}\left[\sum_{\ell=1}^2\h{F}^\ell_{1j}(T'_{1,j}-T'_{1j-1}) \right. \nob
&& \left.  +\h{F}^3_{1j}(P'_{1j}-P'_{1j-1}) +\sum_{\ell=4}^6 \h{F}^{\ell}_{1,j}(r'_{1j}-r'_{1j-1})\right],
\ea
\end{mathletters}
where j = 2 to $N$. Another $2(N-1)$ additional 
equations are supplied by the boundary conditions at 
the surface. From Eqs.~(\ref{eq:sb1}-\ref{eq:sb2}), we can define
\begin{mathletters}
\ba
  F^{M+1j}_R &\equiv & R'_{Mj} - a_1 P'_{Mj} - a_2 T'_{Mj} - a_3, \\
  F^{M+1j}_L &\equiv & L_{Mj}' (\ln L_{Mj}- a_4 P'_{Mj} - a_5 T'_{Mj} - a_6),
\ea
\end{mathletters}
where j = 2 to $N$. The $4M$ additional equations are 
supplied by the polar boundary conditions,
\begin{mathletters}
\ba
 F^{i1}_P &\equiv & P'_{i1}-P'_{i2}, \label{eq:fi1p} \\
 F^{i1}_T &\equiv & T'_{i1}-T'_{i2}, \label{eq:fi1t} \\
 F^{i1}_R &\equiv & R'_{i1}-R'_{i2}, \label{eq:fi1r} \\
 F^{i1}_L &\equiv & L_{i1}-L_{i2}, \label{eq:fi1l}
\ea
\end{mathletters}
where i = 1 to $M$. The F equations are linearized,
\be
  \sum_{l=1}^{M}\sum_{k=1}^{N}\left(\p{F^{ij}_w}{R'_{lk}}\delta R'_{lk} +\p{F^{ij}_w}{L_{lk}}\delta L_{lk}
 +\p{F^{ij}_w}{P'_{lk}}\delta P'_{lk} +\p{F^{ij}_w}{T'_{lk}}\delta T'_{lk}\right)=-F^{ij}_w, \label{eq:linear}
\ee
where $w = P, T, R, L$; $i = 1$ to $M$; and j = 1 to $N$. 
The summation over l has non-zero terms only for l = i-1, i;
the summation over k has non-zero terms only for k = j-1, j. 
See appendix A for the matrix coefficients.

\subsection{Solution of the linearized equations}

\begin{figure}
\plotone{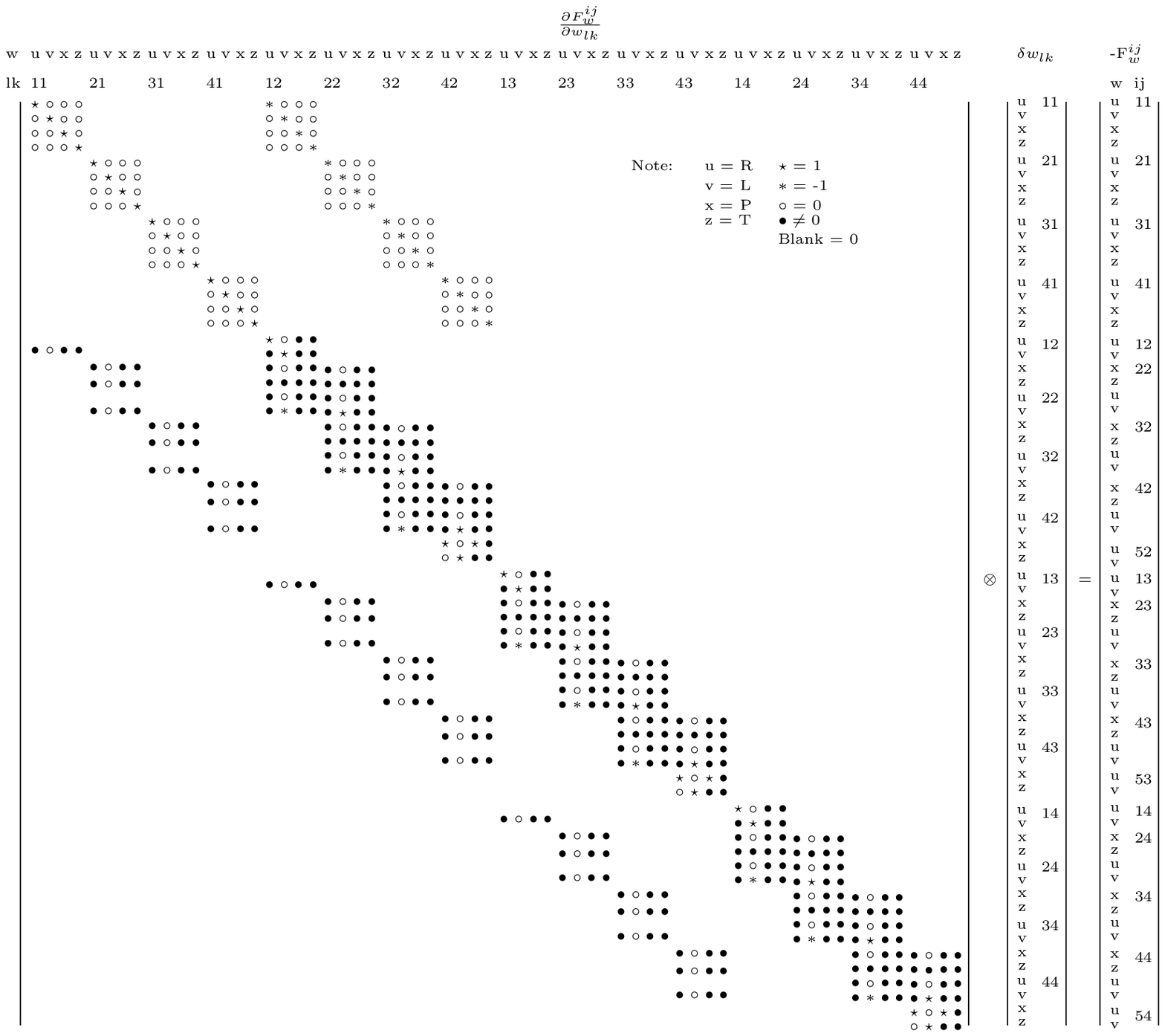}
\caption{Linearization equation for a $4\times4$-point star.}\label{fig:coeff}
\end{figure}

Rather than solving the $(4MN)^2$ system of equations directly, 
we takes advantage of the specific form of the equations and especially 
of the large number of zero elements in the matrix. From Fig.~\ref{fig:coeff} 
we can see that only 12 by $4MN$ elements are non-zero at most. 

\begin{figure}
\plotone{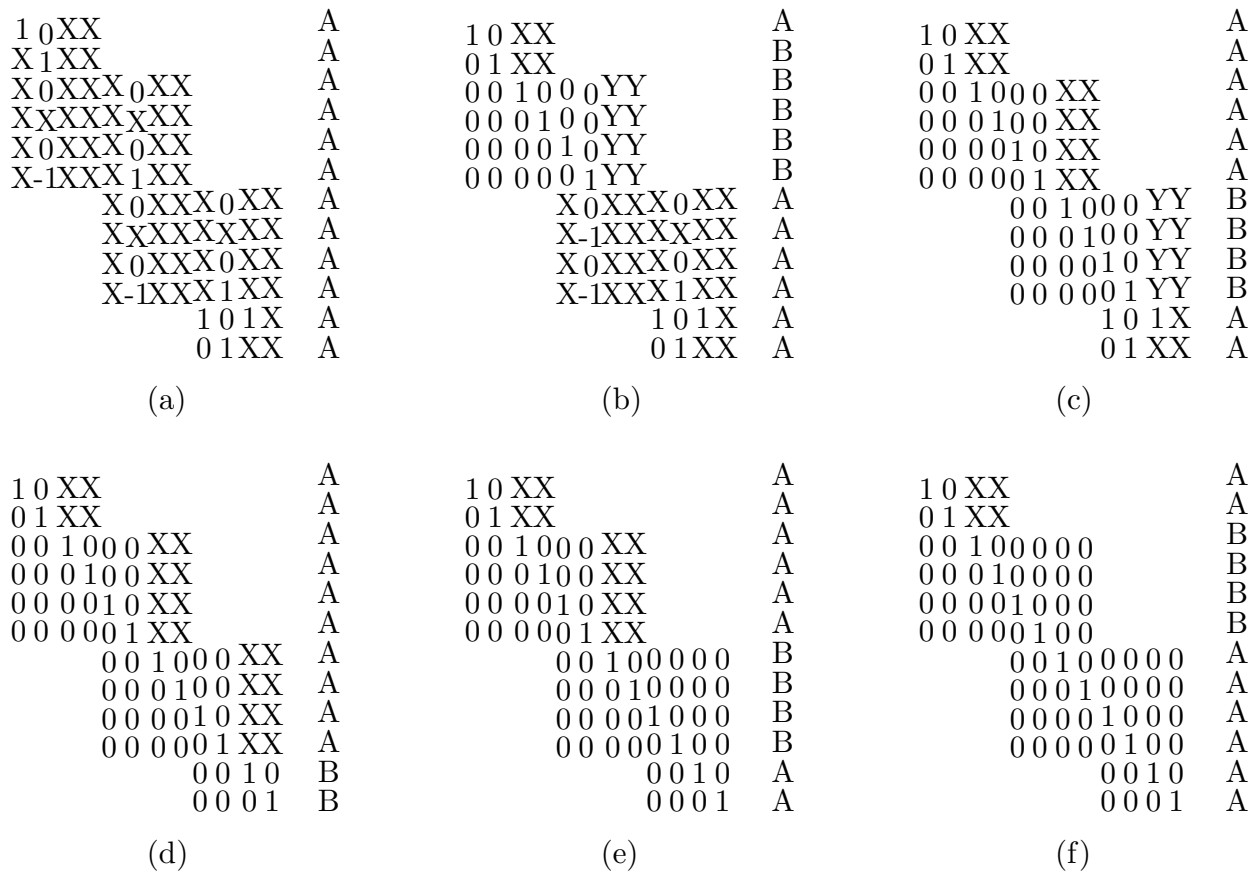}
\caption{Schematic Henyey solution for a 4-point star. The matrix block
is denoted by 0's, 1's, and X's are non-zero. The right hand side is denoted by A, 
the elements changed through pivoting, by Y and B. The final reduction to the
identity matrix is not shown.}\label{fig:heyey}
\end{figure}

The matrix is reduced in a forward direction (i = 2 to $M$) as the
coefficients are defined and is then solved in the backward direction
(i = $M$ to 1) for the corrections ($\delta P'_{ij}$, $\delta T'_{ij}$,
$\delta R'_{ij}$, $\delta L_{ij}$) for j = 2 to $N$. 
The reduction procedure begins:
(i) For j = 2, to use the polar conditions at $\theta=0$ to eliminate those elements 
with subscriptions l = i, k = j - 1 (i.e., Block III defined in Appendix A), 
which can be done by simply adding Block III to Block II.
At the end of this step, the matrix equation for a specified j 
looks like Fig.~\ref{fig:heyey}a for a 4-point star in the mass coordinate 
(including the center and surface boundaries);
(ii) to use the central boundary conditions to eliminate the first two columns 
in Block I for i = 2;
(iii) to continue diagonalizing the four bottom rows for i = 2;
(iv) to store the right-hand side and the elements in the rightmost 
columns, see Fig.~\ref{fig:heyey}b.
After this reduction is completed, the bottom two rows of the first 
part of the coefficient matrix become the ``central boundary equations'' 
for the F equations of the next pair of mass points. 
The method is repeatedly applied until the surface is reached,
whereupon the surface boundary conditions complete the set of $4M$
equations, see Fig.~\ref{fig:heyey}b-c. For the back solution (i) the values
of ($\delta P'_{\s{M2}}$, $\delta T'_{\s{M2}}$) are first calculated, (ii)
then the values of ($\delta R'_{i2}$, $\delta L_{i2}$, $\delta P'_{i-1\,2}$, $\delta
T'_{i-1\,2}$) for i = $M$ to 2 are calculated using the stored elements of the
array and ($\delta P'_{i2}$, $\delta T'_{i2}$), (iii) and finally the values of
($\delta R'_{12}$, $\delta L_{12}$) are computed from the central boundary
conditions and the values of ($\delta P'_{12}$, $\delta T'_{12}$), see
Figure.~\ref{fig:heyey}d-f. Since the sub-matrix with j = 2 has been diagonalized, 
we can use it to diagonalize the sub-matrix with j = 1 and 3. For j = 3, we use 
j = 2 as the ``polar boundary conditions'', and so forth. Finally, we solve the 
matrix equation, the results are stored in the right column.

\subsection{Advancing the model}

These routines are based on the work of Prather (1976) and their revised 
implementations in YREC (Pinsonneault 1988; Guenther et al. 1992; 
Guenther \&\ Demarque 1997).

\subsection{Time steps}

In this section we use the cgs unit for luminosity $L$ (erg/s) and use $X$ 
$(Y)$ to represent the mass fraction of hydrogen (helium). The angular zone 
index (i.e., j) is 2.

The timing routine calculates the time steps based upon a hydrogen- 
or helium-burning source. Let $L_{\s{H}}$ (erg/s) be the total 
hydrogen-burning luminosity, and $L_{\s{He}}$, the helium luminosity. 
There are two time steps,
\Ba
   \Delta t_{\s{H}} &=& \mbox{ hydrogen-burning time step}, \\
   \Delta t_{\s{t}} &=& \mbox{ total time step (i.e., for entropy and helium)},
\Ea
where $\Delta t_{\s{t}} \ne \Delta t_{\s{H}}$ only if the hydrogen shell is 
being shifted outward. If $L_{\s{H}} = 0 $, the following section for 
hydrogen burning is skipped.

For hydrogen-core burning ($X_{\s{core}} > X_{\s{c}}^{\s{min}}$), 
a time step corresponding to a set of reduction in $X_{\s{core}}$ 
is calculated. Let i be the innermost point if the core is radiative 
(i = 1) or the outermost convective point if the core is convective. 
Then, the change in $X_{\s{core}}$ is computed,
\[
  \Delta X_{\s{core}} = \min\{\Delta X_{s{c}}^{\s{max}},\Delta f_{\s{X}}^{\s{max}}\cdot X^i\},
\]
and the time step is
\[
  \Delta t_i = \Delta t_{\s{H}} = 6\cdot 10^{18}\cdot\Delta X_{\s{core}}\cdot m^i/L^i,
\]
where $m^i$ is the mass of the core (gm) and $L^i$ is the luminosity of the core 
(erg/s, assumed to be mainly hydrogen burning).

When the core-burning criterion no longer applies ($X_{\s{core}}<X_{\s{c}}^{\s{min}}$), 
a limit is placed on the total amount of mass that may be burned,
\Ba
  \Delta m &=& \Delta f_{\s{m}}\cdot M_{\sun}\cdot X_{\s{env}}, \\
  \Delta t^{\s{m}}_{\s{H}} &=& 6\cdot 10^{18}\cdot \Delta m/L_{\s{H}}.
\Ea
If there is a hydrogen-burning shell ($X_{\s{core}}=0$), the timing 
routine locates it. Let the subscript $_0$ denote the inner edge of 
the shell (first point where $X>0$); let $_{\s{1/2}}$ denote the 
mid-point of the shell ($X=\frac{1}{2} X_{\s{env}}$); and let $_1$ 
denote the end of the shell ($L^i-L^{i-1} < 10^{-4}\cdot L$ or 
$X=X_{\s{env}}$ or $\epsilon_{\s{H}}=0$). There is a limitation 
set on the maximum depletion at the mid-point of the shell,
\[
  \Delta t^{\s{1/2}}_{\s{H}} = \Delta X_{\s{1/2}}.
\]
With the exception of the core-burning phase, the new hydrogen 
burning time step is limited by the previous total time step,
\[
  \Delta t_{\s{H}} \mbox{( new)} = min\{1.5\cdot\Delta t_{\s{t}}
   \mbox{(old)}, \Delta t_{\s{H}}^m,\Delta t_{\s{H}}^{\s{1/2}}\}.
\]
If there is to be no shell shifting then one sets $\Delta t_t = 
\Delta t_{\s{H}}$. If the hydrogen shell is to be shifted outward 
through $\Delta m_s$ in mass, then the shift time step is computed as
\[
  \Delta t_{\s{shift}} = 6\cdot 10^{18}\cdot X_1\cdot \Delta m_s/L_{\s{H}},
\]
and the total time step is
\[
  \Delta t_t = \Delta t_{\s{H}} + \Delta t_{\s{shift}}.
\]

If there is a hydrogen shell ($X_{\s{core}}=0$), the helium 
burning is examined. For helium-core burning ($Y_{\s{core}}> 
X_{\s{c}}^{\s{min}}$ and $L_{\s{core}}>0.1 L_{\sun})$, the
 maximum helium depletion is
\[
\Delta Y_{\s{core}} = \min\{\Delta Y_{\s{c}}^{\s{min}},
  \Delta f_{\s{Y}}^{\s{max}}\cdot Y_{\s{core}}\},
\]
and the helium time step is
\[
  \Delta t_{\s{He}}=5.85\cdot 10^{17}\cdot \Delta Y_{\s{core}}\cdot M_\sun/L_{\s{core}}.
\]
For helium-shell burning ($Y_{\s{core}}< X_{\s{c}}^{\s{min}}$), 
the amount of mass burned through by the helium shell is limited,
\[
  \Delta t_{\s{He}} = 5.85\cdot 10^{17}\cdot\Delta f_{s{m}}\cdot M_\sun/L_{\s{He}}.
\]
The helium time step places an upper limit on the previously 
computed hydrogen time step,
\Ba
  \Delta t_t &=& \min\{\Delta t_t,\Delta t_{\s{He}}\}, \\
  \Delta t_{\s{H}} &=& \min \{\Delta t_t,\Delta t_{\s{H}}\}.
\Ea

The following parameters used in the determination of the time 
step are read in at the beginning of each model run. Their 
typical values are given as follows:
\Ba
\begin{array}{lll}
X_{\s{c}}^{\s{min}} = 0.001, & \Delta X_{\s{c}}^{\s{max}} = 0.04, & \Delta Y_{\s{c}}^{\s{max}} = 0.02, \\
\Delta f_{\s{X}}^{\s{max}} = 0.5, & \Delta f_{\s{Y}}^{\s{max}} = 0.3, & \Delta f_{\s{m}} = 0.0015 M_\sun, \\
\Delta X_{\s{1/2}}^{\s{max}} = 0.10, & \Delta m_s = 5\cdot 10^{-4} M_\sun.  
\end{array}
\Ea

Of course, we can also use a fixed time step to advance the model.

\subsection{Composition advance}

The mixing routine performs all the operations on the model that are 
needed by the application of the time step to increase the age of the model. 
The routine first checks that there is no mixing within the hydrogen shell if the 
shell is supposed to be shifted. If there is such mixing, the shifting is suppressed 
(i.e., set $\Delta t_t=\Delta t_{\s{H}}$).

Each mass element (the mass contained in the zone defined by $m\in [m_{i-1}, m_{i}]$ and 
$\theta\in[\theta_{j-1},\theta_j]$) is burned individually by computing the energy generation 
rates for the physical conditions existing in that mass element from the previously converged 
model. Since the program stores only the values of hydrogen, total metal and oxygen abundance,
the change in these quantities is computed as
\Ba
  X(\mbox{new}) &=& X(\mbox{old}) - (dX/dt)\cdot \Delta t, \\
  Z(\mbox{new}) &=& Z(\mbox{new}) + (dY/dt)\cdot \Delta t, \\
  X_{\s{16}}(\mbox{new}) &=& X_{\s{16}}(\mbox{old}) - (dX_{\s{O}}/dt)\cdot \Delta t,
\Ea
where $\Delta t = \Delta t_t$ inside the hydrogen shell ($X=0$) and $\Delta t = \Delta t_{\s{H}}$ 
elsewhere.

The routine then mixes those zones that it is instructed to do by being given a set of 
indices (i = $i_1$ to $i_2$ and j = $j_1$ to $j_2$),
\[
  X_{ij} = \left( \sum_{k=i_1}^{k=i_2} \sum_{l=j_1}^{l=j_2} a_{kl} 
    \cdot X_{kl}\right)\cdot\left( \sum_{k=i_1}^{k=i_2} \sum_{l=j_1}^{l=j_2} a_{kl}\right)^{-1}.
\]
The weights $a_{kl}$ are proportional to the amount of mass associated with zone kl are set 
up in the point readjustment routine.

If the hydrogen shell is to be shifted, the routine calculates
\[
  \Delta s_{\s{shift}} = (\delta -\delta^2/2 + \delta^3/3-\delta^4/4)/\ln 10,
\]
where $\delta\equiv \Delta m_s/m_{\s{1/2}}<<1$. The points in the hydrogen shell are 
shifted by $\Delta s_{\s{shift}}$,
\[
  s_0 \le s_i \le s_1 \rightarrow s_i(\mbox{new}) = s_i(\mbox{old}) + \Delta s_{\s{shift}},
\]
where $s_i=\log m_i$. The points up to a distance $f_s\cdot\Delta s_{\s{shift}}$ in front 
of the shell are squeezed together,
\[
  s_1<s_i<s_{\s{end}} \rightarrow s_i(\mbox{new}) = s_i(\mbox{old}) 
    + s_{\s{end}}-s_i(\mbox{old})]/f_s,
\]
where $s_{\s{end}}\equiv s_i+f_s\cdot \Delta s_{\s{shift}}$.

For all of these shifted and squeezed points the changes in $P'$ and $T'$ 
must be preserved for the calculation of the entropy energy term in the subsequent 
model. Thus for every $s_i(\mbox{new})$, one must locate $s_l(\mbox{old})$ such 
that $s_l(\mbox{old})\le s_i(\mbox{new})<s_{l+1}(\mbox{old})$ and then interpolate
linearly in $s$ to get the old values of $P'$ and $T'$ which correspond to
the new value of s. Then the effective changes are stored,
\Ba
  \Delta P'_{ij} &=& P'_{ij}(\mbox{new s}) - P'_{ij}(\mbox{pre-shift s}), \\
  \Delta T'_{ij} &=& T'_{ij}(\mbox{new s}) - T'_{ij}(\mbox{pre-shift s}).
\Ea
For the region in front of the shell that is squeezed, it is desirable to 
preserve the original composition gradient if such a gradient exists. The 
values of $X$, $Z$ and $X_{\s{16}}$ are interpolated linearly in s as are 
$P'$ and $T'$. Note that the shifting process affects only the value of s 
and not the values of (P, T, R, L, X, Z, X$_{16}$) with the exception of 
(X, Z, X$_{16}$) in the squeezed region.

The mixing routine finally checks on the physical sense of the new composition 
at all of the points,
\Ba
  X &=& \max\{X, 0\}, \\
  Z &=& \min\{ Z, 1-X\}, \\
  X_{16} &=& \max\{X_{16}, 0.99\cdot 10^{-3}\cdot Z_{\s{CNO}}\}.
\Ea
The first two requirements are obvious; the third requirement brings 
the value of $X_{16}$ up to the approximate equilibrium value while 
turning off the $X_{16}$ burning rate that is calculated if 
$X_{16}>10^{-3}\cdot Z_{\s{CNO}}$. The value of $Z_{\s{CNO}}=Z-Z_m^0$ 
where $Z_m^0$ is the original weight abundance of all non-CNO metals. 
This method allows for the enrichment of CNO elements from the helium 
burning.

\subsection{Mixing zones}

Consecutive mass shells, which are determined to be convective 
($\nabla_{\s{rad}}>\nabla_{\s{ad}}$) in the previously converged model, 
are mixed together.

If there is a helium-burning convective zone, the semi-convective 
instability is treated as an over-shooting (Castellani et al 1971).
The composition is first burned and mixed according to the standard 
convection zones. At the first radiative point outside a helium convective 
zone, the quantity $f\equiv \nabla_{\s{rad}}^{\s{int}}/\nabla_{\s{rad}}^{\s{ext}}$ 
is defined where the radiative gradient is computed with the (s, P, T, r, L) values 
of the radiative point and with the composition of both the radiative point ($^{\s{ext}}$)
 and the interior convective zone ($^{\s{int}}$). The original convective zone is extended 
outward through the radiative region for all the points at which 
$f\nabla_{\s{rad}}>\nabla_{\s{ad}}$. 

This over-shooting region is restricted to the helium core (X = 0) and is limited by 
the condition of Castellani et al (1971) that defines a maximum radius $R_{\s{max}}$ of 
the over-shooting mixing,
\[
  \int^{R_{\s{max}}}_{R_c}(1-\mu(r)/\mu_c^{\s{int}})dr
<(1- \nabla_{\s{ad$_c$}}^{\s{int}}/ \nabla_{\s{rad$_c$}}^{\s{int}})L_c\Delta t/(40\pi P_c R_c^2), 
\]
where the subscript $_c$ refers to the (s, P, T, r, L) values at the edge of the original 
convective zone. Here $\mu$ is the mean molecular weight. The composition is then 
re-mixed from the beginning of the convective zone 
to the maximum extent of the over-shoot region.

\subsection{Point readjustment}

The point readjustment routine reflects all of the points between
successive models. This routine starts with the central point and 
places each subsequent new point i so that all of the following criteria are met:
\Ba
 && s_i-s_{i-1}\le \Delta s_{\s{max}}, \\
 && P'_{i2}-P'_{i-12} \le \Delta P'_{\s{max}}, \\
 && L_{i2} - L_{i-12} \le \Delta f_{\s{L}}\cdot L_{M2}.
\Ea
All of the new values are interpolated linearly in s by locating the old 
point l such that $s_l(\mbox{old})\le s_i(\mbox{new})<s_{l+1}(\mbox{old})$.
The fundamental variables (s, $P_T$, T, R, L), the composition (X, Z, X$_{16}$),
the density and the entropy terms ($\Delta P'$, $\Delta T'$) are relocated between 
the center and outermost points for all angular zones. These variables are stored 
in temporary arrays and are transferred to the original arrays once the process 
is completed. In addition to the $1^{\s{st}}$ and $M^{\s{th}}$ points remaining 
fixed, other points may be retained:
\begin{description}
\item{1)} the first radiative point (outer edge of convective zone),
\item{2)} the innermost point of the convective envelope,
\item{3)} the edge of the helium core (X = 0),
\item{4)} composition discontinuities, $X_{l2}-X_{l-12}>\Delta X_{\s{disc}}$ or
  $Z_{l2}-Z_{l-12}>\Delta Z_{\s{disc}}$.
\end{description}

The point routine then recalculates the weights assigned to each mass shell based 
upon the mass values at the preceding and following mid-points,
\Ba
 m_i &=& 10^{s_i}, \\
 a_1 &=& \frac{1}{2} (m_1+m_2), \\
 a_i &=& \frac{1}{2} (m_{i+1}-m_{i-1}) \mbox{ for i = 2 to $M-1$}, \\
 a_{M} &=& M_{\s{tot}} - \frac{1}{2}(m_{M} + m_{M-1}).
\Ea
The value $m_i$ defines the location of the i$^{\s{th}}$ shell, and $a_i$ 
is the number of grams contained in the shell.

Additionally, the point routine adjusts the temperature of the outermost 
$M^{\s{th}}$ point by adding a new point or deleting some old points.
Given the desired temperature range $T_{\s{min}}$ to $T_{\s{max}}$, then 
if $T_{M}<T_{\s{min}}$ the outermost point $j<M$ such that $T_l > 
\overline{T}\equiv \frac{1}{2} (T_{\s{min}}+T_{\s{max}})$ is selected 
as the new surface point. The points $l+1$ to $M$ are deleted.
If $T_{M2}>T_{\s{max}}$ the process is more complicated. The 
last atmosphere that was integrated will have stored the values of 
(s$_{\s{atm}}$, $P_{\s{atm}j}$, $T_{\s{atm}j}$, $r_{\s{atm}j}$) for the 
first inward integration step in which $T_{\s{atm}j}>\overline{T}$. 
The new point $M+1$ is added with the following values,
\Ba
  s_{\s{M+1}} = s_{\s{atm}} &&  P'_{\s{M+1j}} = P'_{\s{atm}j} \\
  T'_{\s{M+1j}} = T'_{\s{atm}j} &&  r'_{\s{M+1j}} = r'_{\s{atm}j} \\
  L_{\s{M+1j}} = L_{\s{atm}j} &&  X_{\s{M+1j}} = X_{\s{Mj}} \\
  Z_{\s{M+1j}} = Z_{\s{Mj}} &&  X_{16_{\s{M+1j}}} = X_{16_{\s{Mj}}}
\Ea
 
\subsection{Model calculation sequence}

The following list describes the sequence of calculations that is used in 
computing a series of stellar models. This sequence is the same for both one- 
and two-dimensional model calculations.
\begin{description}
\item{(0)} Input a model and compute a time step.
\item{(1)} Locate the mixing zones and advance the composition and 
  hydrogen shell for the given time step.
\item{(2)} Calculate element diffusion for the given time step.
\item{(3)} Readjust the points in the mass coordinate in the model. 
  This step is the main source of numerical errors and should be 
  switched off for high precision calculations such as solar variability applications.
\item{(4)} Calculate the entropy terms ($\Delta P'$ and $\Delta T'$). 
  Just zero them at the beginning, and give an estimate using their 
  temporal change rate times the given time step.
\item{(5)} Add the predictable corrections to ($P',\ T',\ r',\ L$) if 
 their temporal change rates are available (after advancing one time step). 
 This allows us to use a much larger time step and save a lot of computation time.
\item{(6)} Specify the magnetic field configuration by selecting the functions 
 $\chi(m,\theta)$, $\vartheta(m,\theta)$ and $\gamma(m,\theta)$. 
\item{(7)} Retaining the old surface (or envelope) triangle and surface 
 boundary conditions, do 2 iterations for corrections to the dependent 
 variables ($P',\ T',\ r',\ L$) and apply a given fraction ($\le 100\%$) of the 
 corrections.
\item{(8)} If necessary, relocate the surface triangle for the partially 
 converged model and compute new atmospheres and surface boundary conditions.
\item{(9)} Iterate until the model converges. 
\item{(10)} Refine composition and iterate until the model converges for 
 solar applications that need a high precision.
\item{(11)} Repeat (9) once for solar applications.
\item{(12a)} If the corrections are excessively large at any time 
 or if the model does not converge after many iterations (say 20), 
 then retain the previous model that has been stored on disk and stop.
\item{(12b)} If the model has converged,
\begin{description}
  \item{(i)} compute a new time step,
  \item{(ii)} perform the requested printing,
  \item{(iii)} store the model temporarily on disk, overwrite the previous model,
  \item{(iv)} return to step (1).
\end{description}
\end{description}

\section{Test 1: Two-dimensional standard solar model}\label{sec:ssm}

In this test, we investigate how different resolutions and different 
boundary conditions affect the two-dimensional solar models in the 
standard case (zero-magnetic field). 

Starting from a one-dimensional ZAMS (Zero Age Main Sequence) model, 
we move the fit point to the surface where the mass coordinate 
$s=1\cdot 10^{-14}$ from the usual location $s=1\cdot 10^{-5}$ 
in a stair stepping way. The (ZAMS and the advanced) models are 
determined by the following parameters: the minimum and maximum 
change in $s$ between Henyey grid points, $1\cdot 
10^{-12}\le\Delta s \le 8\cdot 10^{-2}$; the maximum change in $w'$ (=$P'$, $T'$, 
$r'$ and $L/L_{\sun}$) between Henyey grid points, is $|\delta w'|
\le 5.2834\cdot 10^{-3}$. The convergence criteria for the stellar parameters 
are $|\delta P'|\le 6\cdot 10^{-7}$, 
$|\delta T'|\le 4.5\cdot 10^{-7}$, $|\delta r'|\le 3\cdot 10^{-7}$, and 
$|\delta (L/L_{\sun})|\le 9\cdot 10^{-7}$. The convergence tolerance on the rhs of the $P$ 
and $r$ equations is $3\cdot 10^{-7}$, and the convergence tolerance on the rhs of the $L$ 
and $T$ equations is $2.5\cdot 10^{-7}$. We also require $|\delta P'/P'|\le 9$, $|\delta T'/T'|\le 5$,
$|\delta r'/r'|\le 5$, and $|\delta L/L|\le 90$. All these 
requirements must be satisfied simultaneously when we apply the correction to the model.
This is why we have to move the fit point in a stair stepping way. Otherwise, the correction 
is too large and the solution will diverge. The model has about 2401 grid points in the mass 
coordinate $s$, i.e., $M=2401$. We also test the cases with $M=1201,\ 601,\ 301$.

When this one-dimensional convergence has been obtained, the angular part of the 
two-dimensional grid is selected. Unlike the mass coordinate $s$, which is not uniform, we simply 
equally divide the angular coordinate $\theta$ in the range $\theta\in[0,\pi/2]$,
$\theta_j = (\pi/2)(j-1)/(N-1)$, where j = 1 to $N$. We use the converged one-dimensional 
model for every angular zone. We use $N=10,\ 19$, and $37$ in this test.

The solar mass is $M_\sun=1.9891\cdot 10^{33}$ gm. The initial metal mass fraction is assumed 
to be $Z=0.022$ at ZAMS. The model will evolve from ZAMS to the current age of the Sun 
(4.55 Gyr). The hydrogen mass fraction and mixing length parameter (ratio of the 
mixing length over the pressure scale height) are determined by the requirement that 
the solar model at present reproduce the observed radius ($R_\sun=6.9598\cdot 10^{10}$ cm) 
and luminosity ($L_\sun=3.8515\cdot 10^{33}$ erg/s). We first use the one-dimensional code 
to generate a one-dimensional standard solar model as the reference. We then use the 
two-dimensional code to generate the two-dimensional zero-magnetic field solar models with 
different $M$ and $N$ and different surface boundary conditions and compare 
them with the reference. Our aim is to investigate if we can get a two-dimensional 
high-precision solar model.

\subsection{Convergence}

First of all, convergence is the most important requirement in model calculations. 
There is an intrinsic divergence at the poles in 
Eqs.~(\ref{eq:pp}-\ref{eq:tp}), and (\ref{eq:lp}), which results from the 
terms that contain $\cot\theta$.
In order to solve this intrinsic divergence problem, we require both Eq.~(\ref{eq:pole1}) 
and (\ref{eq:pole2}) at and near the poles. In practice, we zero 
Eqs.~(\ref{eq:fi1p}-\ref{eq:fi1l}), where subscript `1' indicates 
the pole ($\theta=0$) and subscript `2' means the adjacent point to the pole.
The denser the grid in the second dimension, the more severe the intrinsic divergence 
problem. Therefore, it is desirable to use fewer grid points in the second dimension 
for the sake of convergence.

Since we have neglected the second-order derivatives with respect to $\theta$ that are believed to be 
smaller corrections to Eqs.~(\ref{eq:pp}-\ref{eq:tp}), and (\ref{eq:lp}) than the 
first-order derivatives with respect to $\theta$, we neglect those second-order derivatives to remove the divergence due to the numerical errors caused by them. 

There is a numerical divergence problem due to the possible equality between $r_{ij}$, 
$T_{ij}$, $P_{ij}$, and $r_{ij-1}$, $T_{ij-1}$, $P_{ij-1}$, respectively. When, say, $r_{ij}$ 
equals $r_{ij-1}$, the difference between them, $\h{R}'\equiv r_{ij}-r_{ij-1}$, vanishes. 
In this case, the derivative of the difference with respect to $r_{ij}$ ($\partial\h{R}'/
\partial r_{ij}$)or $r_{ij-1}$ ($\partial\h{R}'/\partial r_{ij-1}$) should also vanish (i.e., 
$\partial\h{R}'/\partial r_{ij}=0$, or $\partial\h{R}'/\partial r_{ij-1}=0$, when $r_{ij}=r_{ij-1}$). 
If one sets $\partial\h{R}'/\partial r_{ij}=1$ and $\partial\h{R}'/\partial r_{ij-1}=-1$ no matter 
whether $r_{ij}$ equals $r_{ij-1}$ or not, one will run into a numerical divergence problem.
We introduce the $\delta_R$, $\delta_P$, and $\delta_T$ functions in the Appendix A to solve 
this divergence problem.

The fourth divergence problem is due to the numerical error caused 
by numerical integration of $\rho_m$ that affects the ratio $\rho/\rho_m$, 
which is a two-dimensional correction factor that appears in all the stellar 
structure equations, Eqs.~(\ref{eq:pp}-\ref{eq:lp}), noticing that the intrinsic singularity requires that the fewer the grid points at $\theta$ the better. The numerical integral is usually made in terms of the trapezoidal rule, which is of the second-order in accuracy. Deupree (1990) adds more grid points to increase the integration precision when the numerical integral is performed. We find that it is more efficient to introduce a normalization factor in the integral, as shown in Appendix A.2.

When the radiative diffusion approxiamation (i.e., $\lambda=0$) is used, the code converges very well. This approxiamtion is not valid near the surface. If we use the temperature gradient at the surface to replace the actual gradient $\nabla_s$, 
the code also converges well. However, if we use the exact expression given in Eq.~(\ref{eq:ll0}), we cannot get a converged model. The main cause is due to the numerical errors in the numerical derivatives associated with $\lambda_0$.

\subsection{Resolution}

If the convergence solves the internal- or self-consistency problem, then model resolution 
will address the external-consistency issue. Our reference model, i.e., the one-dimensional 
standard solar model, is almost the same as the best model described by Winnick et 
al. (2002), who emphasize its comparison with various observations.

\begin{figure}
\plotone{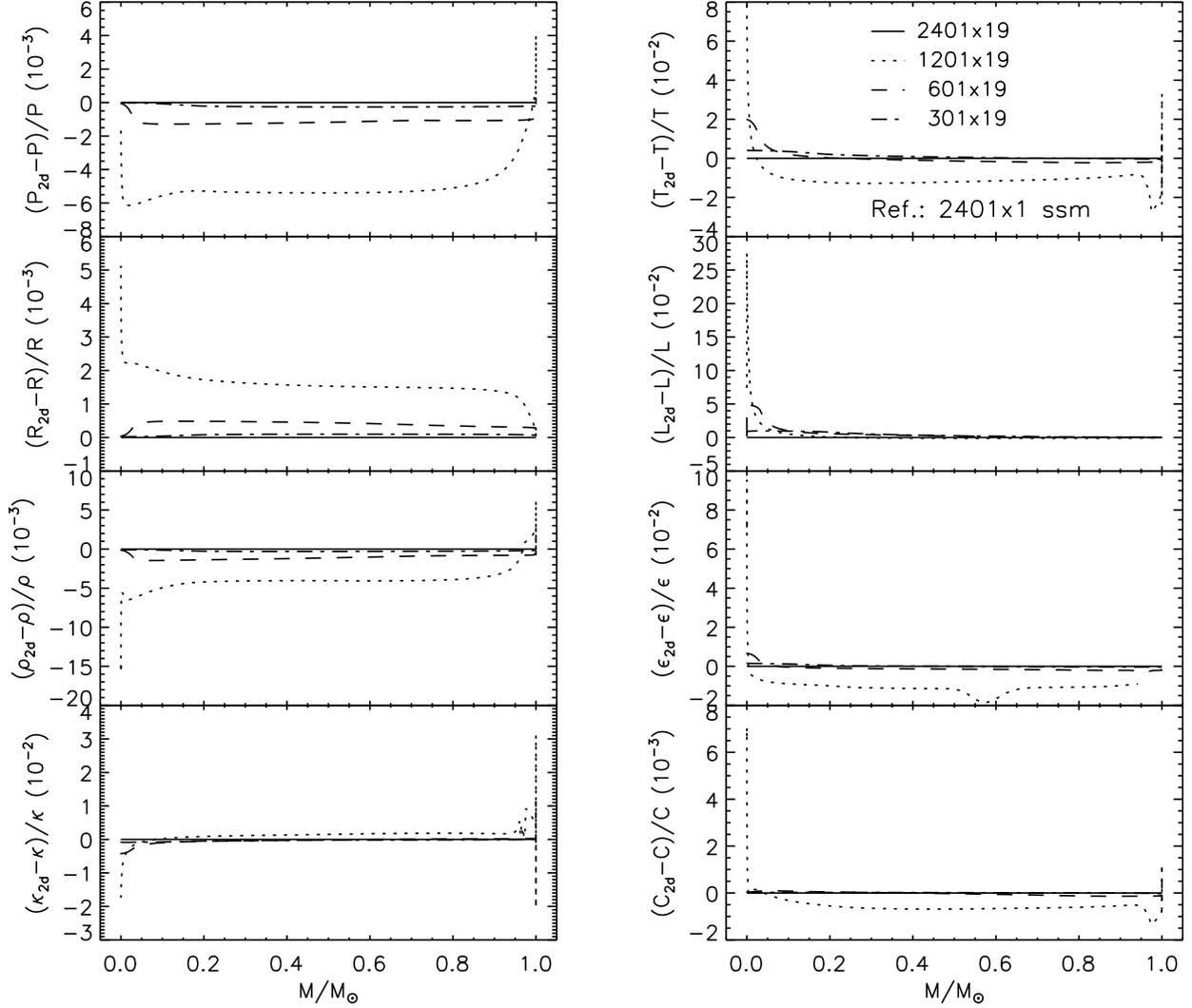}
\caption{Relative changes of pressure $P$, temperature $T$, radius $R$, luminosity $L$, 
density $\rho$, nuclear energy generation rate $\epsilon$, opacity $\kappa$, and sound speed $C$ 
in two-dimensional solar models with different mass-coordinate resolutions (M = 2401, 1201, 601, 301)
with respect to a one-dimensional standard solar model as functions of mass 
coordinate.}\label{fig:kappa1}
\end{figure}

\begin{figure}
\plotone{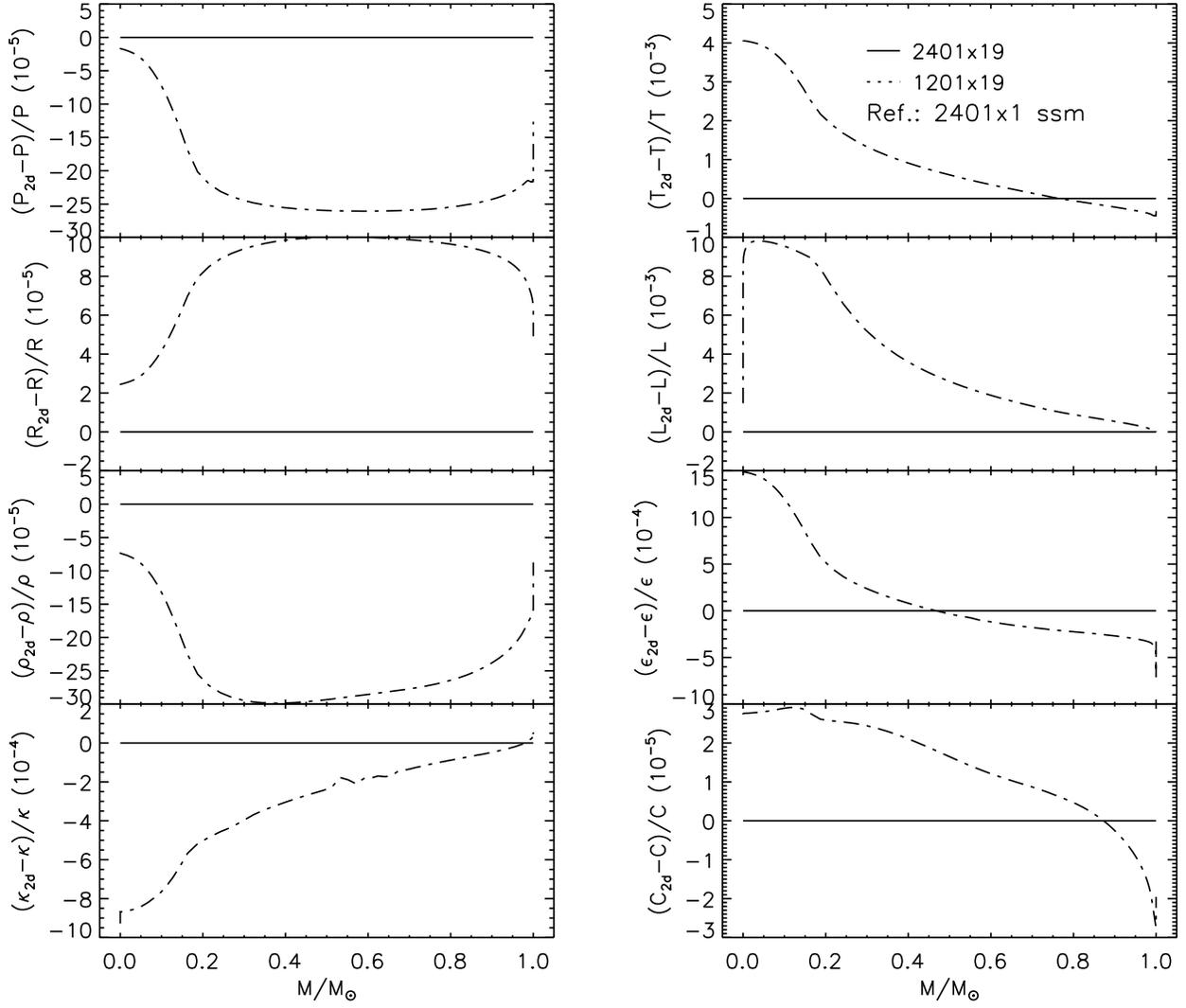}
\caption{Same as Fig.~\protect\ref{fig:kappa1}, but only for M = 2401 and 1201.}\label{fig:kappa2}
\end{figure}

From numerical experiments using different resolutions in both dimensions, we find 
that the model is not sensitive to the resolution in the angular coordinate, but 
very sensitive to the mass coordinate, see Fig.~\ref{fig:kappa1}. This figure compares 
4 mass resolutions, in which the lower resolution is obtained by taking out one 
mass point every two points from the adjacent higher resolution model. 
Fig.~\ref{fig:kappa2} zooms in to compare the models with the highest and 
second highest resolutions.

\begin{figure}
\plotone{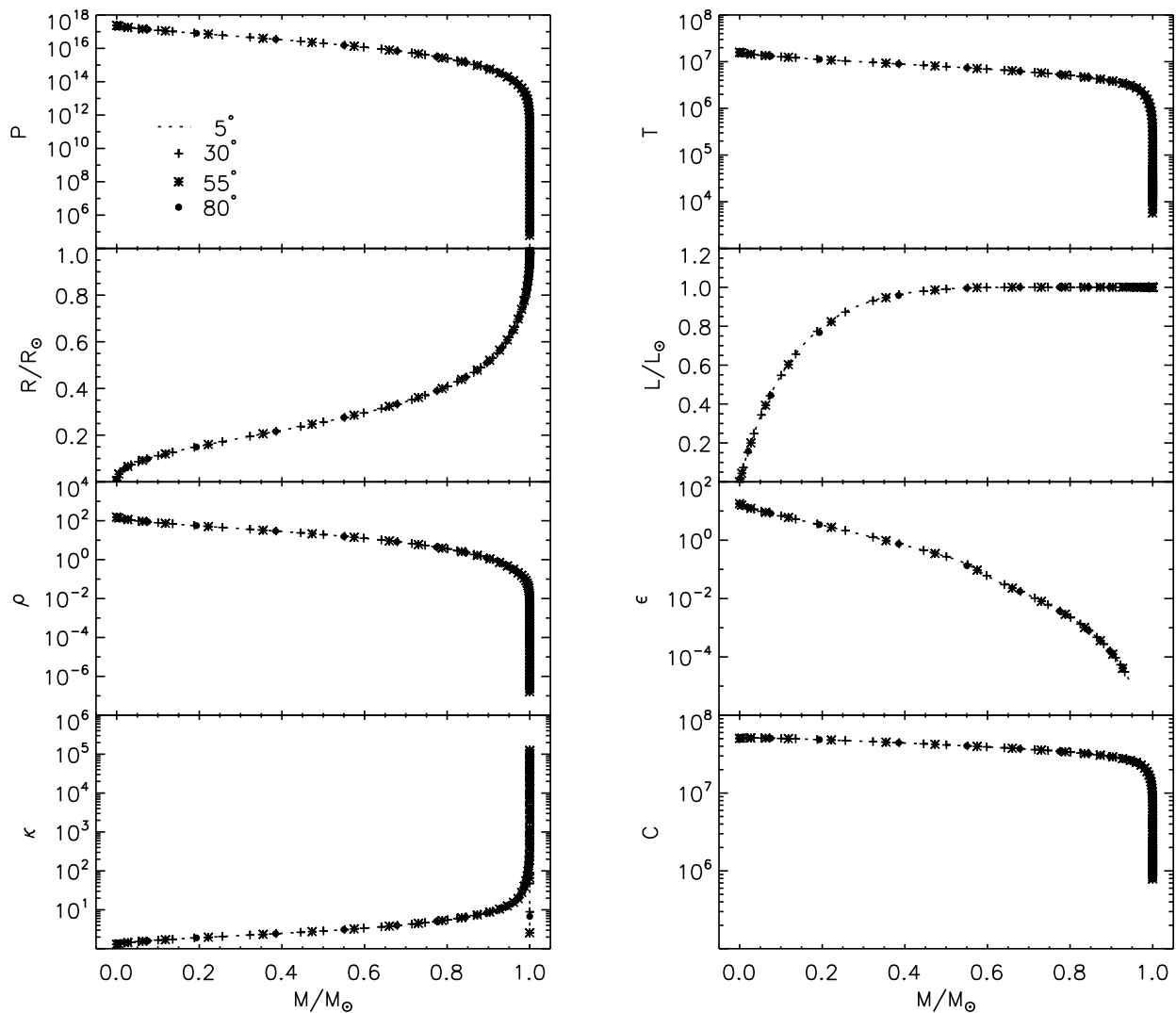}
\caption{Pressure $P$, temperature $T$, radius $R$, luminosity $L$, 
density $\rho$, nuclear energy generation rate $\epsilon$, opacity $\kappa$, and sound speed $C$ 
at different angular coordinates as functions of mass coordinate.}\label{fig:kappak}
\end{figure}

\begin{figure}
\plotone{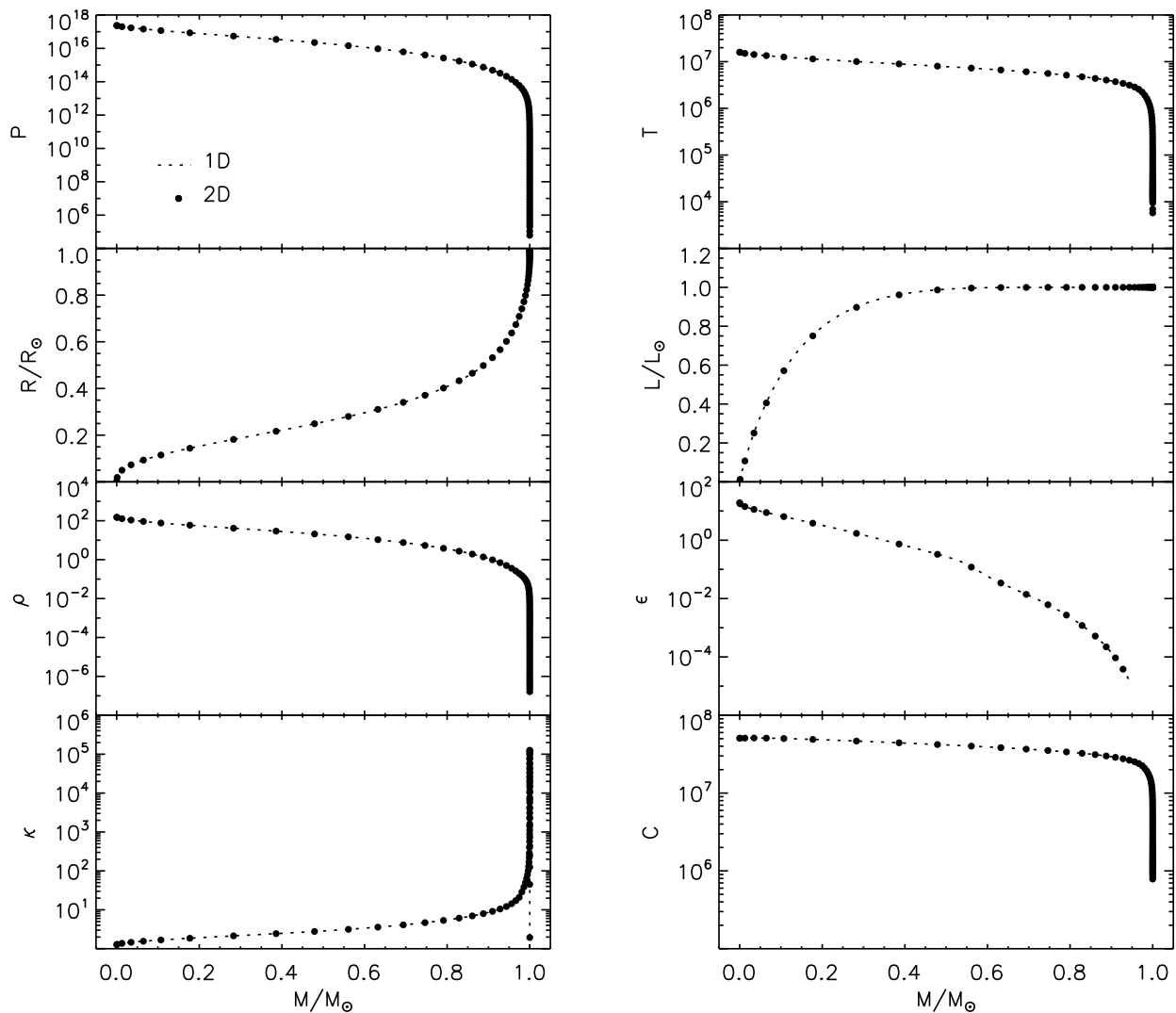}
\caption{Pressure $P$, temperature $T$, radius $R$, luminosity $L$, 
density $\rho$, nuclear energy generation rate $\epsilon$, opacity $\kappa$, and sound speed $C$ 
in both one- and two-dimensional solar models that have the same mass resolution as functions 
of mass coordinate.}\label{fig:kappa}
\end{figure}

We compare different angular zones in Fig.~\ref{fig:kappak} to make sure that 
the two-dimensional model is self-consistent in the angular direction.
Fig.~\ref{fig:kappa} shows that the two- and one-dimensional solar models with 
the same mass resolution are in very good agreement.

\subsection{Surface boundary conditions}

\begin{figure}
\plotone{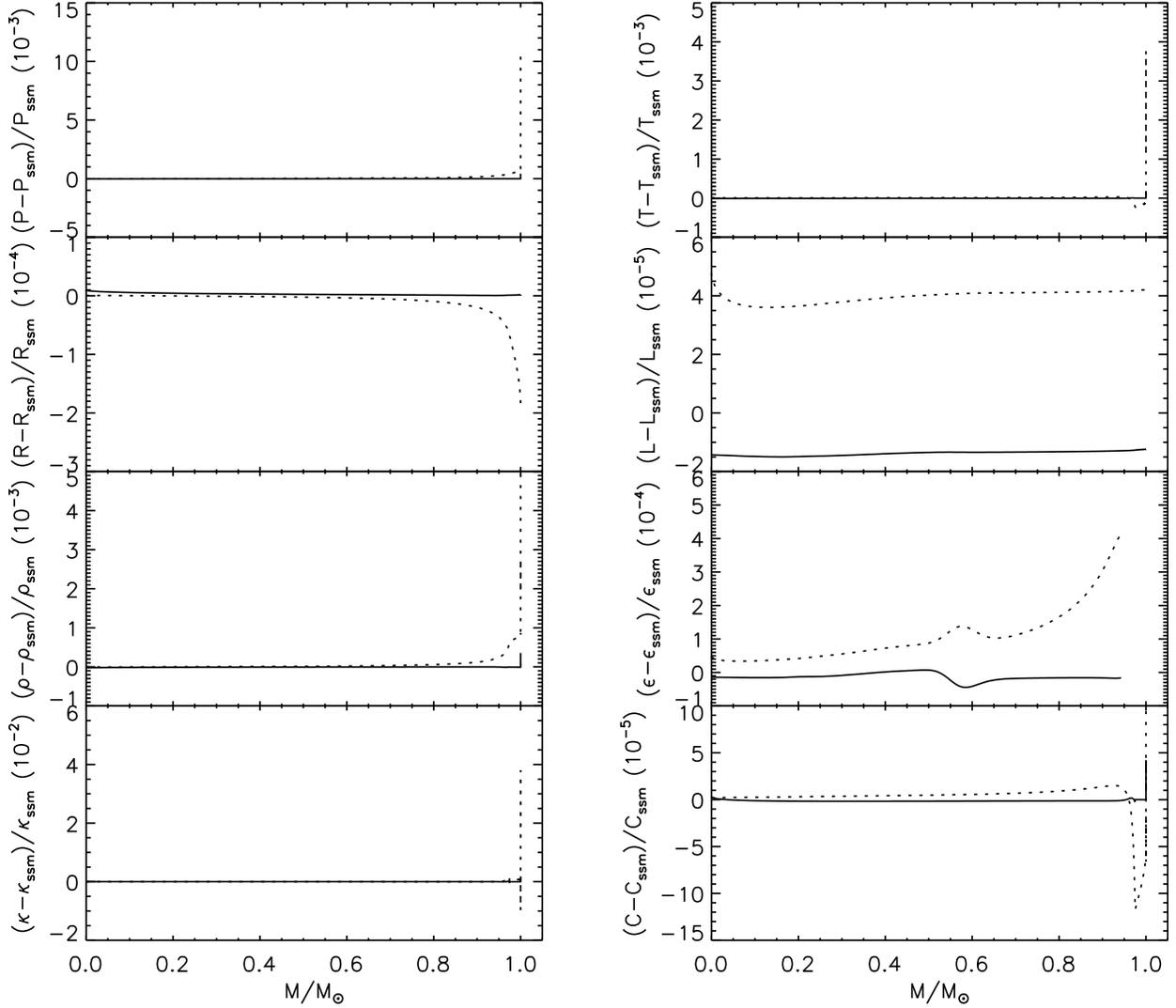}
\caption{Relative changes of pressure $P$, temperature $T$, radius $R$, luminosity $L$, 
density $\rho$, nuclear energy generation rate $\epsilon$, opacity $\kappa$, and sound speed $C$ 
in the two-dimensional solar model with Deupree's surface boundary conditions [solid line: 
Eqs.~(\protect\ref{eq:bob1}-\protect\ref{eq:bob2}); dotted line: $1.01\cdot$Eq.~(\protect\ref{eq:bob1}), 
$1.001\cdot$Eq.~(\protect\ref{eq:bob2})] with respect to a one-dimensional standard solar model 
as functions of mass coordinate.}\label{fig:bob1}
\end{figure}

Until now we have used only the standard surface boundary condition used 
in YREC (Pinsonneault 1988; Guenther et al. 1992). If we use these standard model surface 
values of pressure and temperature as Deupree's reference values, as indicated in section \ref{sec:bob}, 
we obtain the same results, as seen in Fig.~\ref{fig:bob1}. The solid lines use Eqs.~(\ref{eq:bob1}-\ref{eq:bob2}).
In order to investigate how errors 
in the reference pressure and temperature affect the result, we add 1\% to $P_{\s{ref}}$ given in 
Eq.~(\ref{eq:bob1}) and 0.1\% to $T_{\s{ref}}$ given in Eq.~(\ref{eq:bob2}). The result is shown 
by the dotted lines in Fig.~\ref{fig:bob1}. From the dotted lines we can see that errors in the 
surface boundary condition have larger influence on the outer layer than on the deep part of the 
model. 

It is inevitable to introduce some errors when $P_{\s{ref}}$ and $T_{\s{ref}}$ are selected in 
model calculations. Nevertheless, Deupree did not need to worry much about it, since his interest focused 
on the core convection. In contrast, we should be cautious to use Deupree's surface boundary condition,
because we want to apply to solar variability that takes place in the convective envelope.

The model is less sensitive to the error in the reference pressure than to that in the 
reference temperature.

\section{Test 2: Shell-like magnetic fields}\label{sec:smf}

Shell-like magnetic fields depend upon only the radial coordinate $r$. Any physical magnetic field 
should be free of divergence. The following magnetic fields are both radius-dependent and divergence-free:
\Ba
   \vb{B}&=&(0,0,f(r)),\\
   \vb{B}&=&(C/r^2,0,0),\\
   \vb{B}&=&(C/r^2,0,f(r)),
\Ea
where $f(r)$ is an arbitrary function of $r$, and $C$ is an arbitrary constant. If we assume that there is no magnetic field in the radiative zone of the Sun, we have $C=0$. Consequently, the unique physical shell-like magnetic field is
\be
  \vb{B}=(0,0,f(r)).
\ee
This is the case described in section 4.2, in which
\Ba
  \h{M} &=& -\h{B}^4(1+\frac{1}{2}\cot^2\theta)\left(1-\frac{\cot\theta}{2}\p{r'}{\theta}\right)^{-1},\\
  \h{B}^4 &=& \frac{m}{4\pi r^3\rho_m} \frac{\chi\rho}{P_T}.
\Ea
Comparing the 2-D stellar structure equations (Eqs.~\ref{eq:st1}-\ref{eq:st4}) with their 1-D counterparts (e.g., Li et al. 2003), we can see that the terms and/or factors in the brace are due to 2-D effects.

In the solar variability applications, we use a standard solar model at the current age (t=4.55 Gyr)
as the initial model. We apply a cyclic magnetic field to the model and use one year as the time step 
to advance the model.

As in the one-dimensional case, we specify $\chi$ as functions of time $t$ 
(or sunspot number $R_Z$) and the mass depth $m_D=\log(1-m/M_{\sun})$ as
\begin{equation}
  \chi(m_D,R_Z) = \chi_0(R_Z)\exp[-\case{1}{2}(m_D-m_{\mbox{\scriptsize{Dc}}})^2/\sigma^2], \label{eq:chi}
\end{equation}
where $m_{\mbox{\scriptsize{Dc}}}$ specifies the location and $\sigma$
specifies its width. $\chi_0$ is determined by
\begin{equation}
  \chi_0(R_Z) = \frac{B^2_0}{8\pi\rho_c} \{140 + [1 + \log(1+R_Z)]^5\}^2, \label{eq:chi0}
\end{equation}
where $B_0$ is an adjustable parameter (unit: gauss), 
$\rho_c$ is the density at the mass depth of $m_{\s{Dc}}$. 
In this case the magnetic variable-related derivatives reduce to
\begin{eqnarray*}
  \nu &=& \chi\rho/P_T, \\
  \nabla_\chi &=& \p{\ln\chi}{\ln m_{\s{D}}} \cdot \p{\ln m_{\s{D}}}{\ln m}\cdot\p{\ln m}{\ln P_T} \\
     &=& -\frac{m_{\s{D}}(m_{\s{D}}-m_{\s{Dc}})}{\sigma^2\ln 10}\frac{1-10^{m_{\s{D}}}}{10^{m_{\s{D}}}}
         \frac{4\pi P_T r^4}{G M_\sun^2(1-10^{m_{\s{D}}})^2}.
\end{eqnarray*}
In this test, $\chi$ does not depend upon the angular coordinate $\theta$, as required by a shell-like field.
The resultant models should be the same as we obtained in the one-dimensional counterparts (Li et al. 2003).
The method of solution used in this study guarantees this test, as confirmed by
actual model calculations.

\section{Conclusions}

A high-precision two-dimensional framework for treating stellar 
evolution with magnetic fields has been developed and successfully tested. The required high 
precision is achieved by
\begin{description}
\item{(1)} using the mass coordinate to replace the radial coordinate,
\item{(2)} including the convection instability,
\item{(3)} including a stellar atmosphere,
\item{(4)} allowing element diffusion,
\item{(5)} using fixed and adjustable time steps,
\item{(6)} adjusting grid points.
\end{description}

The code has the potential to include rotation and turbulence, 
but does not have the potential to generate them like a fully hydrodynamic
code.

\acknowledgements 
We thank R.G. Deupree for many discussions and constructive 
suggestions during his stay in the department. We also want to thank Dr. 
Christian Straka for helpful discussions on many aspects of this paper. 
We wish to acknoledge the anonymous referee, whose efforts have resulted 
in considerable improvements to the paper.
This work was supported in part by NSF grants ATM 0206130 and ATM 0348837 
to SB. SS and PD were supported in part by NASA grant NAG5-13299. P. V. was supported by
Regione Lazio funds.

\appendix

\section{Coefficient matrix}

Eq.~(\ref{eq:linear}) consists of a set of non-homogeneous linear algebraic 
equations. If we use $\h{A}$ to represent the coefficient matrix, and use 
$\h{B}$ to represent the non-homogeneous term, this equation can be 
written down as follows:
\be
  \h{A}\cdot \delta w = \h{B},
\ee
where
\be
  \delta w = \left(\begin{array}{c}
    \delta r'_{11} \\
    \delta L'_{11} \\
    \delta P'_{11} \\
    \delta T'_{11} \\
     \vdots \\
    \delta r'_{M1} \\
    \delta L'_{M1} \\
    \delta P'_{M1} \\
    \delta T'_{M1} \\
     \vdots \\
    \delta r'_{1N} \\
    \delta L'_{1N} \\
    \delta P'_{1N} \\
    \delta T'_{1N} \\
     \vdots \\
    \delta r'_{MN} \\
    \delta L'_{MN} \\
    \delta P'_{MN} \\
    \delta T'_{MN}
\end{array}\right)
\ee
is a column matrix. $\h{B}$ is also a column matrix,
\be
\h{B} = -\left(\begin{array}{c}
  F^{11}_R \\
  F^{11}_L \\
  F^{11}_P \\
  F^{11}_T \\
  \vdots \\
  F^{M1}_R \\
  F^{M1}_L \\
  F^{M1}_P \\
  F^{M1}_T \\[6pt]
  F^{12}_R \\
  F^{12}_L \\
  F^{22}_P \\
  F^{22}_T \\
  \vdots \\
  F^{M2}_R \\
  F^{M2}_L \\
  F^{M+12}_P \\
  F^{M+12}_T \\
  \vdots \\
  F^{1N}_R \\
  F^{1N}_L \\
  F^{2N}_P \\
  F^{2N}_T \\
  \vdots \\
  F^{MN}_R \\
  F^{MN}_L \\
  F^{M+1N}_P \\
  F^{M+1N}_T
\end{array}
\right)
\ee
The coefficient matrix $\h{A}$ has elements $\p{F^{ij}_w}{w_{lk}}$. 
Only those elements with $l = i-1,\, i$ and $k = j-1,\, j$ 
may be nonzero, as shown in Fig.~\ref{fig:coeff}. We work 
out these nonzero elements in this appendix.

\subsection{Useful partial derivatives}

The partial derivatives of the differential equations are 
required for the linearization. By defining the shorthand 
notation $\partial_X Y=\partial Y/\partial\log X$, we can 
calculate the useful derivatives as follows.

The following derivatives are almost the same as in the 
one-dimensional case (see Prather 1976) except for those terms due to $\rho/\rho_m$. These 
derivatives are nonzero for l = i-1, i and k = j.
If k = j-1, they vanish.
\begin{eqnarray*}
  \partial_{R}\h{P} &=&-4\cdot\h{P}\nonumber \\
  \partial_{L}\h{P}&=& \partial_{T}\h{P}=0 \nonumber \\
  \partial_{P}\h{P}&=&-\h{P} \\
  \partial_{R}\h{R}&=& -3\cdot\h{R}\nonumber\\
  \partial_{L}\h{R}&=& 0 \\
  \partial_{P}\h{R}&=& -\alpha_m\cdot\h{R} \nonumber\\
  \partial_{T}\h{R}&=& \delta_m \cdot\h{R} \nonumber\\
  \partial_{P}\h{L}&=&\frac{M_r}{L_{\sun}}\left[\left(\p{\epsilon}{\ln P}\right)_T
     +\left(\p{\tilde{S}}{\ln P}\right)_T/\Delta t\right]\frac{\rho}{\rho}_m \nonumber\\
  \partial_{R}\h{L}&=& \partial_{L}\h{L}= 0  \\
  \partial_{T}\h{L}&=&\frac{M_r}{L_{\sun}}\left[\left(\p{\epsilon}{\ln T}\right)_P
     +\left(\p{\tilde{S}}{\ln T}\right)_P/\Delta t\right]\frac{\rho}{\rho_m}
\end{eqnarray*}
In the convective zone, we have
\begin{eqnarray*}
  \partial_{R}\h{T}_c&=&(\partial\ln\nabla_c/\partial\ln r-4)\cdot\h{T}_c\nonumber \\
  \partial_{L}\h{T}_c&=& 0 \\
  \partial_{P}\h{T}_c&=&(\partial\ln\nabla_c/\partial\ln P_T-1)\cdot\h{T}_c  \nonumber \\
  \partial_{T}\h{T}_c&=&(\partial\nabla_c/\partial\ln T)\cdot\h{T}_c 
\end{eqnarray*}
In the radiative zone, we have
\begin{eqnarray*}
  \partial_{R}\h{T}_r&=&-4\cdot\h{T}_r \nonumber \\
  \partial_{L}\h{T}_r&=& \h{T}_r/L \\
  \partial_{P}\h{T}_r&=& (\partial\ln\kappa/\partial\ln P_T)_T\cdot\h{T}_r \nonumber \\
  \partial_{T}\h{T}_r&=& [(\partial\ln\kappa/\partial\ln T)_P-4]\cdot\h{T}_r \nonumber
\end{eqnarray*}
The formulas for the various partial derivatives of the physical quantities
will be presented in the following subsections. The equation of state
calculates $\rho$, $\alpha$, $\delta$, $c_p$, $\nabla_{\s{ad}}$ and the pressure and
temperature derivatives of these quantities (see section \ref{sec:eos}). Energy
generation rate $\epsilon$ is a function of $\rho$ and $T$, too. So
$(\partial\epsilon/\partial T)_P$ and $(\partial\epsilon/\partial P)_T$ 
can be expressed by $(\partial\epsilon/\partial\ln
T)_\rho$ and $\partial\epsilon/\partial\ln \rho)_T$ (see
section \ref{sec:energy}):
\begin{eqnarray*}
  (\partial{\epsilon}/\partial{\ln T})_P &=& (\partial{\epsilon}/\partial{\ln T})_\rho 
     +(\partial{\epsilon}/\partial{\ln\rho})_T(\partial{\ln\rho}/\partial{\ln T})_P \\
  (\partial{\epsilon}/\partial{\ln P_T})_T &=&(\partial{\epsilon}/\partial{\ln\rho})_T (\partial{\ln\rho}/
    \partial{\ln T})_P
\end{eqnarray*}
The derivatives of the convective gradient $\nabla_c$ which are presented in section \ref{sec:tg}.

The entropy term contains the only explicit reference
to any time-dependence in the stellar structure equations. It can be
reformulated as follows:
\begin{eqnarray*}
  \tilde{S} &=& -(P_T\delta/\rho)\cdot(\Delta T'/\nabla_{\s{ad}}-\Delta P') \\
  (\partial \tilde{S}/\partial \ln T)_P &=& \tilde{S}[\delta+(\partial\ln\delta/\partial\ln T)_P]
    -(P\delta/\rho\nabla_{\s{ad}})[1-(\partial{\ln\nabla_{\s{ad}}}/\partial\ln T)_P\cdot\Delta T'] \\
  (\partial \tilde{S}/\partial \ln P_T)_T &=& \tilde{S}[1-\alpha+(\partial\ln\delta/\partial\ln P_T)_T] \\
&&       +(P\delta/\rho)[1+(\partial{\ln\nabla_{\s{ad}}}/\partial\ln P_T)_T\cdot\Delta T'/\nabla_{\s{ad}}]
\end{eqnarray*}
where ($\Delta P'$, $\Delta T'$) are the changes between successive models.

The following derivatives are new. Similarly, these derivatives are nonzero for l = i-1, i and k = j.
When $\ell=1$, we have
\begin{eqnarray*}
\partial_R\h{B}^1 &=& -4\cdot\h{B}^1 \\
\partial_L\h{B}^1 &=& \partial_T\h{B}^1 = 0  \\
\partial_P\h{B}^1 &=& -\h{B}^1 \\
\partial_R\h{T}^1_c &=& \h{T}^1\cdot(\pl{\ln\nabla_c}{\ln r}-4) \\
\partial_L\h{T}^1_c &=& 0 \\
\partial_P\h{T}^1_c &=& \h{T}^1_c\cdot[\pl{\ln\nabla_c}{\ln P_T}-1] \\
\partial_T\h{T}^1_c &=& \h{T}^1_c\cdot(\pl{\ln\nabla_c}{\ln T}) \\
\partial_R\h{T}^1_r &=& -4\cdot\h{T}^1_r \\
\partial_L\h{T}^1_r &=& \h{T}^1_r/L \\
\partial_P\h{T}^1_r &=& \h{T}^1_r\cdot(\pl{\ln\kappa}{\ln P_T})_T   \\
\partial_T\h{T}^1_r &=& \h{T}^1_r\cdot[(\pl{\ln\kappa}{\ln T})_{P_T}-4]
\end{eqnarray*}
When $\ell=2$, we have
\begin{eqnarray*}
  \partial_{R}\h{B}^2 &=& -\h{B}^2 \nonumber \\
  \partial_{L}\h{B}^2&=& \partial_{T}\h{B}^2= 0 \nonumber \\
  \partial_{P}\h{B}^2&=& -\h{B}^2    \\
  \partial_{R}\h{T}^2_c &=& \h{T}^2\cdot(\pl{\ln\nabla_c}{\ln R}-1)  \nonumber \\
  \partial_{L}\h{T}^2_c&=& 0  \nonumber \\
  \partial_{P}\h{T}^2_c&=& \h{T}^2\cdot[\pl{\ln\nabla_c}{\ln P}-1]    \\
  \partial_{T}\h{T}^2_c&=& \h{T}^2\cdot(\pl{\ln\nabla_c}{\ln T})  \nonumber \\
  \partial_{R}\h{T}^2_r &=& -\h{T}^2_r  \nonumber \\
  \partial_{L}\h{T}^2_r&=& \h{T}^2_r/L  \nonumber \\
  \partial_{P}\h{T}^2_r&=& \h{T}^2_r\cdot(\pl{\ln\kappa}{\ln P_T})_T  \\
  \partial_{T}\h{T}^2_r&=& \h{T}^2_r\cdot[(\pl{\ln\kappa}{\ln T})_{P_T}-4] 
\end{eqnarray*}
When $\ell=3$, we have
\begin{eqnarray*}
  \partial_{R}\h{B}^3 &=& -3\cdot\h{B}^3 \nonumber \\
  \partial_{L}\h{B}^3&=& 0 \nonumber \\
  \partial_{P}\h{B}^3&=&-\h{B}^3\cdot\alpha_m  \\
  \partial_{T}\h{B}^3&=&  \h{B}^3\cdot\delta_m  \nonumber \\[6pt]
  \partial_{R}\h{T}^3_c &=& \h{T}^3\cdot(\pl{\ln\nabla_c}{\ln R}-3)  \nonumber \\
  \partial_{L}\h{T}^3_c&=& 0  \nonumber \\
  \partial_{P}\h{T}^3_c&=& \h{T}^3\cdot(\pl{\ln\nabla_c}{\ln P}-\alpha_m)  \\
  \partial_{T}\h{T}^3_c&=& \h{T}^3\cdot(\pl{\ln\nabla_c}{\ln T}+\delta_m)   \nonumber \\[6pt]
  \partial_{R}\h{T}^3_r &=& -3\cdot \h{T}^3_r  \nonumber \\
  \partial_{L}\h{T}^3_r&=& \h{T}^3_r/L  \nonumber \\
  \partial_{P}\h{T}^3_r&=& \h{T}^3_r\cdot(\pl{(\ln\kappa}{\ln P_T})_T-\alpha_m+1)  \\
  \partial_{T}\h{T}^3_r&=& \h{T}^3_r\cdot(\pl{\ln\kappa}{\ln T})_{P_T}+\delta_m-4)
\end{eqnarray*}
When $\ell=10,11,13,14$, we have
\begin{eqnarray*}
\partial_R\h{B}^{\ell} &=& \partial_L\h{B}^{\ell} = 0  \\
\partial_P\h{B}^{\ell} &=& \h{B}^{\ell}\cdot(\alpha-1) \\
\partial_T\h{B}^{\ell} &=& -\h{B}^{\ell}\cdot \delta \\[6pt]
\partial_R\h{T}^{\ell}_c &=& \h{T}^{\ell}_c\cdot(\pl{\ln\nabla_c}{\ln r}) \\
\partial_L\h{T}^{\ell}_c &=& 0 \\
\partial_P\h{T}^{\ell}_c &=& \h{T}^{\ell}_c\cdot(\pl{\ln\nabla_c}{\ln P}+\alpha-1) \\
\partial_T\h{T}^{\ell}_c &=& \h{T}^{\ell}_c\cdot(\pl{\ln\nabla_c}{\ln T}-\delta) \\
\partial_R\h{T}^{\ell}_r &=& 0 \\
\partial_L\h{T}^{\ell}_r &=& \h{T}^{\ell}_r/L \\
\partial_P\h{T}^{\ell}_r &=& \h{T}^{\ell}_r\cdot[(\pl{\ln\kappa}{\ln P_T})_T+\alpha] \\
\partial_T\h{T}^{\ell}_r &=& \h{T}^{\ell}_r\cdot[(\pl{\ln\kappa}{\ln T})_{P_T}-\delta-4] 
\end{eqnarray*}
When $\ell=12,15,17$, we have
\begin{eqnarray*}
\partial_R\h{B}^{\ell} &=& -3\cdot\h{B}^{\ell} \\
\partial_L\h{B}^{\ell} &=& \partial_T\h{B}^{\ell} = 0  \\
\partial_P\h{B}^{\ell} &=& -\h{B}^{\ell} \\
\partial_R\h{T}^{\ell}_c &=& \h{T}^{\ell}_c\cdot(\pl{\ln\nabla_c}{\ln r}-3) \\
\partial_L\h{T}^{\ell}_c &=& 0 \\
\partial_P\h{T}^{\ell}_c &=& \h{T}^{\ell}_c\cdot(\pl{\ln\nabla_c}{\ln P_T}-1) \\
\partial_T\h{T}^{\ell}_c &=& \h{T}^{\ell}_c\cdot(\pl{\ln\nabla_c}{\ln T}) \\
\partial_R\h{T}^{\ell}_r &=& -3\cdot\h{T}^{\ell}_r \\
\partial_L\h{T}^{\ell}_r &=& \h{T}^{\ell}_r/L \\
\partial_P\h{T}^{\ell}_r &=& \h{T}^{\ell}_r\cdot(\pl{\ln\kappa}{\ln P_T})_T   \\
\partial_T\h{T}^{\ell}_r &=& \h{T}^{\ell}_r\cdot[(\pl{\ln\kappa}{\ln T})_{P_T}-4]
\end{eqnarray*}
When k = j-1, all derivatives of $\h{B}'s$ and $\h{T}'s$ vanish.

We also need similar derivatives of $\h{D}'s$. When l = i-1, k=j, all 
derivatives of $\h{D}^1$, $\h{D}^2$, $\h{D}^3$, $\h{D}^{12}$, $\h{D}^{15}$, and $\h{D}^{17}$ are zero. The nonzero derivatives are:
\Ba
\partial_{P_{i-1j}}\h{D}^{10} &=& \partial_{P_{i-1j}}\h{D}^{13} = - \h{D}^2 \alpha_{i-1j}/\Delta s_i, \\
\partial_{P_{i-1j}}\h{D}^{11} &=& -\partial_{P_{i-1j}}\h{D}^{10}\cdot(r'_{ij}-r'_{ij-1}) \Delta\theta_j, \\
\partial_{P_{i-1j}}\h{D}^{14} &=& \h{D}^1 \alpha_{i-1j}/\Delta s_i, \\
\partial_{T_{i-1j}}\h{D}^{10} &=&  \partial_{T_{i-1j}}\h{D}^{13} = \h{D}^2 \delta_{i-1j}/\Delta s_i, \\
\partial_{T_{i-1j}}\h{D}^{11} &=& -\partial_{T_{i-1j}}\h{D}^{10}\cdot(r'_{ij}-r'_{ij-1})/\Delta\theta_j, \\
\partial_{T_{i-1j}}\h{D}^{14} &=& -\h{D}^1 \delta_{i-1j}/\Delta s_i.
\Ea
When l = i, k=j, the nonzero derivatives are
\Ba
\partial_{R_{ij}}\h{D}^1 &=& \left(\frac{\cot\theta_j}{2\Delta\theta_j}\right)^2(r'_{ij}-r'_{ij-1}) \left[1
  -\frac{\cot\theta_j}{2\Delta\theta_j}(r'_{ij}-r'_{ij-1})\right]^{-2}\delta_R  \\
&&   +\frac{\cot\theta_j}{2\Delta\theta_j}\left[1
  -\frac{\cot\theta_j}{2\Delta\theta_j}(r'_{ij}-r'_{ij-1})\right]^{-1}\delta_R, \\
\partial_{R_{ij}}\h{D}^2 &=&  \frac{\cot\theta_j}{2\Delta\theta_j} \left[1
  -\frac{\cot\theta_j}{2\Delta\theta_j}(r'_{ij}-r'_{ij-1})\right]^{-2}\delta_R, \\
\partial_{R_{ij}}\h{D}^3 &=&  \left(\frac{\cot\theta_j}{2\Delta\theta_j}\right)^2(P'_{ij}-P'_{ij-1}) \left[1
  -\frac{\cot\theta_j}{2\Delta\theta_j}(r'_{ij}-r'_{ij-1})\right]^{-2}\delta_R, \\
\partial_{R_{ij}}\h{D}^{10} &=&  \frac{1}{\Delta s_i} \partial_{R_{ij}}\h{D}^2\cdot [(\chi'_{ij}-\chi'_{i-1j}) + (\rho'_{ij}-\rho'_{i-1j}) + (\gamma''_{ij}-\gamma''_{i-1j})], \\
\partial_{R_{ij}}\h{D}^{11} &=&  -\frac{\delta_R}{\Delta\theta_j}\h{D}^{10} - \frac{1}{\Delta\theta_j} \partial_{R_{ij}}\h{D}^{10}\cdot (r'_{ij}-r'_{ij-1}), \\
\partial_{R_{ij}}\h{D}^{12} &=&  \frac{1}{\Delta\theta_j} \partial_{R_{ij}}\h{D}^{2}\cdot [(\chi'_{ij}-\chi'_{ij-1}) + (\rho'_{ij}-\rho'_{ij-1}) + (\gamma''_{ij}-\gamma''_{ij-1})], \\
\partial_{R_{ij}}\h{D}^{13} &=&  \frac{1}{\Delta s_i} \partial_{R_{ij}}\h{D}^{2}\cdot [(\chi'_{ij}-\chi'_{i-1j}) + (\rho'_{ij}-\rho'_{i-1j}) + (\vartheta''_{ij}-\vartheta''_{i-1j})], \\
\partial_{R_{ij}}\h{D}^{14} &=&  -\frac{1}{\Delta s_i} \partial_{R_{ij}}\h{D}^{1}\cdot [(\chi'_{ij}-\chi'_{i-1j}) + (\rho'_{ij}-\rho'_{i-1j}) + (\vartheta''_{ij}-\vartheta''_{i-1j})], \\
\partial_{R_{ij}}\h{D}^{15} &=&  \frac{\cot\theta_j}{2\Delta\theta_j} \partial_{R_{ij}}\h{D}^{2} \cdot [(\chi'_{ij}-\chi'_{ij-1}) + (\rho'_{ij}-\rho'_{ij-1}) + (\vartheta''_{ij}-\vartheta''_{ij-1})], \\
\partial_{R_{ij}}\h{D}^{17} &=& - \partial_{R_{ij}}\h{D}^{2}, \\  
\partial_{P_{ij}}\h{D}^{3} &=& \frac{\cot\theta_j}{2\Delta\theta_j} \left[1
    -\frac{\cot\theta_j}{2\Delta\theta_j}(r'_{ij}-r'_{ij-1})\right]^{-1}\delta_P ,  \\
\partial_{P_{ij}}\h{D}^{10} &=& \h{D}^2\alpha_{ij}/\Delta s_{i}, \\
\partial_{P_{ij}}\h{D}^{11} &=& -\partial_{P_{ij}}\h{D}^{10}(r'_{ij}-r'_{ij-1})/\Delta \theta_j, \\
\partial_{P_{ij}}\h{D}^{12} &=& \h{D}^2\alpha_{ij}\delta_\rho/\Delta \theta_j, \\
\partial_{P_{ij}}\h{D}^{13} &=&  \h{D}^2\alpha_{ij}/\Delta s_i, \\
\partial_{P_{ij}}\h{D}^{14} &=& -\h{D}^1\alpha_{ij}/\Delta s_i, \\
\partial_{P_{ij}}\h{D}^{15} &=& \h{D}^2\frac{\cot\theta_j}{2\Delta\theta_j}\alpha_{ij}\delta_\rho, \\
\partial_{T_{ij}}\h{D}^{10} &=& -\h{D}^2\delta_{ij}/\Delta s_{i}, \\
\partial_{T_{ij}}\h{D}^{11} &=& -\partial_{T_{ij}}\h{D}^{10}(r'_{ij}-r'_{ij-1})/\Delta \theta_j, \\
\partial_{T_{ij}}\h{D}^{12} &=& -\h{D}^2\delta_{ij}\delta_\rho/\Delta \theta_j, \\
\partial_{T_{ij}}\h{D}^{13} &=&  -\h{D}^2\delta_{ij}/\Delta s_i, \\
\partial_{T_{ij}}\h{D}^{14} &=& \h{D}^1\delta_{ij}/\Delta s_i, \\
\partial_{T_{ij}}\h{D}^{15} &=& -\h{D}^2\frac{\cot\theta_j}{2\Delta \theta_j}\delta_{ij}\delta_\rho.
\Ea
where $\delta_R=1$ when $r_{ij}-r_{ij-1}\ne 0$, and $\delta_R=0$ when $r_{ij}-r_{ij-1}= 0$. $\delta_P$ and $\delta_\rho$ have the similar meaning. When l = i, k=j-1, the nonzero derivatives are
\[
\partial_{R_{ij-1}}\h{D}^{\ell} = -\partial_{R_{ij}}\h{D}^{\ell}
\]
for $\ell=1,2,3,10,\cdots,15,17$, and
\[
\partial_{P_{ij-1}}\h{D}^{\ell} = -\partial_{P_{ij}}\h{D}^{\ell}
\]
for $\ell=3,12,15$, and
\[
\partial_{T_{ij-1}}\h{D}^{\ell} = -\partial_{T_{ij}}\h{D}^{\ell}
\]
for $\ell=12,15$.

We calculate the derivatives of $\h{F}^2$ and $\h{F}^3$ by taking the advantage of $l_mv_{\s{conv}}\sim$ const. The nonzero derivatives are listed as follows for $l=i-1,i$ and $k=j$.
\Ba
\partial_{R}\h{F}^1 &=& -2\cdot \h{F}^1, \\
\partial_{L}\h{F}^1 &=& 0, \\
\partial_{P}\h{F}^1 &=& -\h{F}^1\cdot[(\pl{\ln\kappa}{\ln P_T})_T+\alpha+\alpha_m], \\
\partial_{T}\h{F}^1 &=& \h{F}^1\cdot[4-(\pl{\ln\kappa}{\ln T})_{P_T}+\delta+\delta_m], \\
\partial_{R}\h{F}^2 &=& -2\cdot \h{F}^2, \\
\partial_{L}\h{F}^2 &=& 0, \\
\partial_{P}\h{F}^2 &=& \h{F}^2\cdot\{(\pl{\ln C_p}{\ln P_T})_T \nob
&& -\beta[2\alpha+(\pl{\ln\kappa}{\ln P_T})_T+(\pl{\ln C_p}{\ln P_T})_T]\}, \\
\partial_{T}\h{F}^2 &=& \h{F}^2\cdot\{1+(\pl{\ln C_p}{\ln T})_{P_T} \nob
&&  +\beta[3+2\delta-(\pl{\ln\kappa}{\ln T})_{P_T}-(\pl{\ln C_p}{\ln T})_{P_T}]\}, \\
\partial_{R}\h{F}^3 &=& -2\cdot \h{F}^3, \\
\partial_{L}\h{F}^3 &=& 0, \\
\partial_{P}\h{F}^3 &=& -\partial_{P}\h{F}^2\cdot\nabla'_{\s{ad}} 
  + \h{F}^3\cdot(\pl{\ln\nabla_{\s{ad}}}{\ln P_T})_T, \\
\partial_{T}\h{F}^3 &=& -\partial_{T}\h{F}^2\cdot\nabla'_{\s{ad}} 
  + \h{F}^3\cdot(\pl{\ln\nabla_{\s{ad}}}{\ln T})_{P_T}, \\
\partial_{R}\h{F}^4 &=& \partial_{R}\h{F}^1\cdot \left\{ \begin{array}{ll}
 \frac{Gm\rho\nabla_c}{rP_T} 
   + (\partial\ln\nabla_c/\partial\ln r - 1) \cdot \h{F}^4 & (\mbox{convective}), \\
  \frac{Gm\rho\nabla_{\s{rad}}}{rP_T}  - \h{F}^4 & (\mbox{radiative}), \\
 \end{array} \right. \\
\partial_{L}\h{F}^4 &=& \left\{ \begin{array}{ll}
   0 & (\mbox{convective}) , \\
   \h{F}^4/L & (\mbox{radiative}), \\
\end{array} \right. \\
\partial_{P}\h{F}^4 &=& \partial_{P}\h{F}^1\cdot \left\{ \begin{array}{ll}
\frac{Gm\rho\nabla_c}{rP_T} + 
  (\partial\ln\nabla_c/\partial\ln P_T +\alpha- 1)\cdot\h{F}^4 & (\mbox{convective}), \\
\frac{Gm\rho\nabla_{\s{rad}}}{rP_T} + 
  [(\partial\ln\kappa/\partial\ln P_T)_T+\alpha]\cdot\h{F}^4 & (\mbox{radiative}), \\
\end{array} \right. \\
\partial_{T}\h{F}^4 &=& \partial_{T}\h{F}^1 \cdot\left\{ \begin{array}{ll}
\frac{Gm\rho\nabla_c}{rP_T} + 
   (\partial\ln\nabla_c/\partial\ln T-\delta)\cdot\h{F}^4 & (\mbox{convective}), \\
\frac{Gm\rho\nabla_{\s{rad}}}{rP_T} + 
   [(\partial\ln\kappa/\partial\ln T)_P-\delta-4]\cdot\h{F}^4 & (\mbox{radiative}), \\
\end{array} \right. \\
\partial_{R}\h{F}^5 &=& \partial_{R}\h{F}^2\cdot\frac{Gm\rho\nabla_c}{rP_T}  
+ (\partial\ln\nabla_c/\partial\ln r - 1) \cdot \h{F}^5, \\
\partial_{L}\h{F}^5 &=&   0, \\
\partial_{P}\h{F}^5 &=& \partial_{P}\h{F}^2\cdot\frac{Gm\rho\nabla_c}{rP_T}  +  (\partial\ln\nabla_c/\partial\ln P_T +\alpha- 1)\cdot\h{F}^5, \\
\partial_{T}\h{F}^5 &=& \partial_{T}\h{F}^2\cdot\frac{Gm\rho\nabla_c}{rP_T}  + (\partial\ln\nabla_c/\partial\ln T-\delta)\cdot\h{F}^5, \\
\partial_{R}\h{F}^6 &=& \partial_{R}\h{F}^3\cdot\frac{Gm\rho}{rP_T}  -\h{F}^6, \\
\partial_{L}\h{F}^6 &=& 0, \\
\partial_{P}\h{F}^6 &=& \partial_{P}\h{F}^3\cdot\frac{Gm\rho}{rP_T}  +  (\alpha-1)\cdot\h{F}^6, \\
\partial_{T}\h{F}^6 &=& \partial_{T}\h{F}^3\cdot\frac{Gm\rho}{rP_T} - \delta\cdot\h{F}^6,
\Ea
where $\beta=(v_{\s{conv}}/v_0)/(1+v_{\s{conv}}/v_0)$.

\subsection{Numerical integrals}

The quantities $\rho_m$, $\alpha_m$ and $\delta_m$ are integrals over $\theta$:
\Ba
\rho_m(m,\theta) &\equiv& \frac{1}{r^2}\cdot\frac{1}{2} \int^\pi_0 d\theta r^2(m,\theta)\rho(m,\theta)\sin\theta, \\
\alpha_m(m)&\equiv&\left(\p{\ln\rho_m}{\ln P_T}\right)_T 
 = \frac{\int^\pi_0 d\theta r^2(m,\theta)\rho(m,\theta)\alpha(m,\theta)\sin\theta}{\int^\pi_0 d\theta r^2(m,\theta)\rho(m,\theta)\sin\theta}, \\
\delta_m(m)&\equiv& -\left(\p{\ln\rho_m}{\ln T}\right)_{P_T} 
 = \frac{\int^\pi_0 d\theta r^2(m,\theta)\rho(m,\theta)\delta(m,\theta)\sin\theta}{\int^\pi_0 d\theta r^2(m,\theta)\rho(m,\theta)\sin\theta}.
\Ea
Of course, the luminosity $L$ is an integral, too:
\[
L(m) \equiv 2\pi\int^\pi_0 d\theta r^2(m,\theta) F_r(m,\theta)\sin\theta = \frac{1}{2}\int^\pi_0 d\theta L'(m,\theta)L_\sun\sin\theta,
\]
where $L'=4\pi r^2 F_r/L_\sun$. In the one-dimensional case, we know the relationship on the solar surface:
\be
  L=4\pi R^2\sigma T_{\s{eff}}^4. \label{eq:lrt}
\ee
If we define
\Ba
 R^2 &\equiv& \frac{1}{2}\int^\pi_0d\theta r^2(M_{\s{tot}},\theta)\sin\theta, \\
 T^4_{\s{eff}} &\equiv& \frac{1}{2R^2}\int^\pi_0d\theta r^2 T^4(M_{\s{tot}},\theta)\sin\theta,
\Ea
Eq.~(\ref{eq:lrt}) holds well in the two-dimensional case, where $M_{\s{tot}}$ is the total mass of the star.

We use the trapezoidal rule to compute these integrals. For example,
\Ba
\rho^{ij}_m &=& \frac{ \frac{1}{r_{ij}^2} \sum_{\ell=2}^{N} \frac{1}{2}(r^2_{i\ell}\rho_{i\ell}\sin\theta_{\ell}
    +r^2_{i\ell-1}\rho_{i\ell-1}\sin\theta_{\ell-1})(\theta_\ell-\theta_{\ell-1})}{
\sum_{\ell=2}^{N} \frac{1}{2}(\sin\theta_{\ell}
    +\sin\theta_{\ell-1})(\theta_\ell-\theta_{\ell-1})}, \\
\frac{L}{L_\sun} &=&  \frac{ \sum_{\ell=2}^{N} \frac{1}{2}(L'_{M\ell}\sin\theta_{\ell}
    +L'_{M\ell-1}\sin\theta_{\ell-1})(\theta_\ell-\theta_{\ell-1})}{
\sum_{\ell=2}^{N} \frac{1}{2}(\sin\theta_{\ell}
    +\sin\theta_{\ell-1})(\theta_\ell-\theta_{\ell-1})}, \\
R^2 &=& \frac{ \sum_{\ell=2}^{N} \frac{1}{2}(r^2_{M\ell}\sin\theta_{\ell}
    +r^2_{M\ell-1}\sin\theta_{\ell-1})(\theta_\ell-\theta_{\ell-1})}{
\sum_{\ell=2}^{N} \frac{1}{2}(\sin\theta_{\ell}
    +\sin\theta_{\ell-1})(\theta_\ell-\theta_{\ell-1})},
\Ea
where $N$ is the total grid number in the second dimension $\theta$. We have introduced the normalization factor
$[\sum_{\ell=2}^{N} \frac{1}{2}(\sin\theta_{\ell}+\sin\theta_{\ell-1})(\theta_\ell-\theta_{\ell-1})]^{-1}$ to remove the discrete error. The other three integrals do not need the normalization factor:
\Ba
\alpha^i_m &=& \frac{ \sum_{\ell=2}^{N} \frac{1}{2}(r^2_{i\ell}\rho_{i\ell}\alpha_{i\ell}\sin\theta_{\ell}
    +r^2_{i\ell-1}\rho_{i\ell-1}\alpha_{i\ell-1}\sin\theta_{\ell-1})(\theta_\ell-\theta_{\ell-1})}{
\sum_{\ell=2}^{N} \frac{1}{2}(r^2_{i\ell}\rho_{i\ell}\sin\theta_{\ell}
    +r^2_{i\ell-1}\rho_{i\ell-1}\sin\theta_{\ell-1})(\theta_\ell-\theta_{\ell-1})}, \\
\delta^i_m &=& \frac{ \sum_{\ell=2}^{N} \frac{1}{2}(r^2_{i\ell}\rho_{i\ell}\delta_{i\ell}\sin\theta_{\ell}
    +r^2_{i\ell-1}\rho_{i\ell-1}\delta_{i\ell-1}\sin\theta_{\ell-1})(\theta_\ell-\theta_{\ell-1})}{
\sum_{\ell=2}^{N} \frac{1}{2}(r^2_{i\ell}\rho_{i\ell}\sin\theta_{\ell}
    +r^2_{i\ell-1}\rho_{i\ell-1}\sin\theta_{\ell-1})(\theta_\ell-\theta_{\ell-1})}, \\
T_{\s{eff}}^4 &=& \frac{ \sum_{\ell=2}^{N} \frac{1}{2}(r^2_{M\ell}T^4_{M\ell}\sin\theta_{\ell}
    +r^2_{M\ell-1}T^4_{M\ell-1}\sin\theta_{\ell-1})(\theta_\ell-\theta_{\ell-1})}{
\sum_{\ell=2}^{N} \frac{1}{2}(r^2_{M\ell}\sin\theta_{\ell}
    +r^2_{M\ell-1}\sin\theta_{\ell-1})(\theta_\ell-\theta_{\ell-1})},
\Ea
because they have already had their own normalization factors.

\subsection{Interior points}

\subsubsection{w=P}\label{sec:wp}

There are three blocks in this group. They are:
\begin{description}
  \item{(I)} l = i-1, k = j;
  \item{(II)} l = i, k = j;
  \item{(III)} l = i, k = j-1.
\end{description}
We present the results one block by one block using the derivatives given above.

\begin{description}
\item{Block I:}
\begin{eqnarray*}
\p{F^{ij}_P}{R'_{i-1j}} &=& -\frac{1}{2}\Delta s_i \partial_R\h{P}_{i-1j} 
 -\frac{1}{2}\Delta s_i\sum_{\ell=1}^{3,12,15,17}\partial_R\h{B}^{\ell}_{i-1j}\cdot\h{D}^{\ell} \\
\p{F^{ij}_P}{L_{i-1j}} &=& 0 \\
\p{F^{ij}_P}{P'_{i-1j}} &=& -\frac{1}{2}\Delta s_i \partial_P\h{P}_{i-1j}-1  \nob
&& - \frac{1}{2} \Delta s_i \left[\sum_{\ell=1,2,3,10}^{15,17} \partial_P\h{B}^{\ell}_{i-1j}\cdot\h{D}^{\ell} 
 + \sum_{\ell=10,11,13,14}(\h{B}^{\ell}_{i-1j}+\h{B}^{\ell}_{ij})\partial_{P_{i-1j}}\h{D}^{\ell}\right]  \\
\p{F^{ij}_P}{T'_{i-1j}}&=&  -\frac{1}{2}\Delta s_i \partial_T\h{P}_{i-1j} \\
&& -\frac{1}{2}\Delta s_i\left[\sum_{3,10,11,13,14}\partial_T\h{B}^{\ell}_{i-1j}\cdot\h{D}^\ell
 + \sum_{\ell=10,11,13,14}(\h{B}^{\ell}_{i-1j}+\h{B}^{\ell}_{ij})\partial_{T_{i-1j}}\cdot\h{D}^{\ell}\right]
\end{eqnarray*}

\item{Block II:}

\begin{eqnarray*}
\p{F^{ij}_P}{R'_{ij}} &=& -\frac{1}{2}\Delta s_i\partial_R\h{P}_{ij}  \\
&&  -\frac{1}{2}\Delta s_i \left[\sum_{\ell=1}^{3,12,15,17}\partial_R\h{B}^{\ell}_{ij}\cdot\h{D}^{\ell} 
  +  \sum_{\ell=1,2,3,10}^{15,17}(\h{B}^{\ell}_{i-1j}+\h{B}^{\ell}_{ij})\partial_{R_{ij}}\h{D}^{\ell}\right] \\
\p{F^{ij}_P}{L_{ij}} &=& 0 \\
\p{F^{ij}_P}{P'_{ij}} &=& -\frac{1}{2}\Delta s_i \partial_P\h{P}_{ij} +1  \\
&& - \frac{1}{2} \Delta s_i\left[\sum_{\ell=1,2,3,10}^{15,17} \partial_P\h{B}^{\ell}_{ij}\cdot\h{D}^{\ell}
  +             \sum_{\ell=3,10}^{15}(\h{B}^{\ell}_{i-1j}+\h{B}^{\ell}_{ij})\partial_{P_{ij}}\h{D}^{\ell}\right]  \\
\p{F^{ij}_P}{T'_{ij}} &=&  -\frac{1}{2}\Delta s_i \partial_T\h{P}_{ij}  \\
&& - \frac{1}{2} \Delta s_i\left[\sum_{\ell=3,10,11,13,14} \partial_T\h{B}^{\ell}_{ij}\cdot\h{D}^{\ell}
  +             \sum_{\ell=10}^{15}(\h{B}^{\ell}_{i-1j}+\h{B}^{\ell}_{ij})\partial_{T_{ij}}\h{D}^{\ell}\right]
\end{eqnarray*}

\item{Block III:}
\begin{eqnarray*}
\p{F^{ij}_P}{R'_{ij-1}} &=& -\frac{1}{2} \Delta s_i \sum_{\ell=1,2,3,10}^{15,17}(\h{B}^\ell_{i-1j}+\h{B}^\ell_{ij})\partial_{R_{ij-1}}\h{D}^\ell \\
\p{F^{ij}_P}{L_{ij-1}} &=& 0 \\
\p{F^{ij}_P}{P'_{ij-1}} &=& -\frac{1}{2} \Delta s_i \sum_{\ell=3,12,15}(\h{B}^{\ell}_{i-1j}+\h{B}^{\ell}_{ij})\partial_{P_{ij-1}}\h{D}^{\ell} \\
\p{F^{ij}_P}{T'_{ij-1}} &=&  -\frac{1}{2} \Delta s_i \sum_{\ell=12,15}(\h{B}^{\ell}_{i-1j}+\h{B}^{\ell}_{ij})\partial_{T_{ij-1}}\h{D}^{\ell} 
\end{eqnarray*}

\end{description}

\subsubsection{w=T}

There are three blocks in this group, too.

\begin{description}
\item{Block I:}
\begin{eqnarray*}
\p{F^{ij}_T}{R'_{i-1j}} &=& -\frac{1}{2}\Delta s_i \partial_R\h{T}_{i-1j}
  -\frac{1}{2}\Delta s_i \sum_{\ell=1,2,3,10}^{15,17} \partial_R\h{T}^\ell_{i-1j}\cdot\h{D}^\ell \\
\p{F^{ij}_T}{L_{i-1j}} &=& -\frac{1}{2}\Delta s_i \partial_L\h{T}_{i-1j} 
  -\frac{1}{2}\Delta s_i  \sum_{\ell=1,2,3,10}^{15,17}\partial_L\h{T}^\ell_{i-1j}\cdot\h{D}^\ell \\
\p{F^{ij}_T}{P_{i-1j}} &=& -\frac{1}{2}\Delta s_i \partial_P\h{T}_{i-1j}  \\
&&  -\frac{1}{2}\Delta s_i \left[\sum_{\ell=1,2,3,10}^{15,17}\partial_P\h{T}^\ell_{i-1j}\cdot\h{D}^\ell 
 + \sum_{\ell=10,11,13,14}(\h{T}^{\ell}_{i-1j}+\h{T}^{\ell}_{ij})\partial_{P_{i-1j}}\h{D}^{\ell}\right] \\
\p{F^{ij}_T}{T_{i-1j}} &=&-\frac{1}{2}\Delta s_i \partial_T\h{T}_{i-1j}-1 \\
&&  -\frac{1}{2}\Delta s_i \left[\sum_{\ell=1,2,3,10}^{15,17}\partial_T\h{T}^\ell_{i-1j}\cdot\h{D}^\ell 
 + \sum_{\ell=10,11,13,14}(\h{T}^{\ell}_{i-1j}+\h{T}^{\ell}_{ij})\partial_{T_{i-1j}}\h{D}^{\ell}\right] 
\end{eqnarray*}

\item{Block II:}
\begin{eqnarray*}
\p{F^{ij}_T}{R'_{ij}} &=& -\frac{1}{2}\Delta s_i \partial_R\h{T}_{ij} \\
&&  -\frac{1}{2}\Delta s_i \left[\sum_{\ell=1,2,3,10}^{15,17}\partial_R\h{T}^\ell_{ij}\cdot\h{D}^\ell 
 + \sum_{\ell=1,2,3,10}^{15,17}(\h{T}^{\ell}_{i-1j}+\h{T}^{\ell}_{ij})\partial_{R_{ij}}\h{D}^{\ell}\right]  \\
\p{F^{ij}_T}{L_{ij}} &=&  -\frac{1}{2}\Delta s_i \partial_L\h{T}_{ij}  
 -\frac{1}{2}\Delta s_i \sum_{\ell=1,2,3,10}^{15,17}\partial_L\h{T}^\ell_{ij}\cdot\h{D}^\ell \\
\p{F^{ij}_T}{P_{ij}} &=&  -\frac{1}{2}\Delta s_i \partial_P\h{T}_{ij}  \\
&&  -\frac{1}{2}\Delta s_i \left[\sum_{\ell=1,2,3,10}^{15,17} \partial_P\h{T}^\ell_{ij}\cdot\h{D}^\ell 
    +\sum_{\ell=3,10}^{15}(\h{T}^{\ell}_{i-1j}+\h{T}^{\ell}_{ij})\partial_{P_{ij}}\h{D}^{\ell} \right] \\
\p{F^{ij}_T}{T_{ij}} &=&  -\frac{1}{2}\Delta s_i \partial_T\h{T}_{ij}+1 \\
&&  -\frac{1}{2}\Delta s_i \left[\sum_{\ell=1,2,3,10}^{15,17} \partial_T\h{T}^\ell_{ij}\cdot\h{D}^\ell 
    +\sum_{\ell=10}^{15}(\h{T}^{\ell}_{i-1j}+\h{T}^{\ell}_{ij})\partial_{T_{ij}}\h{D}^{\ell} \right]
\end{eqnarray*}

\item{Block III:}
\begin{eqnarray*}
  \p{F^{ij}_T}{R'_{ij-1}} &=& -\frac{1}{2}\Delta s_i \sum_{\ell=1,2,3,10}^{15,17}(\h{T}^\ell_{i-1j}+\h{T}^\ell_{ij})\partial_{R_{ij-1}}\h{D}^\ell \\
  \p{F^{ij}_T}{L_{ij-1}} &=&  0  \\
\p{F^{ij}_T}{P'_{ij-1}} &=& -\frac{1}{2} \Delta s_i \sum_{\ell=3,12,15}(\h{T}^{\ell}_{i-1j}+\h{T}^{\ell}_{ij})\partial_{P_{ij-1}}\h{D}^{\ell} \\
\p{F^{ij}_T}{T'_{ij-1}} &=&  -\frac{1}{2} \Delta s_i \sum_{\ell=12,15}(\h{T}^{\ell}_{i-1j}+\h{T}^{\ell}_{ij})\partial_{T_{ij-1}}\h{D}^{\ell} 
\end{eqnarray*}

\end{description}

\subsubsection{w=R}

In this group only the first two blocks are nonzero.
\begin{description}
\item{Block I:}
\begin{eqnarray*}
  \p{F^{ij}_R}{R'_{i-1j}} &=& -\frac{1}{2}\Delta s_i \partial_R\h{R}_{i-1j}-1 \\ 
  \p{F^{ij}_R}{L_{i-1j}} &=&  0 \\ 
  \p{F^{ij}_R}{P'_{i-1j}} &=& -\frac{1}{2}\Delta s_i \partial_P\h{R}_{i-1j} \\ 
  \p{F^{ij}_R}{T'_{i-1j}} &=& -\frac{1}{2}\Delta s_i \partial_T\h{R}_{i-1j} \\
\end{eqnarray*}

\item{Block II:}
\begin{eqnarray*}
  \p{F^{ij}_R}{R'_{ij}} &=& -\frac{1}{2}\Delta s_i \partial_R\h{R}_{ij} +1 \\
  \p{F^{ij}_R}{L_{ij}} &=&  0 \\
  \p{F^{ij}_R}{P'_{ij}} &=& -\frac{1}{2}\Delta s_i \partial_P\h{R}_{ij} \\
  \p{F^{ij}_R}{T'_{ij}} &=& -\frac{1}{2}\Delta s_i \partial_T\h{R}_{ij}
\end{eqnarray*}
\end{description}

\subsubsection{w = L}

Similarly, in this group only the first two blocks are nonzero.
\begin{description}
\item{Block I:}
\begin{eqnarray*}
\p{F^{ij}_L}{R'_{i-1j}} &=& -\frac{1}{2}\frac{\Delta s_i\cot\theta_j}{\Delta\theta_j} \left[
  \sum_{\ell=1}^2\partial_R\h{F}^\ell_{i-1j}\cdot(T'_{ij}-T'_{ij-1}) \right. \\
&& \left.  + \partial_R\h{F}^3_{i-1j}\cdot(P'_{ij}-P'_{ij-1})
  +\sum_{\ell=4}^6\partial_R\h{F}^\ell_{i-1j}\cdot(r'_{ij}-r'_{ij-1})       \right] \\
\p{F^{ij}_L}{L_{i-1j}} &=&  -1 -\frac{1}{2}\frac{\Delta s_i\cot\theta_j}{\Delta\theta_j} \partial_L\h{F}^4_{i-1j}\cdot(r'_{ij}-r'_{ij-1})\\
\p{F^{ij}_L}{P'_{i-1j}} &=& -\frac{1}{2}\Delta s_i \partial_P\h{L}_{i-1j} \\
&& -\frac{1}{2}\frac{\Delta s_i\cot\theta_j}{\Delta\theta_j}\left[\sum_{\ell=1}^2\partial_P\h{F}^\ell_{i-1j} 
  \cdot(T'_{ij}-T'_{ij-1}) \right.\\
&& \left. + \partial_P\h{F}^3_{i-1j}\cdot(P'_{ij}-P'_{ij-1}) +\sum_{\ell=4}^6 \partial_P\h{F}^\ell_{i-1j}\cdot(r'_{ij}-r'_{ij-1})  \right]  \\
\p{F^{ij}_L}{T'_{i-1j}} &=& -\frac{1}{2}\Delta s_i \partial_T\h{L}_{i-1j} \\
&& -\frac{1}{2}\frac{\Delta s_i\cot\theta_j}{\Delta\theta_j}\left[\sum_{\ell=1}^2\partial_T\h{F}^\ell_{i-1j} 
  \cdot(T'_{ij}-T'_{ij-1}) \right.\\
&& \left. + \partial_T\h{F}^3_{i-1j}\cdot(P'_{ij}-P'_{ij-1}) +\sum_{\ell=4}^6 \partial_T\h{F}^\ell_{i-1j}\cdot(r'_{ij}-r'_{ij-1})  \right]  
\end{eqnarray*}

\item{Block II:}
\begin{eqnarray*}
\p{F^{ij}_L}{R'_{ij}} &=& -\frac{1}{2}\frac{\Delta s_i\cot\theta_j}{\Delta\theta_j} \left[
  \sum_{\ell=1}^2\partial_R\h{F}^\ell_{ij}\cdot(T'_{ij}-T'_{ij-1}) \right. \\
&&\left. + \partial_R\h{F}^3_{ij}\cdot(P'_{ij}-P'_{ij-1})  
  +\sum_{\ell=4}^6 [\partial_R\h{F}^\ell_{ij}\cdot(r'_{ij}-r'_{ij-1}) 
 +(\h{F}^{\ell}_{ij}+\h{F}^{\ell}_{i-1j})\cdot\delta_R] \right]  \\
\p{F^{ij}_L}{L_{ij}} &=&  1-\frac{1}{2}\frac{\Delta s_i\cot\theta_j}{\Delta\theta_j} \partial_L\h{F}^4_{ij}\cdot(r'_{ij}-r'_{ij-1}) \\
\p{F^{ij}_L}{P'_{ij}} &=& -\frac{1}{2}\Delta s_i \partial_P\h{L}_{ij} \\
&& -\frac{1}{2}\frac{\Delta s_i\cot\theta_j}{\Delta\theta_j} \left[
  \sum_{\ell=1}^2\partial_P\h{F}^\ell_{ij}\cdot(T'_{ij}-T'_{ij-1}) \right. \\
&& + \partial_P\h{F}^3_{ij}\cdot(P'_{ij}-P'_{ij-1}) +(\h{F}^3_{ij}+\h{F}^3_{i-1j})\cdot\delta_P 
 \left. +\sum_{\ell=4}^6 \partial_P\h{F}^\ell_{ij}\cdot(r'_{ij}-r'_{ij-1}) \right]  \\
\p{F^{ij}_L}{T'_{ij}} &=& -\frac{1}{2}\Delta s_i \partial_T\h{L}_{ij} \\
&& -\frac{1}{2}\frac{\Delta s_i\cot\theta_j}{\Delta\theta_j} \left\{
  \sum_{\ell=1}^2[\partial_T\h{F}^\ell_{ij}\cdot(T'_{ij}-T'_{ij-1})+(\h{F}^{\ell}_{ij}+\h{F}^{\ell}_{i-1j})\cdot\delta_T]\right. \\
&&  + \partial_T\h{F}^3_{ij}\cdot(P'_{ij}-P'_{ij-1}) 
\left. + \sum_{\ell=4}^6 \partial_T\h{F}^\ell_{ij}\cdot(r'_{ij}-r'_{ij-1}) \right\} 
\end{eqnarray*}
where $\delta_P=1$ when $P_{ij}\ne P_{ij-1}$, and $\delta_P=0$ when $P_{ij}=P_{ij-1}$. The definition of $\delta_T$ is similar.

\item{Block III:}
\begin{eqnarray*}
\p{F^{ij}_L}{R'_{ij-1}} &=& \frac{1}{2}\frac{\Delta s_i\cot\theta_j}{\Delta\theta_j} \sum_{\ell=4}^6(\h{F}^\ell_{ij}+\h{F}^\ell_{i-1j})\cdot\delta_R  \\
\\
\p{F^{ij}_L}{L_{ij-1}} &=&  0 \\
\p{F^{ij}_L}{P'_{ij-1}} &=& \frac{1}{2}\frac{\Delta s_i\cot\theta_j}{\Delta\theta_j} (\h{F}^3_{ij}+\h{F}^3_{i-1j})\cdot\delta_P  \\
\p{F^{ij}_L}{T'_{ij-1}} &=& \frac{1}{2}\frac{\Delta s_i\cot\theta_j}{\Delta\theta_j} \sum_{\ell=1}^2(\h{F}^\ell_{ij}+\h{F}^\ell_{i-1j})\cdot\delta_T
\end{eqnarray*}
where $\delta_R=1$ when $r_{ij}\ne r_{ij-1}$, and $\delta_R=0$ when $r_{ij}=r_{ij-1}$. 
\end{description}

\subsection{Boundary points}

\subsubsection{Center: $\omega=r$}

Central boundary points have only block II for $w=r$:
\begin{eqnarray*}
  \p{F^{1j}_R}{R'_{1j}} &=& 1 \\
  \p{F^{1j}_R}{L_{1j}} &=&  0 \\ 
  \p{F^{1j}_R}{P'_{1j}} &=& \frac{1}{3}\alpha_{m1j} \\
  \p{F^{1j}_R}{T'_{1j}} &=& -\frac{1}{3}\delta_{m1j}
\end{eqnarray*}

\subsubsection{Center: $\omega=L$}

Central boundary points have blocks II and III for $\omega=L$:
\begin{description}
\item{Block II:}
\begin{eqnarray*}
\p{F^{1j}_L}{R'_{1j}} &=& -\frac{\cot\theta_j}{\Delta\theta_j} \left\{
  \sum_{\ell=1}^2\partial_R\h{F}^\ell_{1j}\cdot(T'_{1j}-T'_{1j-1}) \right. \\
&&\left. + \partial_R\h{F}^3_{1j}\cdot(P'_{1j}-P'_{1j-1})  
  +\sum_{\ell=4}^6 [\partial_R\h{F}^\ell_{1j}\cdot(r'_{1j}-r'_{1j-1}) 
 +\h{F}^{\ell}_{1j}\cdot\delta_R] \right\}  \\
\p{F^{1j}_L}{L_{1j}} &=&  1 \\
\p{F^{1j}_L}{P'_{1j}} &=& -\partial_P\h{L}_{1j} \\
&& -\frac{\cot\theta_j}{\Delta\theta_j} \left[
  \sum_{\ell=1}^2\partial_P\h{F}^\ell_{1j}\cdot(T'_{1j}-T'_{1j-1}) \right. \\
&& + \partial_P\h{F}^3_{1j}\cdot(P'_{1j}-P'_{1j-1}) +\h{F}^3_{1j}\cdot\delta_P 
 \left. +\sum_{\ell=4}^6 \partial_P\h{F}^\ell_{1j}\cdot(r'_{1j}-r'_{1j-1}) \right]  \\
\p{F^{1j}_L}{T'_{1j}} &=& -\partial_T\h{L}_{1j} \\
&& -\frac{\cot\theta_j}{\Delta\theta_j} \left\{
  \sum_{\ell=1}^2[\partial_T\h{F}^\ell_{1j}\cdot(T'_{1j}-T'_{1j-1})+\h{F}^{\ell}_{1j}\cdot\delta_T]\right. \\
&&  + \partial_T\h{F}^3_{1j}\cdot(P'_{1j}-P'_{1j-1}) 
\left. + \sum_{\ell=4}^6 \partial_T\h{F}^\ell_{1j}\cdot(r'_{1j}-r'_{1j-1}) \right\} 
\end{eqnarray*}

\item{Block III:}
\begin{eqnarray*}
\p{F^{1j}_L}{R'_{1j-1}} &=& \frac{\cot\theta_j}{\Delta\theta_j} \sum_{\ell=4}^6\h{F}^\ell_{1j}\cdot\delta_R  \\
\\
\p{F^{1j}_L}{L_{1j-1}} &=&  0 \\
\p{F^{1j}_L}{P'_{1j-1}} &=& \frac{\cot\theta_j}{\Delta\theta_j} \h{F}^3_{1j}\cdot\delta_P  \\
\p{F^{1j}_L}{T'_{1j-1}} &=& \frac{\cot\theta_j}{\Delta\theta_j} \sum_{\ell=1}^2\h{F}^\ell_{1j}\cdot\delta_T
\end{eqnarray*}

\end{description}

\subsubsection{Surface}

\paragraph{Standard}
The surface boundary conditions are linearized as follows:
\Ba
\delta r'_{Mj} +0\cdot \delta L_{Mj} - a_1 \delta P'_{Mj} - a_2 \delta T'_{Mj} 
  &=& - F_R^{M+1j}, \\
0\cdot \delta r'_{Mj} +\delta L_{Mj} - L_{Mj}\cdot a_4 \delta P'_{Mj} - 
  L_{Mj}a_5 \delta T'_{Mj} &=& - F_L^{M+1j}.
\Ea

\paragraph{Deupree's} His surface boundary equations are simpler:
\Ba
0\cdot \delta r'_{Mj} +0\cdot \delta L_{Mj} 
  + 1\cdot \delta P'_{Mj} +0\cdot \delta T'_{Mj} &=& - F_R^{M+1j}, \\
0\cdot \delta r'_{Mj} +0\cdot \delta L_{Mj} + 0\cdot \delta P'_{Mj} 
  +1\cdot \delta T'_{Mj} &=& - F_L^{M+1j},
\Ea
where
\Ba
   F_R^{M+1j} &=& P'_{M+1j} - P'_{\s{ref}}, \\
   F_L^{M+1j} &=& T'_{M+1j} - T'_{\s{ref}}.
\Ea

\subsubsection{Pole}

The polar boundary equations are extremely simple:
\Ba
   \delta P'_{i1}-\delta P'_{i2} &=& 0 \\
   \delta T'_{i1}-\delta T'_{i2} &=& 0 \\
   \delta r'_{i1}-\delta r'_{i2} &=& 0 \\
   \delta L_{i1}-\delta L_{i2} &=& 0
\Ea

\section{Input physics}

\subsection{The equations of state}\label{sec:eos}

When a magnetic field is present, the equation of state relates the density
$\rho$ to the pressure $P$, temperature $T$, magnetic energy per unit mass
$\chi$, the ratio of specific heats $\gamma$, and the chemical composition:
\[
  \rho=\rho(P_T,T,\chi;X,Z),
\]
where $P=P_0+P_r+P_m$ is the total pressure, $P_0$ the gas pressure,
$P_r=aT^4/3$ the radiative pressure, $P_m=\chi\rho$ the magnetic
pressure, $X$ the mass fraction of hydrogen, $Z$ the mass fraction of
elements heavier than helium (the so-called metal mass fraction). Its 
differential form is
\[
  \frac{d\rho}{\rho}=\alpha\frac{d P_T}{P}-\delta\frac{dT}{T}-\nu\frac{d\chi}{\chi},
\]
where
\begin{eqnarray*}
   \alpha &=& (\partial \ln\rho/\partial\ln P) \mbox{ at constant T, $\chi$}, \\
   \delta &=& -(\partial \ln\rho/\partial\ln T) \mbox{ at constant P, $\chi$}, \\
   \nu &=& -(\partial\ln \rho/\partial\ln \chi) \mbox{ at constant P, T},
\end{eqnarray*}
here $X$ and $Z$ are assumed to be constant.

Since it is tedious to accurately calculate the equation of state from
first principles, the equations of state are usually provided by the
numerical tables as functions of $(\rho,T,X,Z)$ for $P_0$, $S$ (entropy),
$U$ (internal energy), $(\partial U/\partial\rho)_T$, $c_v=(\partial
U/\partial T)_\rho$, $\chi_\rho=(\partial\ln P_0/\partial \rho)_T$,
$\chi_T=(\partial\ln P_0/\partial T)_\rho$, $\Gamma_1=(\partial\ln
P_0/\partial\ln\rho)_S$,
$\Gamma'_2=\Gamma_2/(1-\Gamma_2)=1/\nabla_{\s{ad}}$, and
$\Gamma'_3=(\partial\ln T/\partial\ln\rho)_{P_0}-1$. The equation of state for the 
gas is taken from Rogers, Swenson, \&\ Iglesias (1996). In order to take into
account a magnetic field based on the EOS tables, one can use the following
correction method:
\begin{description}
\item{(i)} Using the total pressure $P=P_0+P_r+P_m$, the total internal energy
$U=U_0+3P_r/\rho+\chi$, and the total entropy $S=S_0+4P_r/\rho T+\chi/T$ to
replace the gas pressure $P_0$, the gas internal energy $U_0$, and the gas
entropy $S_0$ respectively when interpolating to obtain the density for the
given $P$ and $T$;
\item{(ii)} Using $(P_0+P_m)/P$ to rescale $\chi_\rho$;
\item{(iii)} Using $P_0/P$ to rescale $\chi_T$ from the EOS tables and add
$4P_r/P$;
\item{(iv)} Adding $12P_r/T$ to $c_v$ from the EOS tables;
\item{(v)} Computing $\Gamma'_3=P\chi_T/c_v\rho T$,
$\Gamma_1=\chi_\rho+\chi_T\Gamma'_3$, and $\Gamma'_2=\Gamma_1/\Gamma'_3$.
\end{description}
Taking these as known, we can calculate
\[
 \alpha = 1/\chi_\rho, \hspace{3mm} \delta = \chi_T/\chi_\rho, \hspace{3mm} \nu = P_m/P, \hspace{3mm} 
\nabla_{\s{ad}}=1/\Gamma'_2, \hspace{3mm} c_p=P \delta/\rho T \nabla_{\s{ad}}.
\]
 These quantities are used in calculating the convective
gradient $\nabla_c$.

\subsection{Energy generation}\label{sec:energy}

The calculation of the energy generation includes the individual rates for
the PP-chain (PPI, PPII, PPIII), the CNO-cycle with a simplified NO approach
to equilibrium. The coefficients of all of the reaction rates and the
formulae for most of them are taken from Fowler, Caughlan and Zimmerman
(1975).

The reaction rate for the PP-chain is actually that for the H$^1$(p,
e$^+\nu$)D$^2$ reaction and assumes that all the other reactions in the
chain are relatively instantaneous. The burning rate is then
\Ba
   (dX/dt)_{\s{PP}}&=&4.181\cdot 10^{-15}\rho X^2 T_9^{-2/3}
\exp(-3.380/T_9^{1/3}) \\
  && \cdot\phi(\alpha')\cdot(1.0+0.123T_9^{1/3}+1.09T_9^{2/3}+0.938 T_9) \hspace{2mm}
\mbox{  sec$^{-1}$}, \nonumber
\Ea
where $T_9=T/10^9\,^\circ$\,K, the screening factor $f_s$ is set equal to 1,
\begin{eqnarray*}
\phi(\alpha') &=& 1+\alpha'[(1+2/\alpha')^{1/2}-1] \\
\alpha' &=& 1.93\cdot10^{17}(Y/2X)^2\exp(-10.0/T_9^{1/3}).
\end{eqnarray*}
The total energy of the PP-chain (subtracting the energy of the neutrinos
which are produced) is
\[
  \epsilon_{\s{PP}}=6.398\cdot10^{18}\psi(dX/dt)_{\s{PP}} \hspace{2mm}\mbox{
erg/gm/sec},
\]
where
\begin{eqnarray*}
  \psi &=& 0.979 f_{\s{I}} + 0.960 f_{\s{II}} + 0.721 f_{\s{III}} \\
  f_{\s{I}} &=& [(1+2/\alpha')^{1/2}-1]/[(1+2/\alpha')^{1/2}+3] \\
  f_{\s{II}} &=& (1-f_{\s{I}})/(1+\Gamma) \\
  f_{\s{III}} &=& 1-f_{\s{I}} - f_{\s{II}} \\
  \Gamma &=& 10^{15.6837} [X/(1+X)]T_9^{-1/6}\exp(-10.262/T_9^{1/3}).
\end{eqnarray*}

The derivatives of $\epsilon_{\s{PP}}$ can be found directly:
\Ba
(\partial \ln \epsilon_{\s{PP}}/\partial \ln\rho)_T &=& \epsilon_{\s{PP}} \\
(\partial \ln \epsilon_{\s{PP}}/\partial \ln T)_\rho &=& \epsilon_{\s{PP}}[-2/3
  +1.1267/T_9^{1/3}+(\partial \ln\phi/\partial \ln T)_\rho 
  +(\partial \ln\psi/\partial \ln T)_\rho \nonumber\\
&&     + (0.041T_9^{1/3}+0.727 T_9^{2/3} +0.938 T_9) \nonumber \\
&&     /(1+0.123T_9^{1/3}+1.09T_9^{2/3}+0.938T_9)] \nonumber \\
(\partial \ln\phi/\partial \ln T)_\rho &=& (2/\phi-1)(1+2/\alpha')^{-1/2}3.333/T^{1/3} \nonumber \\
(\partial \ln\psi/\partial \ln T)_\rho &=& \psi^{-1}\{[0.258-0.239/(1+\Gamma)] 
  (\partial f_{\s{I}}/\partial \ln T)_\rho \nonumber\\
&& -0.239f_{\s{III}}/(1
  +\Gamma)(\partial \ln\Gamma/\partial \ln T)_\rho \} \nonumber \\
(\partial \ln\Gamma/\partial \ln T)_\rho &=& -1/6+3.4207/T_9^{1/3} \nonumber \\
(\partial f_{\s{I}}/\partial \ln T)_\rho &=& -4\{\alpha'(1+2/\alpha')^{1/2}
  [(q+2/\alpha')^{1/2} + 3]^2\}^{-1} \cdot 3.333/T_9^{1/3} \nonumber
\Ea
In the calculation of the CNO bi-cycle, CN equilibrium is assumed and the CN cycle 
is assumed to be the only source of energy. The hydrogen-burning rate due to the CN 
cycle is then
\begin{eqnarray*}
 (dX/dt)_{\s{CN}} = 1.202\cdot10^7\rho X X_{\s{N}} T_9^{-2/3} \exp(-15.228/T_9^{1/3})   \mbox{ sec$^{-1}$}
\end{eqnarray*}
and the energy produced is 
\[
  \epsilon_{\s{CN}}=5.977\cdot10^{18} (dX/dt)_{\s{CN}} \mbox{ erg/gm/sec}.
\]
The value of X$_{\s{N}}$ (N$^{14}$ abundance by weight) assumes that all 
the carbon and nitrogen is in the form of N$^{14}$,
\[
  X_{\s{N}}=Z-Z_{\s{m}}-X_{\s{O}},
\]
where Z is the total metal abundance by weight, Z$_{\s{m}}$ is 
the weight abundance of all non-CNO metals, and X$_{\s{O}}$ is 
the weight abundance of O$^{16}$. The approach to NO equilibrium 
is taken as a simple burning rate of O$^{16}$ assuming O$^{17}$ 
equilibrium,
\Ba
 dX_{\s{O}}/dt=9.54\cdot10^7\rho X X_{\s{O}} T_9^{-17/21} \exp(-16.693/T_9^{1/3}) \\
   -1.6\cdot10^{-3}(dX/dt)_{\s{CN}} \mbox{ sec$^{-1}$} \nonumber
\Ea
Between successive models the value of X$_{\s{O}}$ is decreased 
at a rate of (dX$_{\s{O}}$/dt) per second, and thus the value 
of X$_{\s{N}}$ is correspondingly increased. Here are the 
derivatives of the CN energy production:
\begin{eqnarray*}
 (\partial\epsilon_{\s{CN}}/\partial\ln\rho)_T &=& \epsilon_{\s{CN}}  \\
 (\partial\epsilon_{\s{CN}}/\partial\ln T)_\rho &=& \epsilon_{\s{CN}}(-2/3+5.076/T_9^{1/3} .
\end{eqnarray*}

\subsection{Radiative opacities}

An estimate of magnetic effects on the radiative opacities [$\kappa=\kappa(T,\rho,X,Z)$] 
can be found in Li and Sofia (2001). Since they are small, we use only the OPAL opacities 
tables (Iglesias and Rogers 1996) together with the low-temperature opacities from Alexander 
and Ferguson (1994). For X and Z the linear interpolation is used, but for $T$ and $\rho$ the 
cubic splint interpolation is used. The cubic splint interpolation scheme allows one to 
obtain the derivatives of $\kappa$ with respect to $T$ and $\rho$. These derivatives are 
needed in the linearization of the equations of energy transport.

\subsection{The convective gradient and its linearization}\label{sec:tg}

The calculation of the convective temperature gradient $\nabla_c$ in the envelope of 
the stellar models employs the mixing length theory (Henyey et al. 1965; Lydon and 
Sofia 1995) when magnetic fields are taken into account.

Defining $\delta'\equiv \nabla_{\s{rad}}-\nabla'_{\s{ad}}$, the Schwarzschild 
(1906) criterion is used to determine convection: $\delta>0$ means convective. 
In the deep interior convection zones $\nabla_c$ is set  equal to the adiabatic 
gradient $\nabla'_{\s{ad}}$.

In the envelope the evaluation of $\nabla_c$ is more complex, and we solve
\[
   F(x) \equiv a_3x^3+x^2+a_1 x-1=0
\]
for $y>0$ such that $F(y)=0$, where
\Ba
  a_1 &\equiv& V = \alpha_3\phi(\kappa T^3/C_p)(H_P/g\delta\delta')^{1/2} \\ 
  a_3 &\equiv& \frac{3}{4}\phi \omega^2/V.
\Ea
We have defined $\delta'\equiv (\nabla_{\s{rad}}-\nabla'_{\s{ad}})$, 
$\phi\equiv (1+\frac{1}{3}\omega^2)^{-1}$, $\omega=\rho\kappa l_m$, 
and $\alpha_3\equiv  16\sqrt{2}\,\sigma$.
The root $y$ is guaranteed to lie in the interval $(0,+1)$ since $F(0)=-1<0$ 
and $F(1)=a_1+a_2>0$. Further, this root $y$ is unique since the derivative of $F$, 
$F'(x)=3a_3 x^2+2x+a_1$, is positive definite for $x>0$. An initial estimate of 
the root $y$ is made and a second-order Newton-Raphson correction is applied:
\[
  \Delta y = -F(y)/F'(y)-\frac{1}{2}[F(y)/F'(y)]^2\cdot F''(y)/F'(y).
\]
The initial estimate of $y$ is $y=1/a_1$, unless $a_3>10^3$ in which case 
$y=(1/a_3)^{1/3}$ that follows the asymptotic behavior of the solution in 
either limit. Given the solution $y$, the convective gradient is computed:
\[
  \nabla_c =\nabla'_{\s{ad}} + (\nabla_{\s{rad}}-\nabla'_{\s{ad}})y(y+a_1).
\]

The linearization of the convective gradient is cumbersome but can be calculated. 
We consider derivatives with respect to $\ln P_T$, $\ln T$, $\ln R$ and $L$.
\[
   \p{\nabla_c}{\ln x} = \p{\nabla_{\s{ad}}}{\ln x} + y(y+a_1)\p{\delta'}{\ln x}
     + \delta \left[ (2y+a_1)\p{y}{\ln x} + a_1 y\p{\ln a_1}{\ln x}\right],
\]
where
\[
 \p{\delta'}{\ln x}= \p{\nabla'_{\s{rad}}}{\ln x} - \p{\nabla_{\s{ad}}}{\ln x}.
\]
The derivatives of $\nabla'_{\s{ad}}$ come from the equation of state, and are 
non-vanishing only for $x=P_T$ or $T$. The derivatives of 
\[
  \nabla_{\s{rad}}=(3/16\pi ac)(\kappa L L_\sun P_T)/(GMT^4)
\]
are non-vanishing for $x=P_T$, $T$, or $L$,
\Ba
 \p{\nabla_{\s{rad}}}{\ln P_T} &=& \nabla_{\s{rad}} \left(1+\p{\ln \kappa}{\ln P_T}\right)_T \\
 \p{\nabla_{\s{rad}}}{\ln P_T} &=& \nabla_{\s{rad}} \left(-4+\p{\ln \kappa}{\ln T}\right)_P \\
 \p{\nabla_{\s{rad}}}{L} &=& \nabla_{\s{rad}}/L.
\Ea
In the radiative zone, the actual temperature gradient is equal to the radiative temperature gradient 
$\nabla=\nabla_{\s{rad}}$, its derivatives are given here. The opacity tables provide $\log\kappa$
vs. ($\log\rho$, $\log T$). In order to calculate
\begin{eqnarray*}
(\partial{\ln\kappa}/\partial{\ln T})_P &=& (\partial{\ln\kappa}/\partial{\ln T})_\rho 
  +(\partial{\ln\kappa}/\partial{\ln\rho})_T (\partial{\ln\rho}/\partial{\ln T})_P, \\
(\partial{\ln\kappa}/\partial{\ln P})_T &=& (\partial{\ln\kappa}/\partial{\ln\rho})_T 
  (\partial{\ln\rho}/\partial{\ln T})_P,
\end{eqnarray*}
one needs $(\partial\ln\kappa/\partial\ln T)_\rho$ and
$(\partial\ln\kappa/\partial\ln\rho)_T$ (see Iglesias \&\ Rogers 1996, and Alexander \&\ Ferguson 1994). The derivatives of $y$ are functions of $a_1$ and $a_3$,
\[
   \p{y}{\ln x} = -\frac{1}{F'(y)}\left(a_1y\p{\ln a_1}{\ln x} + a_3 y^3\p{a_3}{\ln x}\right),
\]
which need the derivatives of $a_1$ and $a_3$,
\Ba
  \p{\ln a_1}{\ln x} &=& \p{\ln\phi}{\ln x}+\p{\ln \kappa}{\ln x} + 3\p{\ln T}{\ln x} - \p{\ln C_p}{\ln x}
+\frac{1}{2}\p{\ln H_P}{\ln x} - \frac{1}{2\delta'}\p{\delta'}{\ln x} \\
&&  - \frac{1}{2}\p{\ln g}{\ln x}  - \frac{1}{2}\p{\ln \delta}{\ln x} \\
  \p{\ln a_3}{\ln x} &=& 2\p{\ln\omega}{\ln x} + \p{\ln\phi}{\ln x} - \p{\ln a_1}{\ln x}.
\Ea
The derivative of $\delta \equiv -(\partial \ln \rho/\partial T)_P$ and $C_p$ are computed by the 
equation of state. By calculating the derivative of $\phi$,
\[
  \p{\ln \phi}{\ln x} = -\frac{2}{3}\omega^2\phi \p{\ln\omega}{\ln x},
\]
and by expressing $H_P$ and $g$ explicitly, one can obtain
\Ba
   \p{\ln a_1}{\ln x} &=& -\frac{2}{3}\omega^2\phi \p{\ln\omega}{\ln x} 
                          -\p{\ln C_p}{\ln x}
       -\frac{1}{2\delta}\p{\delta}{\ln x} -\frac{1}{2}\p{\ln\delta}{\ln x}\\
    && -\frac{P_T}{\ln x} + 3\p{\ln T}{\ln x} + 2\p{\ln r}{\ln x} 
       + \p{\ln\kappa}{\ln x}
-\frac{1}{2}\p{\ln \rho}{\ln x} \\
  \p{\ln a_3}{\ln x} &=& 2\phi \p{\ln\omega}{\ln x} -\p{\ln a_1}{\ln x}.
\Ea
The derivatives of $\omega$ with respect to $P_T$, T, r are straightforward,
\Ba
  \p{\ln \omega}{\ln P_T} &=& 1 + \left(\p{\ln\kappa}{\ln P_T}\right)_T \\
  \p{\ln \omega}{\ln T} &=& \left(\p{\ln\kappa}{\ln T}\right)_P \\
  \p{\ln \omega}{\ln r} &=& 2.
\Ea

{}

\clearpage

\end{document}